\documentclass[12pt,tightenlines,eqsecnum,floatfix,superscriptaddress,showpacs,amssymb,amsmath,amsfonts,aps,altaffilletter,nofootinbib,longbibliography,showpacs]{revtex4-2}

\usepackage{graphicx}
\usepackage{graphicx,verbatim}
\usepackage{hyperref}
\usepackage{cancel}
\usepackage{amsfonts}
\usepackage{amssymb}
\usepackage{mathrsfs}  
\usepackage{multirow}
\usepackage{mathrsfs}
\usepackage{amssymb,latexsym}
\usepackage{psfrag}
\usepackage{accents}
\usepackage{caption}
\usepackage{subcaption}
\usepackage{placeins}

\usepackage{tikz}
\usetikzlibrary {decorations.pathmorphing}

\usepackage[colorinlistoftodos]{todonotes}
\usepackage{epstopdf}

\usepackage[colorinlistoftodos]{todonotes}
\usepackage{epstopdf}
\usepackage[latin1]{inputenc}
\usepackage{soul}
\usepackage{appendix}
\usepackage{amssymb}
\usepackage{calrsfs}
\usepackage{color}
\usepackage[mathscr]{euscript}

\usepackage{caption}
\newcommand{\ttouch}{\ensuremath{T_{\rm touch}}}

\newcommand{\Munit}[1][\,]{\ensuremath{#1\mathcal{M}}}

\def\be{\begin{equation}}
\def\ee{\end{equation}}
\def\ba{\begin{eqnarray}}
\def\ea{\end{eqnarray}}

\def\qo{\mathring{q}}

\def\omegao{\mathring\omega}

\def\SU(2){\rm SU(2)}
\def\su(2){\rm su(2)}

\def\D{\mathcal{D}}

\def\DH{\mathcal{H}}

\newcommand{\pb}[1]{\hbox{\lower0.5ex\hbox{${}_{\leftarrow}$}}\kern-1.9ex{#1}}

\def\Lie{\mathcal{L}}

\def\IHo{\Delta_{\circ}}

\def\qo{\mathring{q}}
\def\tq{\tilde{q}}
\def\tD{\tilde{D}}

\def\vo{\mathring{v}}
\def\lo{\mathring{\ell}}

\def\ello{\ell_{\circ}}

\def\rmd{\mathrm{d}}

\def\uk{\underline{k}}

\def\ko{\mathring{k}}
\def\uko{\mathring{\underline{k}}}

\def\IH{\Delta}

\def\G{\mathfrak{G}} %

\def\B{\mathfrak{B}} 
\def\a{\mathfrak{a}}

\def\lb{\bar{\ell}}
\def\nb{\bar{n}}
\def\t{\tilde}
\def\h{\hat}

\def\b{\bar}

\def\={\hat{=}}
\def\f{\frac}

\def\bM{\bar{M}}
\def\bg{\bar{g}}
\def\bD{\bar{D}}
\def\bkappa{\bar\kappa}
\def\bomega{\bar\omega}
\def\bq{\bar{q}}
\def\bIH{\bar\IH}
\def\hM{\hat{M}}
\def\hg{\hat{g}}
\def\hD{\hat{D}}
\def\hn{\hat{n}}
\def\hq{\hat{q}}
\def\be{\begin{equation}}
\def\ee{\end{equation}}
\def\ba{\begin{eqnarray}}
\def\ea{\end{eqnarray}}

\def\SO(3){\rm SO(3)}
\def\so(3){\rm so(3)}
\def\SO(4){\rm SO(4)}
\def\so(4){\rm so(4)}
\def\SO(1,4){\rm SO(1,4)}
\def\so(1,4){\rm so(1,4)}
\def\SU(2){\rm SU(2)}

\def\R{\mathcal{R}}

\newcommand{\pullback}[1]{\hbox{\lower0.5ex\hbox{${}_{\leftarrow}$}}\kern-1.9ex{#1}}
\newcommand{\pullbacklong}[1]{\hbox{\lower0.85ex\hbox{${}_{\longleftarrow}$}}\kern-3.0ex{#1}}
\newcommand{\pullbackllong}[1]{\hbox{\lower0.85ex\hbox{${}_{\longleftarrow\!\!-\!\!-\!\!-\!\!-}$}}\kern-6.4ex{#1}}

\def\Lie{\mathcal{L}}
\def\scri{\mathcal{I}}
\def\scrip{\scri^{+}}
\def\scrim{\scri^{-}}

\def\omegao{\mathring\omega}
\def\qo{\mathring{q}}

\def\vo{\mathring{v}}
\def\lo{\mathring{\ell}}

\def\rmd{\mathrm{d}}

\def\Lie{\mathcal{L}}
\def\R{\mathcal{R}}

\def\G{\mathfrak{G}}

\def\be{\begin{equation}}
\def\ee{\end{equation}}
\def\ba{\begin{eqnarray}}
\def\ea{\end{eqnarray}}

\def\SO(3){\rm SO(3)}
\def\so(3){\rm so(3)}
\def\SO(4){\rm SO(4)}
\def\so(4){\rm so(4)}
\def\SO(1,4){\rm SO(1,4)}
\def\so(1,4){\rm so(1,4)}
\def\SU(2){\rm SU(2)}

\def\map{\mathfrak{f}}

\def\b1{b_1}   
\def\o={\mathring{=}}

\begin{document}

\title[]{Quasi-Local Black Hole Horizons: Recent Advances}

\author{Abhay Ashtekar}
\address{Institute for Gravitation and the Cosmos \& Physics Department, Pennsylvania State University, University Park, PA 16802, U.S.A.}
\address{Perimeter Institute for Theoretical Physics, 31 Caroline St N, Waterloo, ON N2L 2Y5, Canada}
\email{ashtekar.gravity@gmail.com}
\author{Badri Krishnan}
\address{Institute for Mathematics, Astrophysics and Particle Physics, Radboud University, Heyendaalseweg 135, 6525 AJ Nijmegen, The Netherlands}
\address{Albert-Einstein-Institut, Max-Planck-Institut f\"ur Gravitationsphysik, Callinstra{\ss}e 38, 30167 Hannover, Germany}
\address{Leibniz Universit\"at Hannover, 30167 Hannover, Germany}
\email{badri.krishnan@ru.nl}

\begin{abstract} 

  While the early literature on black holes focused on event horizons,
  subsequently it was realized that their teleological nature makes
  them unsuitable for many physical applications both in classical and
  quantum gravity. Therefore, over the past two decades, event
  horizons have been steadily replaced by quasi-local horizons which
  do not suffer from teleology. In numerical simulations event
  horizons can be located as an `after thought' only after the entire
  space-time has been constructed. By contrast, quasi-local horizons
  naturally emerge \emph{in the course of} these simulations,
  providing powerful gauge-invariant tools to extract physics from the
  numerical outputs. They also lead to interesting results in
  mathematical GR, providing unforeseen insights. For example, for
  event horizons we only have a qualitative result that their area
  cannot decrease, while for quasi-local horizons the increase in the
  area during a dynamical phase is quantitatively related to
  \emph{local} physical processes at the horizon. In binary black hole
  mergers, there are interesting correlations between observables
  associated with quasi-local horizons and those defined at future
  null infinity. Finally, the quantum Hawking process is naturally
  described as formation and evaporation of a quasi-local
  horizon. This article focuses on the \emph{dynamical} aspects of quasi-local horizons in
  classical general relativity, emphasizing recent results and ongoing
  research.
  
\end{abstract}

\maketitle
\tableofcontents

\section{Introduction} 
\label{s1}

\subsection{Preamble}
\label{s1.1} 
Thanks to the advent of global methods, research in general relativity (GR) had a `renaissance' that began in the 1960s. Among advances  that launched the new epoch, frameworks describing black holes (BHs) and gravitational waves (in full non-linear GR) stand out because they continue to serve as engines driving research at the forefront of the field. In particular, research on BHs dominated several areas of gravitational physics because the mathematical theory \cite{Hawking:1971vc,HawkingEllis:1973,LesHouches:1973} turned out to be extraordinarily rich and full of surprises.  Laws of BH mechanics brought out deep and completely unforeseen connections between classical GR, quantum physics and statistical mechanics 
\cite{Hawking:1971vc,Bardeen:1973gs,Bekenstein:1973ur,Bekenstein:1974ax}. They provided concrete challenges to quantum gravity that drove advances in that area over several decades. In the classical theory, BH uniqueness theorems came as a surprise since they imply that stationary BHs are extraordinarily simple compared to, say, stationary stars (for reviews, see \cite{mh,Chruciel2012}). The subsequent analysis of the detailed properties of Kerr-Newman solutions \cite{bc} and perturbations thereof \cite{Chandrasekhar:1985kt} constituted a large fraction of research in the GR  in the seventies and eighties. The binary-black hole (BBH) problem also served to push the frontiers of GR at the interface of geometric analysis and numerical methods  that had been developed to explore Einstein's vacuum equations \cite{choptuik2015probingstrongfieldgravity}. 
 
In these developments, BHs were generally characterized by their event horizons (EHs). From the causal structure perspective, the use of EHs is very natural and the power of this notion is illustrated by the richness of results it led to in the 1970s. However, as we discuss below, this notion also has some serious limitations. The most important { ones} stem from the fact that the notion is teleological; to determine its location one needs to know space-time geometry to the infinite future! It has been emphasized that one cannot make a priori assumptions about the nature of this geometry in explorations of fundamental issues such as cosmic censorship in classical general relativity \cite{Andersson:2007fh}, or the endpoint of the evaporation process of BHs in quantum gravity \cite{Ashtekar_2020}. Perhaps an even more striking example of limitations of EHs is seen in numerical simulations for BBH mergers. EHs cannot be used in the course of a simulation to locate where the BHs are, nor to say when the merger happens, since they can be found only at the end of the simulation. Because of these limitations of EHs, quasi-local horizons  (QLHs) were introduced over two decades ago (see, e.g., 
\cite{Hayward:1993wb,Ashtekar:1999yj,Ashtekar:2000sz,Ashtekar:2000hw,Ashtekar:2001jb,Ashtekar:2001is,Ashtekar:2002ag,Ashtekar:2003hk,Ashtekar:2004cn,Booth:2005qc,Gourgoulhon:2005ng,Jaramillo:2011zw,Hayward:2009ji}). 
Since they are free of teleology, they are routinely used in numerical simulations of BBH mergers; they led to physically more appropriate laws of BH mechanics; their properties have been analyzed using geometric analysis; and they are being used to investigate the BH evaporation process.  

With the discovery of gravitational { waves a decade ago}, research on BHs has entered the next level of intense activity as well as maturity that observations invariably bring to theoretical investigations. In this phase conceptual, mathematical and numerical issues related to BHs and null infinity  --where waveforms are defined-- have dominated research in GR. In this article we will discuss the role played by the QLHs in these developments. However, the literature on recent  developments itself is substantial, and the original notion of dynamical horizons 
\cite{Ashtekar:2002ag,Ashtekar:2003hk} has been superseded by a more convenient notion of dynamical horizon segments. Therefore, this is a self-contained summary of the current status rather than a traditional update of our previous Living Reviews article on  the subject \cite{Ashtekar:2004cn}. We will simply refer to the earlier reviews \cite{Ashtekar:2004cn,Booth:2005qc,Gourgoulhon:2005ng,Jaramillo:2011zw} that discuss the older material in detail at various junctures. In this article, the primary focus will be more recent findings, in particular:
\vskip0.05cm 
\indent (i) An interplay between geometric analysis and numerical relativity (NR) that has provided qualitatively new information about the dynamics of horizons of progenitors in BBH mergers. Contrary to a common belief that arose from investigations of apparent horizons, QLHs of the progenitors do \emph{not} abruptly jump at the merger to the common horizon surrounding them both. 
{In examples, where detailed numerical simulations are available,} the evolution is much more subtle and the horizons of the progenitors and of the remnant are in fact joined continuously.\\ 
\indent (ii) A close relation between the geometry of QLHs and of null infinity $\scri$ 
and a surprising similarity between expressions and properties of certain fundamental observables defined intrinsically at the QLHs and independently at $\scri$.
These results are analytic. \\
\indent (iii) Existence of correlations between physical fields characterizing the geometry of QLHs and waveforms at $\scrip$ for BBH mergers, despite the fact that no causal signal can propagate between the two! These relations were found by combining analytical and numerical methods.
\vskip0.1cm
\noindent However, before embarking on these recent developments, to make the review self-contained, we will first summarize the central ideas. This material may be of interest also to experts who are familiar with the older reviews because, as indicated above, there have been a few conceptual shifts. They provide a fresh perspective that could be useful for future work as well.

\subsection{From EHs to QLHs}
\label{s1.2}

\begin{figure}[]
  \begin{center}
    \includegraphics[width=0.9\columnwidth]{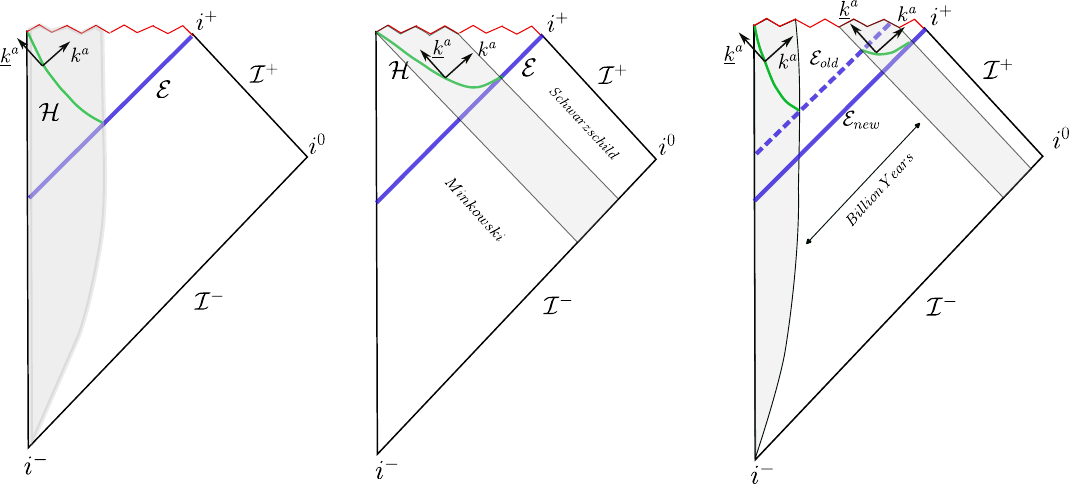}    
  \caption{\footnotesize{\emph{Left Panel: Penrose diagram of an Oppenheimer-Snyder (OS) stellar collapse.} The event Horizon (EH) $\mathcal{E}$ is the future boundary of the part of space-time from which signals can reach future null infinity $\scrip$. The {spherical} dynamical horizon (DH) $\mathcal{H}$ has zero radius at formation and grows {in the past direction} to join the EH.\,\,\, {\emph{Middle Panel: Collapse of a null fluid.}} The fluid (shaded region) is incident from $\scrim$. To the past of the null fluid, space-time is flat. The EH forms and grows already in this flat region although there is nothing happening there. DH forms in the fluid region and grows in area and joins the EH once all of the null fluid has fallen in.\,\,  {\emph{Right Panel: OS stellar collapse followed by a null fluid collapse, say, a billion years later.}} `New EH' is the actual EH of this space-time after the stellar \emph{and} the null fluid collapse while `Old EH' denotes the \emph{would be} EH, had there been no null fluid collapse in the distant future. DH grows in area only when there is flux of matter passing through it --first in the stellar region and then in the null fluid region. In the intervening billion years, it becomes an isolated horizon. {It has been shown \cite{Booth_2010} that the this isolated horizon segment is extremely close to the actual EH, a feature that is difficult to see because distances are not faithfully represented in Penrose diagrams.} }} 
\end{center}
\label{fig1}
\end{figure}

The Event Horizon Telescope (EHT) has provided us with striking images of light rings around two supermassive BHs. Observations of the LIGO-Virgo collaboration are even more impressive because they capture the dynamical phase of BHs. They have revealed that BHs of tens of solar masses are ubiquitous in our universe.  But how does the underlying analysis encode the basic idea that we are dealing with BHs?  What are the disjoint geometric entities that `merge' in a BBH coalescence? Or, more simply, what is the precise feature of space-time geometry that encapsulates the idea that we are dealing with a BH? The common answer is:\, A BH is characterized by an \emph{Event Horizon}(EH). Much of the rationale behind this conviction comes from the fact that this was the underlying scenario that led to the first set of remarkable results in the 1970s, mentioned above. Consequently, EHs are often taken to be synonymous with BHs.  

However, EHs are rather enigmatic. First, they require the presence of an asymptotic region with future infinity, $\scrip$. The EH is the future boundary of the causal past $J^{-}(\scrip)$ of $\scrip$. Consequently, it encloses the BH region $\mathcal{B}$, represented by the complement of $J^{-}(\scrip)$ in the physical space-time. These definitions seem natural because they neatly encapsulate the idea that no causal signals from $\mathcal{B}$ can reach future infinity $\scrip$ (see for example the left panel of Fig. 1 depicting an Oppenheimer-Snyder (OS) stellar collapse). However, they immediately imply that BHs are teleological!

This teleology is illustrated in the middle panel of Fig.~1, representing a Vaidya space-time depicting the spherical collapse of a null fluid incident from $\scrim$. In this case the space-time metric $g_{ab}$ in the triangular region to the past of the null fluid (depicted by the shaded region) is flat, it becomes dynamical in the shaded region, and finally it is a Schwarzschild metric to the future of that region. Note that EH \emph{forms and grows} in the flat region --where nothing  is happening at all-- \emph{in anticipation} of a collapse that is to occur occur sometime in the distant future! Recent results of Kehle and Unger show that the same phenomenon occurs if one uses Vlasov fluids (that emerge from $i^-$ rather than $\scrim$, see Fig. 1 of \cite{kehle2024extremalblackholeformation}); null fluids incident from $\scrim$ are not essential. The right panel of Fig.~1 depicts an OS stellar collapse, followed by a null fluid collapse, say a billion years later {(for details, see \cite{Booth_2010})}. It shows that the location of the EH is shifted in anticipation of a null fluid collapse which is to occur a billion years later. {The shift can be significant during the late phase of the second collapse if the mass in the collapsing fluid is much larger than the mass of the initial star.}
\footnote{The situation with two well separated collapses is more dramatic in presence of a  positive cosmological constant. If the mass at the end of the fist collapse is less than {\smash{$1/3G \sqrt{\Lambda}$} }, the DHS formed during the first collapse again joins on to an IHS which coincides with the would be $\mathcal{E}_{\rm old}$. There are trapped surfaces to the future of the DHS and the IHS. But if the mass at the end of the second collapse then exceeds {\smash{$1/3G \sqrt{\Lambda}$ }}, then there is no EH at all (because $\scrip$ does not exist (although $\scrim$ does)! See Fig. 3 in \cite{Senovilla:2022bsn}.} 

As a general rule such teleology is regarded as spooky, and avoided studiously in physics. For EHs, it occurs because the notion is unreasonably global; one needs to know the space-time structure all the way to the infinite future to determine if it admits an EH. Consequently, as noted above, EHs cannot be used to track the dynamics of the progenitors or the final remnant \emph{during} numerical simulations; EHs can only be identified at the \emph{end} of the simulation, as an after thought. 

Now, the theory of gravitational waves also uses $\scrip$, but there $\scrip$ serves only to specify the appropriate boundary conditions, whence there is no teleology. In the case of EHs, the role is qualitatively different because it refers not just to a small neighborhood of $\scrip$ but to its entire causal past. As a result, as Hajicek has shown, there are models (introduced to study BH evaporation) in which the space-time structure can be manipulated in a Planck size neighborhood of the singularity so that the EH is completely removed \cite{PhysRevD.36.1065}! Since there is every reason to expect that quantum gravity effects will modify the space-time geometry in such a neighborhood, it is far from clear that there will be  EHs in space-times depicting BH evaporation. One would need complete space-time to settle this question, and that can only be provided by a satisfactory quantum theory of gravity. In fact in the GR-17 conference in Dublin, Hawking asserted that ``a true event horizon never forms" in such space-times. 

Over the last two decades, these limitations led the community to replace EHs by QLHs that refer to the space-time geometry just in their vicinity, and are therefore free of teleological pathologies. This transition has led to a number of novel insights in all areas of relativistic gravity in which  black holes feature prominently. Hence, as we will see in sections \ref{s4} - \ref{s6}, the use of QLHs has become ubiquitous in mathematical GR, NR and gravitational wave physics. However, as of now, an important issue remains: What is the best characterization of a BH region and its boundary?  This issue of is discussed in section \ref{s7}.   

\subsection{Outline} 
\label{s1.3}

This review is organized as follows. Section \ref{s2.1} introduces the precise notions of QLHs, emphasizing two classes:\\
\indent (i) \emph{Non-expanding} horizon segments (NEHSs), that are null. 
The Hawking quasi-local mass of any 2-sphere cross-section of an NEHS is the same. In this sense there is no leakage of energy across an NEHS. When endowed with some natural additional structures, they become \emph{weakly isolated} horizon segments (WIHSs) and \emph{isolated} horizon segments (IHSs); and,\\
\indent(ii) \emph{Dynamical} horizon segments (DHSs), that are either space-like or time-like. The Hawking quasi-local mass changes from one cross-section of a DHS to another. This is directly related to the fact that there is flux of energy flowing across DHSs.%
\footnote{Although DHSs are closely related to the Dynamical Horizons (DHs) introduced in \cite{Ashtekar:2002ag,Ashtekar:2003hk,Ashtekar:2004cn}, the specific conditions that appear in the two sets of definitions are a bit different. In particular, while DHs were required to be space-like, as we just noted DHSs can also be time-like, e.g. in the OS dust collapse and, more importantly, in the semi-classical geometry of an evaporating BH. Therefore, DHSs can be used more widely than DHs.} 
\vskip0.1cm
\noindent The rest of section \ref{s2} focuses on the first of these two classes of QLHs and summarizes their basic properties. WIHSs and IHSs feature primarily in the description of BHs in equilibrium.  WIHSs are tailored to the discussion of zeroth and first laws of QLH mechanics and also play a key role in bringing out close similarities between QLHs and null infinity. The most familiar examples of horizons --in particular, those in the Schwarzschild and Kerr solutions-- are IHSs. But IHSs exist more generally even in absence of Killing fields. Their geometrical shape and spin structure is neatly encoded in an invariantly defined set of numbers, representing \emph{source} multipoles of BHs in equilibrium. In the early and late stages of BH mergers, as well as in the semi-classical phase of BH evaporation, DHSs evolve slowly and are well approximated by perturbed IHSs. 

In section \ref{s3} we introduce DHSs from a fresh perspective, provide examples, and discuss their properties. As the name suggests, DHSs feature prominently in dynamical processes involving BH formation as well as BH mergers in classical GR, and in the BH evaporation process in quantum gravity. They provide the natural arena for the discussion of the second law of QLH mechanics. While Hawking's celebrated area theorem is a qualitative statement, guaranteeing only that the area of the EH cannot decrease under physically reasonable assumptions, the second law of DHSs provides a \emph{quantitative} relation between the amount by which area increases and fluxes across the DHS. Source multipoles are also well-defined for DHSs. Now they are time dependent, providing an invariant description of the evolution of the horizon structure. By contrast, one cannot define multipoles on EHs to capture their dynamics in an invariant manner.  

In \ref{s4} we return to MTSs { introduced in section \ref{s2}} and summarize results on their time evolution from mathematical as well as NR perspectives. They pave the way to discuss key conceptual issues: the existence, uniqueness and  dynamics of QLHs. { QLHs don't jump abruptly at the merger; the process is much more subtle. In the simpler cases that have been analyzed in detail numerically, the QLHs of the progenitors and the remnant join the final common DHS, resulting in a continuous QLH.} Together, these results provide a foundation for much of the current intuition on geometrical structures in the strong field regime of BBH space-times. 

In section \ref{s5} we discuss a rather surprising interplay between geometrical structures on WIHSs $\IH$ and null infinity $\scrip$ and observables associated with them. In particular, the familiar Bondi-Metzner-Sachs (BMS) group  at $\scrip$ descends directly from the symmetry group on WIHSs.  In addition there is a striking similarity in the functional forms of certain observables, defined independently on $\scrip$ and on WIHs. 

In section \ref{s6} we continue the discussion of the interplay between QLHs and null infinity. It is well known that quasi-normal frequencies play a key role the analysis of waveforms at $\scrip$ in the ring-down phase of BBH mergers. Interestingly, numerical investigations have shown that these frequencies are also naturally encoded in the way observables associated with DHSs evolve in time. Section \ref{s6.1} presents illustrative results from these investigations. In section \ref{s6.2} we focus on the dynamical phase soon after the merger during which the DHS $\DH$ of the remnant is evolving sufficiently slowly so as to be well approximated by a perturbed IH. In this regime there is a strong correlation between the evolution of multipole moments of the DHS and certain observables  associated with the BMS group, defined independently at $\scrip$. This is quite surprising because there can be no causal communication from $\DH$ to $\scrip$. These results also pave a  way to sharpen the domain of validity of the perturbative regime during and after the merger, a subject on which there has been a lively debate in recent years. In section \ref{s6.3} we consider the IHS of a BH that is tidally perturbed by the presence of a second BH in a BBH coalescence. In this case, the IHS of { each} BH is distorted by the presence of the { other} due to the `Coulombic interaction' between them even when the gravitational radiation falling into the first BH is negligible. Thus one is led to a complementary class of perturbed IHSs, those in which perturbation is time { \emph{independent}}. Using the characteristic initial value problem based on two null surfaces it has been shown that (for slowly rotating BHs) the standard tidal Love number of { each} BH vanishes, in spite of the fact that its horizon \emph{is} distorted. This can occur because the standard Love number refers to the field quadrupole moment, while the distortion refers to the source multipoles! The complementary situations analyzed in sections \ref{s6.2} and \ref{s6.3} strongly suggest that horizon multipoles provide  promising tools to deepen our understanding of the BBH merger process. 

Section \ref{s7} summarizes these recent advances in our understanding of the structure of QLHs and their applications to mathematical GR, NR and gravitational wave science and discusses directions for further work. As indicated in the abstract, QLHs also play a significant role in the quantum gravity investigations of BHs. Discussion of this application would have required major detours into the underlying ideas and mathematical structures from quantum theory. We chose not to undertake this task to keep the review focused; interested readers will find a discussion of these aspects of QLHs in recent reviews \cite{g2022blackholeentropyloop,ashtekar2023regularblackholesloop,afshordi2024blackholesinside2024}.
In addition we chose not to include two topics in classical GR that were discussed at length in our previous review \cite{Ashtekar:2004cn}:  (i) Boundary conditions on Cauchy slices that need to be imposed on the initial data to ensure that the inner boundaries represent a QLH at that instant of time, and ways of extracting mass and spin of BHs using QLHs; and, (ii) Role played by QLHs in extracting properties of hairy black holes, in particular their relation to solitons. This is because there have not been significant new developments in these areas. Interested readers can find a summary in sections 5 and 6 of \cite{Ashtekar:2004cn}.

Our conventions are the following. We work exclusively in 4 space-time dimensions with metrics $g_{ab}$ of signature -,+,+,+. The torsion-free derivative operator compatible with the metric is denotes by $\nabla$ and curvature tensors are defined by $R_{abc}{}^d\, V_d := 2 \nabla_{[a} \nabla_{b]}\, V_c; \, R_{ac} :=  R_{abc}{}^{b}$ and $R= g^{ab}\, R_{ab}$. Expressions of the components of the Weyl tensor in the standard Newman-Penrose tetrad can be found in section \ref{s6.3}. The symbol $\=$ is used to denote equality that holds only at points of the given QLH (which could be $\scrip$). Finally, throughout we will set $c=1$ and, following the convention used in the literature on gravitational waves, in section \ref{s6} we also set $G=1$ and assume vacuum Einstein's equations. 

Finally, this review is addressed to several different communities: Experts in geometric analysis, mathematical GR,  numerical relativity, quantum gravity and gravitational wave astronomy. Secs.~\ref{s2.1}, \ref{s2.2}, \ref{s2.4}, \ref{s3.1} and \ref{s3.3} would be of interest to all these communities as they introduce basic concepts, definitions and notation (although the detailed discussion in section \ref{s3.1} of the various cases listed in Tables 1 and 2 can one skipped without a loss of continuity). Secs.~\ref{s2.3}, \ref{s3.2}, \ref{s3.4}, \ref{s4} and \ref{s5} would be of interest primarily to researchers in mathematical relativity and geometric analysis.  Numerical relativists would be interested especially in Secs.\ref{s4}, \ref{s6.1} and \ref{s6.2}, and experts who focus on gravitational wave astronomy, in Sec.~\ref{s6}.

\section{QLHS I: Basic Definitions and Equilibrium Situations}
\label{s2}

The most familiar examples of horizons come from Schwarzschild and Kerr space-times. These are generally thought of as EHs but it is well known that they are also Killing horizons, a notion that does not refer to $\scrip$. However, they can be characterized as QLHs as well, using space-time geometry in their immediate vicinity, without any reference to $\scrip$ or a Killing vector. But they are very special QLHs in that they represent boundaries of black (and white) holes \emph{in equilibrium}. In dynamical situations, EHs continue to be null, although they would be non-smooth generically because of creases, corners and caustics \cite{gadioux2023creasescornerscausticsproperties}; in fact they can even be nowhere differentiable \cite{Chrusciel:1996tw}! By contrast, dynamical QLHs are quite different: they are neither null nor do they have the EH pathologies because they are smooth.

In the literature to date, there are several closely related but distinct notions of QLHs, especially in the dynamical context. The reason is that it is difficult to strike an ideal balance: there is a tension between the wish to maintain generality (that leads one to impose minimal requirements) and the desire to obtain a rich set of physically interesting results (that emerge only in more restricted settings). In addition, sometimes the same nomenclature has been used to denote technically different notions. In this review, we have reorganized the material to streamline the terminology that is well-suited to both classical and quantum BHs (although it may not be ideal for all cosmological applications). As a result our restructured summary contains new elements and results. We will make brief remarks on differences between our terminology and that in the literature. 

Section \ref{s2.1} introduces the basic definitions. In the rest of this section we focus on properties of QLHSs in equilibrium. Properties of the dynamical QLHSs are discussed in section \ref{s3}.

\subsection{Basic notions}
\label{s2.1}

Consider $S$, a space-like submanifold with topology of a 2-sphere, in a given 4-dimensional space-time $(M, g_{ab})$. $S$ admits precisely two future directed null normal vector fields $k^a$ and $\uk^a$ and without loss of generality we can rescale them so that $k^a\uk_a =-1$. This normalization still allows rescalings $k^a \to \alpha k^a;\,\, \uk^a \to (1/\alpha)\,\uk^a$ for some positive function $\alpha$. Our considerations will be insensitive to this freedom.

Let us begin with a key notion that underlies discussions of QLHs.
\vskip0.05cm
\emph{Definition 1:} $S$ is said to be a \emph{marginally trapped surface} (MTS) if the expansion of one of the null normals, say $k^a$, vanishes, i.e. if $\theta_{(k)} := \tq^{ab} \nabla_a k_b =0$ where $\tq_{ab}$ is the projection operator into $S$ and $\nabla_a$ is the derivative operator defined by $g_{ab}$.\vskip0.05cm
{We use the term MTS rather than the more commonly used term Marginally Outer Trapped Surface (MOTS) because the term MOT carries a connotation of an `outer direction' (versus inner) and, in quasi-local considerations, it is not always possible to single one out. On the spherical MTSs of the OS collapse, for example, there is no unambiguous `outer' direction.}  

With the notion of MTSs at hand, one can define QLHs as follows.\vskip0.05cm
\emph{Definition 2:} A QLH $\mathfrak{H}$ is a 3-dimensional sub-manifold of $(M, g_{ab})$ that admits a foliation by a 1-parameter family of MTSs. (Since QLHs are world-tubes of MTSs,  they have also been called \emph{marginally trapped tubes} (MTTs) in the literature 
\cite{Ashtekar:2005ez,Booth:2005qc,Andersson:2007fh,Bengtsson_2011,senovilla2012stabilityoperatormotscore}.)

Two classes of QLHs are of particular interest in BH physics. Those in the first class  capture dynamical phases of BHs, and those in the second, equilibrium phases.
  \vskip0.05cm
\emph{Definition 3:} A connected component (without boundary) $\DH$ of a QLH $\mathfrak{H}$ is said to be a \emph{dynamical horizon segment} (DHS) if\\
(i) $\DH$ is nowhere null. Being connected, $\DH$ is either a space-like or a time-like component of $\mathfrak{H}$;\\
(ii) The expansion $\theta_{(\uk)}$ of the other null normal is nowhere zero. Again, since $\DH$ is  connected, $\theta_{(\uk)}$ is either positive or negative on each DHS; and\\
(iii) $\mathcal{E} := \t{q}^{ac}\,\t{q}^{bd}\,\sigma_{ab}^{(k)} \sigma_{cd}^{(k)} + R_{ab} k^a k^b$ does not vanish identically on any MTS $S$ in $\DH$. Here $\sigma_{ab}^{(k)} = {\rm TF}\,\t{q}_a{}^c \t{q}_b{}^d\, \nabla_c k_d$ is the shear of the null normal $k^a$ evaluated on $S$, and ${\rm TF}$ refers to the trace-free part of the tensor. 
\vskip0.1cm

\noindent The quantity $\mathcal{E}$ can be thought of as a `null energy flux' crossing the QLH, defined by the vector field $k^a$ (that includes both matter and gravitational wave contributions).
Intuitively, the QLH is dynamical if this flux does not vanish identically across any of its MTSs. In the complementary case --in which $\mathfrak{H}$ is null-- discussed below, this flux vanishes identically. Examples of DHSs appear in each of the three panels of Fig. 1: they are all confined to the \emph{curved region of space-time}, in the support of collapsing matter (in contrast to the EH in the middle panel which first forms and grows in flat space-time). 

\vskip0.1cm 

\emph{Remarks}: 

1. In the original papers \cite{Ashtekar:2002ag,Ashtekar:2003hk}, `dynamical horizons' were required to be space-like  with negative $\theta_{(\uk)}$. This restriction was motivated by some qualitative arguments due to Hayward \cite{Hayward:1993wb} and, more importantly, by the fact that the implicit focus of these papers was on the QLHs representing remnants in BBH mergers which are space-like with negative $\theta_{(k)}$ soon after the merger. Our discussion in section \ref{s3.2} will explain why, more generally, space-like DHSs are ubiquitous in BBH mergers.
More recently, An \cite{An_2020} has used the characteristic initial value formulation (on finite null cones) to investigate the formation of QLHs due to converging gravitational waves \emph{in full, non-linear GR}, without any symmetry assumptions. He shows that for an open set of data, this gravitational collapse creates a \emph{space-like} DHS that grows in area.   

However, DHSs in the OS stellar collapse are time-like \cite{BenDov:2004gh}. Although subsequent systematic analysis of spherically symmetric solutions showed that in a general stellar collapse DHSs are space-like `in most circumstances' \cite{Booth:2005ng,Helou:2016xyu}  
re-enforcing the case to focus on space-like DHSs, time-like DHSs do arise as the initial data becomes closer to being homogeneous. To accommodate these cases and, more importantly, the long semi-classical phase of the BH evaporation process where the DHS is time-like, in this review we allow DHSs to be either space-like to time-like.  Finally, condition (iii) in Definition 3 was absent in the definition of `dynamical horizons' of \cite{Ashtekar:2002ag,Ashtekar:2003hk}. It
serves to remove certain degenerate cases involving space-times in which one has abundant MTSs but no trapped 2-spheres (i.e. 2-spheres on which both expansions are negative). This situation can occur in very special space-times in which all scalars constructed from curvature and its derivatives vanish \cite{Senovilla_2003}. \vskip0.cm

2. For those who have worked exclusively with EHs, it often comes as a surprise that DHSs admit a foliation by MTSs since the expansion $\theta_{(k)}$ of the null normal $k^a$ to EHs is generically non-zero in dynamical situations. Indeed if an EH admitted a foliation with $\theta_{(k)} \=0$, its area would not change! The point is that DHSs are \emph{not} null, whence the fact that $\theta_{(k)} \=0$ for a null normal $k^a$ to a foliation by 2-spheres does \emph{not} imply that their area cannot change. Indeed, Definition 3 immediately implies that the area of the MTSs on $\DH$ changes monotonically.
\vskip0.1cm 

{3. It is also interesting to note that there are solutions to Einstein's equations with a positive cosmological constant in which there is no event horizon but there is a DHS that captures the notion of a BH. These solutions \cite{Senovilla:2022bsn,senovilla2023blackholesuniversalproperties} admit trapped surfaces which lie to the future of a space-like DHS, but there is no event horizon because these space-times fail to admit $\scrip$. (They do admit $\scrim$, initial data on which determines the full space-time.)}\vskip0.2cm

The complementary case to DHSs is provided by segments of  QLHs $\mathfrak{H}$ that are null \cite{Ashtekar:2000sz,Ashtekar:2000hw,Ashtekar:2001jb}. In this case, it is convenient to impose a mild restriction on the space-time Ricci tensor evaluated on $\mathfrak{H}$ \cite{akkl1}.
\vskip0.05cm
\indent \emph{Definition 4:} A connected component $\IH$ of a QLH is said to constitute a \emph{non-expanding horizon segment} (NEHS) of $\mathfrak{H}$ if\\ (i) it is null, with $k^a$ as its normal; and,\\ (ii) the Ricci tensor of the 4-metric $g_{ab}$ satisfies $R_a{}^b k^a \= \alpha k^b$ for some $\alpha$.
\vskip0.05cm
NEHSs occur both as manifolds with and without boundaries. Since $\mathfrak{H}$ admits a foliation by MTSs, and NEHSs are null, it follows that \emph{every} cross-section of an NEHS is an MTS. In the first two panels of Fig. 1, the DHSs join on to an NEHS to the right (which are the future parts of the event horizons $\mathcal{E}$). The figure in the right panel features DHSs (shown in Green) in the two shaded regions depicting collapsing matter. The left DHS lies in the collapsing star, and is time-like. At its right end, it joins an NEHS (part of the `old EH' $\mathcal{E}_{old}$ in between the shaded regions). As we proceed to the right, this NEHS segment joins on to a space-like DHS that lies in the collapsing null fluid region. Finally, the space-like DHS joins an NEHS that coincides with the right piece of the EH. Note that while the EH is null everywhere and passes through both collapsing matter and vacuum regions, the left segment of the DH is time-like, the right segment is space-like, and they both lie in the support of matter. The NEHS in all panels also lie in the curved but vacuum region of the space-time. (In these spherically symmetric space-times there is no gravitational radiation in the vacuum region, whence there is no flux of energy across these NEHSs.) 
\vskip0.05cm

{\emph{Remark:} The original definition of NEHS \cite{Ashtekar:2000hw,Ashtekar:2001jb} used the dominant energy condition (DEC) in place of (ii). Condition (ii) used in the present Definition is weaker: Given condition (i), if the matter field satisfies DEC on $\IH$, then (ii) is automatically satisfied on that $\IH$. Furthermore, this weakening is necessary to probe the interplay between horizons and null infinity (discussed in section \ref{s5}). Therefore, in the more recent literature 
\cite{akkl1,Ashtekar1_2024,Ashtekar2_2024,Ashtekar3_2024} the dominant energy condition was replaced by (ii).} Finally, in GR (ii) has a natural interpretation: It is a necessary and sufficient condition for the fluxes $T_{ab} k^a \xi^a$ across $\IH$, associated with tangent vectors $\xi^a$ (to $\IH$) to vanish, as one would expect of horizons that represent equilibrium states.  

\subsection{Properties of NEHs}
\label{s2.2}
 
Condition (ii) in Definition 4 of NEHSs together with the Raychaudhuri equation leads to a rich geometric structure \cite{Ashtekar:2000hw}:\\
(A) The intrinsic `metric' $q_{ab}$ (of signature (0,+,+)) on $\IH$ satisfies $\Lie_k {q}_{ab}\, \=\,0$. In particular, then, the area of any cross-section $S$ of an NEHS is the same, the intrinsic metric on any NEHS is `time independent'. Recall that Hawking's  quasi-local mass  
associated with any 2-sphere $S$ in the physical space-time is given by \cite{10.1063/1.1664615}
\be \label{MH} M_H[S] := \f{R}{2G}\,\,\Big[ 1\, +\, \f{1}{8\pi}\,{\oint_S} \theta_{(k)}\theta_{(\uk)} \rmd^2 V\,\, \Big], \ee
where $R$ is the areal radius of $S$ and $\theta_{(k)},\,\theta_{(\uk)}$ expansions of the two null normals to $S$ (satisfying $k^a \uk_a = -1$). Therefore, for \emph{any} 2-sphere cross-section $S$ of an NEH $\IH$, we have $M_H [S] = R/2G$. Since $R$ is constant on $\IH$, we are led to the conclusion that there is no flux of energy across $\IH$. The property $\Lie_k {q}_{ab}\, \=\,0$ also implies that not only the expansion $\theta_{(k)}$ but the shear $\sigma_{ab}^{(k)}$ of the null normal $k^a$ also vanishes on any cross-section $S$ of $\IH$. Since condition (ii) in Definition 4 of a NEHS implies  $R_{ab} k^a k^b\, \=\,0$ on $\IH$, $\mathcal{E}$ vanishes identically on $\IH$. This is a  complementary  sense in which there is no flux of energy across $\IH$; NEHSs can be thought of as parts of the BH boundary that are in equilibrium.\\
(B) The space-time derivative operator $\nabla$ induces a derivative operator $D$ on $\IH$ via pull-back. It satisfies $D_a q_{bc}\, \=\,0$ and $D_a k^b = \omega_a k^b$ for some 1-form $\omega_a$ on $\IH$. $\omega_a$ is called `the rotational 1-form' because one can show that it encodes the angular momentum associated with the NEHS $\IH$. (Strictly, $\omega_a$ should be denoted by $\omega_a^{(k)}$ since it depends on the normalization of $k^a$. But the superscript is generally dropped for simplicity.) Finally $\kappa_{(k)} := k^a \omega_a$ is called the \emph{surface gravity} of the NEHS $\IH$.\\
Thus, each NEHS $\IH$ comes with a natural pair $(q_{ab}, D_a)$, referred as the NEH geometry. Note incidentally that, because ${q}_{ab}$ is degenerate, there are infinitely many derivative operators that satisfy $D_a {q}_{bc}\, \=\, 0$. Thus, $D_a$, obtained by pulling back to $\IH$ the space-time derivative operator $\nabla_a$ has  additional information that is not encoded in the metric $q_{ab}$. In particular, while $q_{ab}$ is time independent, as we discuss in section \ref{s5.2}, in general $D_a$ is time \emph{dependent}. 

Now, if $(\IH,k^a)$ is an NEHS, then so is $(\IH, fk^a)$ where $f$ is any (smooth) positive function. 
Under this rescaling the pair $(q_{ab}, D_a)$ on $\IH$ is left invariant but $\omega_a$  changes via
$\omega_a \to \omega_a + D_a \ln f$. The freedom to rescale $k^a$ by an arbitrary positive function can be restricted considerably by imposing a natural requirement on the choice of $k^a$: $\Lie_k \omega_a \,\=\,0$. This condition asks that $\omega_a$ be also \emph{time independent}, and on any NEH one can choose null normals $k^a$ so that it is met. Next, note that $\omega_a$ does not change under $k^a \to k^{\prime a} = c\, k^a$ where $c$ is a constant and if  $\Lie_k \omega_a \,\=\,0$ then $\Lie_{k^\prime} \omega_a \,\=\,0$. Therefore it is natural to consider equivalence classes $[k^a]$ of null normals, where $k^a \approx k^{\prime\,a}$ if $k^{\prime\,a} = c k^a$ for a (positive) constant $c$. An NEHS $\IH$ equipped with an equivalence class $[k^a]$ for which $\omega_a$ is time independent is called  a \emph{weakly isolated horizon segment} (WIHS); an NEHS can always be endowed with the structure of a WIHS.  

Using condition (ii) in Definition 4 one can show that\, $\Lie_k\, \omega_a\, \, \= \,0$\, if and only if the surface gravity $\kappa_{[k]}$ is constant on $\Delta$. Thus, not only does a WIHS represent a horizon in equilibrium in the sense that the intrinsic metric $q_{ab}$ is time independent but also in the the sense that the \emph{zeroth law of BH mechanics also holds on $\IH$.} As discussed in section \ref{s2.3}, WIHs provide a natural setting to establish  the first law as well. Finally, as we will see in section \ref{s5}, somewhat surprisingly, $\scri$ is also a WIHS, albeit in the conformally completed space-time.

While an NEHS can always be given the structure of a WIHS simply by rescaling the null normal, this condition does not determine $[k^a]$ uniquely; an NEHS can be given several distinct WIHS structures $(\Delta, [k^a])$. To further restrict the freedom, let us first note that, on a WIHS, a part of the connection $D$  is Lie-dragged along $k^a$: $[\Lie_k, D_a] k^b\, \=\, -(\Lie_k \omega_a)\,k^b \=\,0$. In this sense, a WIHS is in equilibrium in a stronger sense than an NEHS is. It is natural to attempt to strengthen the equilibrium condition further by requiring $[k^a]$ to Lie drag the \emph{full} derivative operator $D$ so that $[\Lie_k,  D_a] t^b\, \=\,0$ for all vector fields $t^a$ tangential to $\IH$. Then, the full geometry $(q_{ab}, D_a)$ of the NEHS would be time independent. While a given NEH may not admit such a $[k^a]$, generically it does exist (see Remark 2 below) and, furthermore, it is unique \cite{Ashtekar:2001jb}. If $[k^a]$ satisfies this condition, the pair $(\IH, [k^a])$ is called an \emph{isolated horizon segment} (IHS). The NEH segments in all three panels of Fig. 1 are in fact IH segments. Other examples arise in BBH mergers: the DHS of each of the progenitors tends to an IHS in the distant past, and the DHS representing the remnant also tends to an IHS in the distant future. Therefore, there is a long and interesting regime, prior to taking these limits, in which the DHs are well approximated by \emph{perturbed IHs} \cite{Ashtekar:2021kqj}. This property will play a key role in our discussion of `gravitational wave tomography' in section \ref{s6.2}.\vskip0.05cm 

\emph{Remarks:} 

1. The NEHs, WIHs and IHs are defined as sub-manifolds of the given space-time $(M, g_{ab})$ in their own right. They do not involve any extra structure such as the presence of a space-time Killing vector one needs to define a Killing horizon, or a choice of a Cauchy surface that is necessary to define an apparent horizon. All conditions involved refer to properties of fields on these sub-manifolds themselves. 

2. The phrase `generically' in the passage from WIHSs to IHSs refers to a condition that a certain elliptic operator on the space of $L^2$ functions on an MTS does not admit zero as an eigenvalue \cite{Ashtekar:2001jb}. (This condition is automatically satisfied if the MTSs on the WIH are strictly stable, since the elliptic operator is the adjoint of the stability operator for MTSs; see section \ref{s4.1}.)  Again this operator refers only to the intrinsic structures on the NEHS. While this condition would be generically satisfied by the intrinsic geometry $(q_{ab}, D_a)$ of a given NEH, one can construct examples where it fails. Thus, while every NEH can be given the structure of a WIH simply by an appropriate choice of $[k^a]$, it cannot always be given the structure of an IHS; the notion of an IH is genuinely stronger than that of a WIHS. 

3. Nonetheless, the notion of an IHS is considerably weaker than the notion of a Killing horizon, on which the \emph{full space-time geometry} is in equilibrium: the null normal to a Killing horizon Lie-drags not only the intrinsic geometry $(q_{ab}, D_a)$ but also the full $\nabla$ and the full 4-dimensional curvature tensor and all of its derivatives. In particular, the Robinson-Trautmann radiating solutions \cite{pc,Podolsky:2009an} 
admit IHSs that are not Killing horizons.

{ 4. In numerical simulations NEHs and IHSs have been used to specify the horizon boundary conditions on the initial data in approaches that use excision (see  e.g. \cite{PhysRevD.79.087506,Vasset_2009} and earlier references therein.)}

5. Properties of NEHSs listed in this subsection are consequences of geometric conditions in the definitions, without reference to any field equations. Thus, they also hold for metric theories of relativistic gravity beyond GR.

\subsection{The Zeroth and the First Laws of Horizon Mechanics}
\label{s2.3}

In the 1970s the notion of EHs seemed especially compelling because of they are subject to three laws that have close similarity with the three laws of thermodynamics \cite{Bardeen:1973gs}. 
\footnote{\,\,\,This paper was entitled ``The four laws of black hole mechanics'', but it is now known that the fourth of these laws does not hold \cite{kehle2024extremalblackholeformation}: In classical GR, one \emph{can} obtain  extremal BHs (that have `zero temperature') by non-linearly perturbing non-extremal ones.}
A natural question then is whether QLHs are also subject to similar laws. Not only is the answer in the affirmative, but the QLH-laws have been shown to be more physically satisfactory than those associated with EHs \cite{Ashtekar:2000hw,Ashtekar:2001is,Ashtekar:2003hk}. In this subsection we will provide a brief summary of these results for the zeroth and the first law that hold on WIHSs; the second law will be discussed in section \ref{s3.2}. \vskip0.2cm

The zeroth and the first law of BH mechanics were first established for EHs of \emph{stationary, axisymmetric BHs in GR} with matter fields that satisfy the dominant energy condition 
(see, e.g., \cite{wald:1994,tj,Wald2001}). These EHs are also Killing horizons: a null normal $k^a$ to the EH is the restriction to the EH of a Killing field $K^a$,
\be k^a\, \= \,K^a \qquad {\rm with} \qquad K^a\, \= \, t^a + \Omega\, \varphi^a\, , \ee
where $t^a$ is the stationary Killing field (which is time-like and with unit norm at spatial infinity) and $\varphi^a$ is the rotational Killing field. The constant $\Omega$ is called the \emph{angular velocity} of the EH and the acceleration $\kappa$ of $k^a$, given by \, $k^b \nabla_b k^a\, \= \, \kappa\, k^a$,\,  is called the \emph{surface gravity} of the EH. The zeroth law of EH mechanics says that $\kappa$ is constant on the EH, just as the temperature is constant for a thermodynamical system in equilibrium \cite{bc}. This fact led to what was at the time a counter intuitive conjecture that a multiple of $\kappa$ should perhaps be identified with the temperature of a BH. The first law relates the parameters associated with two nearby stationary axisymmetric BHs:
\be \label{1stlaw1} \delta M \,\=\, \f{\kappa}{8\pi G}\, \delta \a + \Omega\, \delta J \ee
where $M, J$ denote the total mass and angular momentum of space-time measured at spatial infinity, $\a$ is the area of any 2-sphere cross-section of the EH. In light of the close resemblance with the first law of thermodynamics, Bekenstein suggested that a suitable multiple of the area $\a$ should be interpreted as the entropy associated with the EH and introduced thought experiments to support this conjecture \cite{Bekenstein:1973ur,Bekenstein:1974ax}. In classical general relativity, constants relating $\kappa$ and temperature, and $\a$ to entropy remain undetermined. They were fixed by Hawking's calculation of quantum radiance of BHs and involve Planck's constant $\hbar$ \cite{Hawking:1974sw}. Eq. (\ref{1stlaw1}) is referred to as the `passive version' of the first law since it relates the parameters associated with two nearby stationary, axisymmetric space-times, rather than a `physical process version' in which a given BH is perturbed and settles down to a nearby equilibrium state. Finally, in the derivation of these zeroth and first laws, it suffices to treat the EH as a Killing horizon in a stationary, axisymmetric, asymptotically flat space-time, without reference to its definition as the future boundary of $J^{-}(\scrip)$.

Let us now consider a  WIHS $(\IH, [k^a])$. We already saw in section \ref{s2.3} that the zeroth law holds: For any $k^a \in [k^a]$, the surface gravity $\kappa_{(k)}$, defined by\, $k^b \nabla_b k^a\, \= \,\kappa_{(k)} k^a$,\, is constant on $\IH$ even though the BH may be distorted, e.g., by matter rings. For the first law, let us consider space-times $(M, g_{ab})$ that are asymptotically flat and admit a WIH $\IH$ and a rotational vector field $\varphi^a$ that is a symmetry \emph{only of the intrinsic metric} $q_{ab}$ on $\IH$. (Space-time need not be either stationary or axisymmetric even in a neighborhood of $\IH$.) Then, it follows that there is a 2-sphere `charge' $J_\IH^{(\varphi)}$ on $\IH$  associated with the intrinsic rotational symmetry $\varphi^a$,
\be \label{J1} J_\IH^{(\varphi)}\, \=\, -\f{1}{8\pi G}\, \oint_{S} (\omega_a \varphi^a)\, \rmd^2 {V} \, ,\ee
which is conserved, i.e., independent of the cross-section $S$ of $\IH$. While $\varphi$ and $J_\IH^{(\varphi)}$ refer only to structures on $\IH$, if $\varphi^a$ were the restriction to $\IH$ of a Killing field on $(M, g_{ab})$, then one can show that $J_\IH^{(\varphi)}$ would agree with the Komar integral, i.e., with the standard angular momentum one would assign to $S$. To define energy one needs a vector field representing a `time translation symmetry'. Since both $k^a$ and $\varphi^a$ are Killing fields of the intrinsic metric $q_{ab}$ of $\Delta$, any linear combination of them is a symmetry. Our experience with the Kerr isolated horizon suggests that we use a general linear combination $t^a \,\=\, c k^a - \Omega \varphi^a$ to represent a `time translation' symmetry on $\IH$ and associate energy $E_\IH^{(t)}$ with it, where $c$ and $\Omega $ are constants on $\IH$. Freedom to rescale $k^a$ by a positive constant $c$ is unavoidable since WIHSs are equipped with only the equivalence class $[k^a]$ of null normals rather than a preferred one. Note also that now the symmetry vector field $t^a$, so defined, refers only to $\IH$. Therefore we cannot fix the freedom in $c$ by fixing the normalization of $t^a$ at spatial infinity as was done to arrive at (\ref{1stlaw1}).

With these structures at hand the first law for WIHSs was established using a Hamiltonian framework (for further details, see \cite{Ashtekar:2001is}). The phase space consists of space-times that are asymptotically flat and admit an inner boundary $\IH$ that is a WIH with $[k^a]$ as its null normals and $\varphi^a$ as a rotational symmetry. The coefficients $c$ and  $\Omega $ in the definition of $t^a$ are constants on the inner boundary $\IH$ of every space-time, but these constants can vary from one space-time to another in the phase space. For example, in a static space-time, the horizon angular velocity $\Omega$ would vanish but in a general stationary space-time, it would not. One extends $t^a$ and $\varphi^a$ away from $\IH$ to vector fields that are \emph{asymptotic} time translation and rotation symmetries at spatial infinity; space-times under consideration \emph{need not admit any Killing fields}. Diffeomorphims along these space-time vector fields induce motions in the phase space. The one induced by $\varphi^a$ is a canonical transformation and $J_\IH^{(\varphi)}$ is the corresponding phase space `horizon charge', defined by the rotational symmetry $\varphi^a$. 

On the other hand, it turns out that the diffeomorphism generated by $t^a$ does not define a canonical transformation on the phase space \emph{unless} the phase space functions $ (c,\,\Omega)$ are so chosen that $ \f{\kappa_{{}_{(c k)}}}{8\pi G}\, \delta \a_\IH\, +\, \Omega\, \delta J_\IH^{(\varphi)}$ is an exact variation.  (Here $\kappa_{{}_{(c k)}}$ is the surface gravity of $ck^a$ and $\a_\Delta$ is the area of any 2-sphere cross section of $\IH$). One can satisfy this condition using a restricted but still a large class of phase space functions $(c, \Omega)$. The resulting vector fields  $t^a = c k^a + \Omega \varphi^a$ are said to be \emph{permissible} (infinitesimal) symmetries. For each permissible $t^a$, we have a Hamiltonian $H^{(t)}$ generating the canonical transformation induced by  $t^a$ representing a time translation symmetry on the boundaries, $\IH$ and spatial infinity. The associated charge at spatial infinity is the ADM energy defined by the asymptotic time translation $t^a$, while the horizon contribution  $E_\IH^{(t)}$ to $H^{(t)}$ --or, the `horizon charge' defined the time-translation symmetry $t^a$-- satisfies
\be \label{1stlaw2} \delta E^{(t)}_\IH \,\=\, \f{\kappa_{{}_{(c k)}}}{8\pi G}\,\, \delta \a_{{}_\IH} + \Omega\,\, \delta J_\IH^{(\varphi)}\, .  \ee
This is the first law of mechanics of WIHSs. It has the same form as the first law (\ref{1stlaw1}) for Killing horizons in stationary axisymmetric space-times when $k^a$ is chosen to be the restriction of a globally defined Killing field in $(M,g_{ab})$. But there are some differences as well as noteworthy features:\vskip0.05cm
(1) While the first law (\ref{1stlaw1}) requires Killing fields defined everywhere on $(M, g_{ab})$, (\ref{1stlaw2}) does not. In the first law for WIHSs $t^a$ and $\varphi^a$ are symmetries only of the intrinsic metric $q_{ab}$ of $\IH$. Thus, we only need $\IH$ to be in equilibrium; there may well be dynamical processes away from $\IH$. This is analogous to the fact that in the first law of thermodynamics only the system under consideration has to be in equilibrium; not the whole universe.\vskip0.05cm
(2) All quantities that enter Eq. (\ref{1stlaw2}) are defined intrinsically on $\IH$. By contrast, in the older first law (\ref{1stlaw1}) for Killing horizons one goes back and forth: There, $\kappa_{(k)}$ and $\a$ refer to the horizon while $M$ and $J$ are defined at spatial infinity. They represent the \emph{total} mass and angular momentum of space-time, including the contributions from matter that resides outside the horizon. \vskip0.05cm
(3) An unforeseen feature of the QLH framework is that we have as many first laws as there are permissible $t^a$\,! Recall that $t^a$ is permissible if and only if the phase space functions $c, \Omega$ are chosen such that $ \f{\kappa_{{}{(ck)}}}{8\pi G}\, \delta \a_\IH\, +\, \Omega\, \delta J_\IH^{(\varphi)}$ is an exact variation. This condition is satisfied if and only if \\
(i) $c$ and $\Omega$ depend only on $a_\IH$ and $J_\IH^{(\varphi)}$ and not on other attributes of the WIHS, e.g., those characterizing distortions in its shape; and,\\
(ii) these phase space functions satisfy $\f{1}{8\pi G}\, \f{\partial \kappa_{(ck)}}{\partial J_\IH^{(\varphi)}}\, \= \, \f{\partial \Omega}{\partial \a_\IH} $. \\
Thus, there is a first law for each pair $(c,\, \Omega)$ of functions of $(\a_\IH,\, J_\IH)$, satisfying this condition.
\vskip0.05cm
(4) In the Kerr family, it is natural to choose for $\varphi^a$ and $t^a$ the rotational and the time-translation Killing fields. Then the corresponding $c$ and $\Omega$ on $\Delta$ can be expressed as functions of $\a_\IH,$ and $J_\IH$ alone and they automatically satisfy the above constraint, leading to the well-known first law for this family. Now, one can also choose the same functions of $(\a_\IH,\, J_\IH)$ for $c$ and $\Omega$ on \emph{any} WIHS.  Then the corresponding energy $E^{(t)}_\IH$ is referred as the \emph{mass} of that WIHS \emph{and denoted by} $M_\IH$. But of course there are many other permissible choices, each providing its own surface gravity $\kappa_{(ck)}$, horizon angular velocity $\Omega$ and horizon energy $E_\IH^{(t)}$, and the corresponding first law in which the notion of energy is tied to the choice of $t^a$.
\vskip0.05cm
(5) As we already mentioned, thanks to the constraint equations of GR, Hamiltonians $H^{(\varphi)}$ and $H^{(t)}$ generating  diffeomorphisms along (the extensions of) $\varphi^a$ and $t^a$ reduce to linear combinations of surface terms. $E_\IH^{(t)}$ and $J_\IH^{(\varphi)}$ are the surface terms at the horizon.  
Those at spatial infinity are the Arnowitt, Deser, Misner (ADM) energy and angular momentum. But in contrast to the first law (\ref{1stlaw1}) for EHs, the ADM charges do not appear in the statement of the first law (\ref{1stlaw2}) for WIHSs. \vskip0.05cm
(6) For simplicity, we restricted ourselves to vacuum GR in the above summary. But the zeroth and the first law continue to hold also in presence of gauge fields with additional terms involving the gauge potentials and charges, both evaluated on $\IH$ \cite{Ashtekar:2000hw,Ashtekar:2001is}. (The zeroth law continues to hold on WIHSs in any relativistic theory of gravity.) Again, while in the standard treatment with Killing horizons, the charge is evaluated at infinity and the gauge potential at the horizon, all quantities that enter the first law for WIHS refer only to the horizon.
\vskip0.05cm
(7) In the QLH framework, so far the 1st law has been obtained only for GR (coupled to matter). There is another framework that encompasses higher derivative theories as well \cite{Iyer:1994ys}. In this sense that framework is much more general. However, there the first law is obtained only for stationary space-time backgrounds and the derivation makes a crucial use of the bifurcate horizon where the stationary Killing field vanishes. In non-stationary contexts, such as those depicted in the right panel of Fig. 1, there are no bifurcate surfaces associated with BHs since they are in equilibrium only for a finite duration during which the QLHS is an IHS. Nonetheless, the first law (\ref{1stlaw2}) for WIHSs applies to these segments as well. Thus, within GR the first law provided by the QLH framework is more satisfactory since it is applicable to BHs that are not eternal but formed in physically realistic processes. \vskip0.15cm
\texttt{Open Issue (OI-1)} Can one extend the first law (\ref{1stlaw2}) to relativistic theories of gravity beyond general relativity? Note that the notion of an IH does extend to these theories and, as noted above, the zeroth law holds on the WIHSs in these theories.
\subsection{Multipole moments of IHSs}
\label{s2.4}

In Newtonian gravity and Maxwell's theory, we have two sets of multipoles: the source multipoles that characterize the spatial distributions of mass/charge and current densities and the field multipoles that characterize the asymptotic behavior of the Newtonian potential/Maxwell field. The two sets are related by field equations. This structure continues to be available in linearized GR. In full, non-linear GR, on the other hand the situation is not so simple. Definitions of field multipoles have been available \cite{Hansen:1974zz,RevModPhys.52.299} and the sense in which they characterize the space-time geometry \emph{in the asymptotic region} has been spelled out \cite{Beig}. For extended bodies in general relativity, a definition of the source multipoles in terms of the stress energy tensor $T_{ab}$ is also available \cite{dixon1974dynamics3,dixon1967description,dixon1970dynamics1,dixon1970dynamics2,Ehlers}. But in practice the procedure to compute them is intricate and furthermore it cannot be used for BHs for which the stress-energy tensor vanishes. It is natural to ask if there is another avenue to define source multipoles for BHs. The answer is in the affirmative:  using physical considerations and geometrical structures of QLHs, one can extract \emph{effective} mass and spin `surface densities' and use them to define mass and angular momentum multipoles \cite{Ashtekar:2004gp, Ashtekar:2013qta}. In this subsection we will summarize the results for IHSs and in section \ref{s3.3} for DHSs.

IHSs of direct physical interest to the binary BH problem are non-extremal and admit a (possibly approximate) rotational Killing field $\varphi^a$. Therefore for simplicity we will focus on IHSs $(\IH, [k^a], q_{ab}, D)$ on which $\kappa_{(k)}$ is non-zero and $\Lie_\varphi q_{ab}\, \=\,0$. {(See remark 4 at the end of this subsection on a definition of multipoles of non axisymmetric IHS.)} We will further assume that vacuum equations hold on $\IH$; for inclusion of gauge fields, see  \cite{Ashtekar:2004gp}. 

Now, thanks to Einstein's equations, one can find freely specifiable data that completely characterizes the isolated horizon geometry $(q_{ab}, D)$. Fix any cross-section $S$ of $\IH$. Then for any non-extremal horizon, the free data consists of $(\t{q}_{ab},\, \t\omega_a)$ where $\t{q}_{ab}$ is the intrinsic metric on $S$ and $\t\omega_a = \t{q}_a{}^b\, \omega_b$, the projection of the rotational 1-form $\omega_a$ to $S$ \cite{Ashtekar:2001jb}. The diffeomorphism invariant information in the 2-metric $\t{q}_{ab}$ is contained in its scalar curvature $\t\R$. It turns out that the information contained in the fields $\t\R$ and $\t\omega_a$ can be encoded in two sets of multipoles.

{Let us then consider IHSs on which the metric $\t{q}_{ab}$ admits a rotational Killing field $\varphi^a$ and choose a cross-section $S$ to which $\varphi^a$ is tangential. Then, as shown in \cite{Ashtekar:2004gp} (section 2.2), $S$ admits invariantly defined coordinates $\varphi \in [0, 2\pi)$ and $\zeta \in [-1, 1]$ such that:\\
(i) $D_a \zeta = \f{1}{R^2}\, \t\epsilon_{ba} \varphi^b$  and $\oint_S  \zeta \t\epsilon =0$ where $R$ is the area radius and $\t\epsilon_{ab}$ the area 2-form defined by $\t{q}_{ab}$ on $S$; \\
(ii) The metric $\t{q}_{ab}$ has the form \begin{equation}
  \label{eq:canonical-2metric}
  {\tq}_{ab} = R^2\,(f^{-1}\tilde{D}_a\zeta\tilde{D}_b\zeta + f\tilde{D}_a\varphi\tilde{D}_b\varphi)\,,
\end{equation}
where $f\,=\, R^{-2}\, \varphi_a \varphi^a$.} It is easy to check that $\tq_{ab}$ is a round 2-sphere metric if  $f =1 - \zeta^2$. Now, by inspection, the area-element corresponding $\tq_{ab}$ of Eq.~(\ref{eq:canonical-2metric}) is independent of $f$ and is therefore the same as that of the round 2-sphere. Multipole moments are defined using the spherical harmonics $Y_{\ell, m}$ of the round 2-sphere metric as weight functions (but because of axisymmetry, only those with $m=0$ are non-vanishing).  The inner-products between the $Y_{\ell, m}$ computed using the physical 2-metric $\tq_{ab}$ are the same as for the standard spherical harmonics on a round 2-sphere, since the two area elements are the same.

With this structure on hand one can define `shape' multipoles $\texttt{I}_\ell$ that capture all the information in $\t\R$ and spin-multipoles $\texttt{S}_\ell$ that contain all the information in $\t\omega_a$ \cite{Ashtekar:2004gp}:
\be \label{dimensionless1} \texttt{I}_\ell\,  -\, i\, \texttt{S}_\ell := {\f{1}{4}}\, \oint_S \big[\,\t\R \,+ \, 2i\, \t\epsilon^{ab} \t{D}_a \t\omega_b\,\big]\,  Y_{\ell,0} (\zeta) \, \rmd^2 {V}\, , \ee
where as usual $\ell$ runs over non-negative integers. The choice of the overall numerical factor is motivated by the fact that on any NEHS, the Newman-Penrose Weyl tensor component $\Psi_2 = \f{1}{2} \big(C_{abcd} +\, i\, {}^\star C_{abcd}\big)\, k^a \uk^b k^c \uk^d$ is given by
$-\f{1}{4}\,(\t\R + 2i\, \t\epsilon^{ab} \t{D}_a \t\omega_b)$. Thus, on any IHS,
\be \label{dimensionless1-psi} \texttt{I}_\ell\,  -\, i\, \texttt{S}_\ell := -\, \oint_S \big[\Psi_2]\,  Y_{\ell,0} (\zeta) \, \rmd^2 {V}\, . \ee
The $\texttt{I}_\ell$ and  $\texttt{S}_\ell$ are dimensionless --they are the \emph{geometrical} multipoles that capture the full invariant content of the horizon geometry in the following sense. If one is given the set $\texttt{I}_\ell, \texttt{S}_\ell$ (and two numbers, values of the area $\a_\IH$ and surface gravity $\kappa_{(k)}$) constructed from any given IHS, one can reconstruct the geometry $(\IH,\, {q}_{ab},\, [k^a],\, D)$ of that IHS up to diffeomorphisms. 

To define the mass and angular momentum multipoles, one introduces an `effective mass surface density' and an `effective spin-current density' on IHS  using an analogy to electrodynamics \cite{Ashtekar:2004gp}. Recall from section \ref{s2.4} that on any axisymmetric IHS, one has well-defined notions of the areal radius $R$ and the total mass $M_\IH$. The mass and angular momentum multipoles $(\texttt{M}_\ell, \texttt{J}_\ell)$ are defined by rescaling the geometrical multipoles by appropriate dimensionfull factors:
\be \label{dimensionfull} \texttt{M}_\ell\, =\, \sqrt{\f{4\pi}{2\ell+1}} \,\, \f{M_\IH\, R^\ell}{2\pi}\,\, \texttt{I}_\ell \qquad {\rm and}\qquad  
\texttt{J}_\ell \,=\,  \sqrt{\f{4\pi}{2\ell+1}}\,\, \f{R^{\ell+1}}{4\pi G}\,\, \texttt{S}_\ell  \ee
(Because of axis-symmetry all multipoles with $m\not=0$ vanish). These multiples have a number of physically expected properties:\\
(i) The mass monopole and the angular momentum dipole agree with the standard $M$ and $\vec{J}$ in the 
Kerr(-Newman) family. \\
(ii) The mass dipole and angular momentum monopole vanish identically on all IHSs.\\
(iii) In spherically symmetric space-times all multipoles except the mass monopole vanish. If the horizon geometry is reflection-symmetric, as in the Kerr family, then  all $\texttt{M}_\ell$ vanish for odd $\ell$ and all $\texttt{J}_\ell$ vanish for even $\ell$.\\
(iv) Recall that in the Hamiltonian framework angular momentum $J_\IH^{(\varphi)}$ arises as the `horizon surface charge' defined by the canonical transformation generated by the symmetry vector field $\varphi^a$. That property is carried over to all higher angular momentum multipoles as well. They are the horizon surface charges corresponding to the canonical transformations generated by vector fields $\t\epsilon^{ab} D_a P_\ell(\zeta)$ (up to overall constants that depend on dimensional factors involving $G$ and $R^{\ell-1}$). However, it is not known if mass multipoles share this feature.
\vskip0.1cm
{ \texttt{Open Issue 2 (OI-2)} Are mass multipoles `horizon surface charges' associated with some natural canonical transformations associated with symmetries?}

\vskip0.2cm

\emph{Remarks:}\vskip0.05cm
1. In Newtonian theory, the source multipoles --constructed from integrals over the source-density $\rho$-- are the same as the field multipoles --constructed from the asymptotic behavior of the potential $\Phi$-- because of the field equation $D^2 \Phi\, =\, - 4\pi G\, \rho$. Due to non-linearities of Einstein's equations this equality no longer holds: Field multipoles, defined at infinity, receive contributions also from the gravitational field outside the support of the sources. This can be seen explicitly in Kerr space-times. Because of the presence of exact isometries, the corresponding charges yield the mass monopole and angular momentum dipole moments both at the horizon and at infinity, whence for these two there is an agreement between source and field multipoles. But already for the mass quadrupole $M_2$ and angular momentum octupole $J_3$ there is a $\sim 10\%$ difference between the source and field moments for large values of the Kerr parameter $a$. More generally, as we will see in section \ref{s6.3}, vanishing of the Love number for BHs does \emph{not} imply that the two coalescing BHs do not distort each other's horizon geometry \cite{RibesMetidieri:2024tpk} because the Love number refers to the field multipoles while the horizon distortions refer to these sources multipoles. For considerations of equations of motion of test bodies that are in a near zone of a given BH, it is the source multipoles of the BH that would be more relevant. 
\vskip0.05cm
2. In our discussion we restricted ourselves to non-extremal IHSs because the situation in the extremal case is surprisingly simple: Every extremal, axisymmetric IHS has the same $(q_{ab},\, \omega_a)$ as the extremal Kerr IHS in vacuum GR { \cite{Hajicek1974,Lewandowski_2003}}. Thus, the multipoles of any extremal IHS are the same as those of extremal Kerr IHS (and in the Einstein-Maxwell theory, the same as those of extremal Kerr-Newman IHS). \vskip0.05cm
3. In stationary space-times the field moments determine the geometry in the asymptotic region \cite{Beig}. Using the  characteristic initial value formulation, one can show that source-multipoles also determine the metric { as a Taylor expansion to any finite order} in a neighborhood of the horizon \cite{Ashtekar:2000sz,Pawlowski:2005}.\vskip0.05cm
4. Note that while multipoles provide set of numbers that suffice to characterize geometry in an invariant fashion, they are not unique in this respect. For example, for the field multipoles, Geroch provided a set for static space-times \cite{Geroch:1970cd} while Hansen provided another set for stationary space-times \cite{Hansen:1974zz}. The two sets of mass multipoles $M_\ell$ do not agree in the static limit (where all $J_\ell$ vanish). Each set suffices to determine  the asymptotic geometry in any static space-time and thus carries the same information; the information is just reshuffled. The situation is the same for source multipoles: One can have two distinct sets of multipoles that characterize the geometry of an IHS in an invariant fashion. The set we presented is the one that has been most widely used in NR, and it has been extended to DHSs. But there is another set for IHSs which has the advantage that it does not require axisymmetry \cite{akkl1}.  While they are distinct sets of numbers --they do not agree in the axisymmetric case-- each suffices to characterize the geometry of IHSs. \vskip0.2cm

\section{QLHS\,II: Dynamical Situations}
\label{s3}

Let us now turn to the physically more interesting DHSs one encounters in BH formation, merger and evaporation. In section \ref{s3.1} we summarize the key properties of DHSs that are then used in the subsequent discussion. (For technical details see, e.g., \cite{Hayward:1993wb} and section 2 of \cite{Ashtekar:2003hk} where, however, the terminology is somewhat different.) Section \ref{s3.2} presents the second law of QLH mechanics in which the change in area of the MTSs in the DHSs is directly related to fluxes of energy across the portion of the DHS bounded by the cross-sections. While we use Einstein's equations (with matter sources) in our main discussion, as we point out at the end, the relation between the change in area and `local happenings' at the DHS holds in all metric theories of gravity. In section \ref{s3.3} we introduce multipole moments of DHSs that characterize the shape and spin-structure of MTSs in an invariant fashion. As discussed in section \ref{s6.2}, these moments have been used in NR to analyze the horizon dynamics in detail and in analytical considerations of tomography that relate the late time horizon dynamics and waveforms at infinity. In section \ref{s3.4} we discuss issues related to the uniqueness of DHSs and their geometrical structure. 

\subsection{Salient properties of DHSs}
\label{s3.1}

As explained in section \ref{s2.1}, DHSs are either space-like or time-like. They can be classified further using the signs of 
\be \theta_{(\uk)} \qquad {\rm and} \qquad \mathcal{E}= \sigma_{ab}^{(k)}\,\sigma^{ab}_{(k)} + R_{ab} k^a k^b,\,\qquad {\rm on}\,\,\, \DH\, \ee
where, as before, $\uk^a$ is the other future directed null normal to the MTSs satisfying $k^a \uk_a = -1$ and, as noted before, $\mathcal{E}$ can be regarded as a `null energy flux' across $\DH$, associated with $k^a$. Tables 1 and 2 present a list of possibilities for DHSs that would be of interest to the mathematical physics community. However, this is not a complete classification of DHSs: To avoid making the discussion excessively long, we focus on segments on which $\mathcal{E}$ has a fixed sign everywhere. Finally, as discussed in section \ref{s3.2} and \ref{s4.1}, the wave form community can focus just on the very first row of Table 1 in the analysis of BH mergers. 

\goodbreak
\vskip0.05cm

\begin{table}
\begin{center}
\footnotesize
\begin{tabular}{|l|c|c|c|}
\hline
\emph{Null Energy Flux}\quad   &\,\,\, \emph{sign of $\theta_{(\uk)}$}\,\,\, & \,\,\,\emph{Transition as one moves along $\uk^a$}\,\,\, &\,\,\, \emph{Area of MTSs}\,\,\,\\ \hline\hline
\multirow{2}{*} \quad $\mathcal{E} >0$ &  - ;\,\,\, T-DHS & untrapped region $\rightarrow$ trapped region & increases \\ \cline{2-4}
&\, {\tiny{+}} ; \,\,\, AT-DHS &  anti-trapped region $\rightarrow$ untrapped region & decreases\\ \hline
\multirow{2}{*}\quad $\mathcal{E} <0$ &  - ; \,\,\, T-DHS & trapped region $\rightarrow$  untrapped region & increases     \\ \cline{2-4}
& \, {\tiny{+}} ; \,\,\, AT-DHS & untrapped region $\rightarrow$ anti-trapped region & decreases    \\ \hline
\end{tabular}
\caption{\footnotesize{\emph{Properties of space-like Dynamical Horizon Segments.} The last column refers to the change in area along the projection of $k^a$ into the DHS.
}}
\label{tab1}
\end{center}
\end{table}
\vskip0.1cm
\emph{Properties:}\vskip0.1cm
\indent\emph{(1)} Conditions on $\theta_{(k)}$ and $\theta_{(\uk)}$ in \emph{Definition 3} immediately imply that the area of MTSs of any DHS $\DH$ is monotonic. This is a simple geometrical consequence. Field equations are not used. \vskip0.1cm

\emph{(2)} Recall, first, that a  2-sphere $S$ is said to be \emph{untrapped} if the two future directed null normals $k^a$ and $\uk^a$ to it satisfy $\theta_{(k)} \theta_{(\uk)} <0$;
\emph{trapped} if $\theta_{(k)}$ and $\theta_{(\uk)}$ are both negative,\, and \emph{anti-trapped} if $\theta_{(k)}$ and $ \theta_{(\uk)}$ are both positive. Recall also that on any DHS $\theta_{(k)} =0$ and $\theta_{(\uk)}$ is either strictly positive or strictly negative. Eq. (\ref{sig1}) below implies that $\Lie_{\uk} \theta_{(k)}$ cannot vanish on $\DH$ (since, by our last assumption, $\mathcal{E}$ is either positive on negative on $\DH$). Therefore DHSs have the following property:\\
(i) If $\theta_{(\uk)} <0$, then small displacements along curves tangential to $\uk^a$ yield  untrapped 2-spheres on one side of $\DH$ and trapped 2-spheres on the other side. In this sense, $\DH$ separates an untrapped region from a trapped region. It will be called a \emph{trapping dynamical horizons segment} (T-DHSs); and,\\
(ii) If $\theta_{(\uk)} >0$, then small displacements along curves tangential to $\uk^a$ yield  untrapped 2-sphere's on one side and anti-trapped 2-spheres on the other. In this sense, $\DH$ separates an untrapped region from an anti-trapped region. It will be called an \emph{anti-trapping dynamical horizons segment} (AT-DHSs).\\
(This classification does not refer to field equations either.) All DHSs in Fig.1 are T-DHSs. In classical GR, AT-DHSs are associated with white holes and therefore rare in physical situations normally considered. But they do arise in the BH evaporation process \cite{Sawayama:2005mw,Hayward_2006,g2022blackholeentropyloop,ashtekar2023regularblackholesloop,ashtekar2025blackholeevaporationloop}.\vskip0.1cm

\emph{(3)} Let us introduce a vector field $V^a$ tangential to $\DH$ and orthogonal to MTSs such that the diffeomorphism it generates preserves the foliation by MTSs. Then, (using the remaining rescaling freedom $k^a \to \alpha\, k^a;\, \uk^a \to \f{1}{\alpha}\, \uk^a$,\, and replacing $V^a$ by $-V^a$ if necessary) one can express $V^a$ as a linear combination: $V^a = k^a - f \uk^a$ for some function $f$. One can readily check that the DHS $\DH$ is space-like if and only if  $f >0$ and time-like if and only if $f <0$. ($f$ cannot vanish anywhere because $\DH$ is nowhere null.) { Since $\Lie_V \theta_{(k)} \,\=\, 0$ on $\DH$, using the Raychaudhuri equation, one has: 
\be \label{sig1} f\,\Lie_{\uk} \theta_{(k)}\, \= \, \Lie_{k} \theta_{(k)} \, \= \, 
- \big(\sigma_{ab}^{(k)} \sigma^{ab}_{(k)} + R_{ab} k^a k^b \big)\, \= \, -\, \mathcal{E}\, \ee
where we have extended $k^a$ to a neighborhood of $\DH$ using the geodesic equation} and used the Raychaudhuri equation in the second equality.  As before,  {\smash{$\sigma_{ab}^{(k)} :\= \,\big(\tq_a{}^c \tq_b{}^d\, -\, \f{1}{2}\tq_{ab} \tq^{cd}\big) \nabla_c k_d $}} denotes the shear of the null normal $k^a$ evaluated on the MTSs. In GR, one can regard $\mathcal{E}$ as \emph{a null energy flux} that includes both matter and gravitational wave contributions, and by inspection it is positive if matter satisfies the dominant  energy condition. We will now explore consequences of this equation. Since every DHS $\DH$ is either space-like or time-like, we can divide the discussion in two parts: 

\begin{table}
\begin{center}
\footnotesize
\begin{tabular}{|l|c|c|c|}
\hline
\emph{Null Energy Flux}\quad   &\,\,\, \emph{sign of $\theta_{(\uk)}$}\,\,\, & \,\,\,\emph{Transition as one moves along $\uk^a$}\,\,\, &\,\,\, \emph{Area of MTSs}\,\,\,\\ \hline\hline
\multirow{2}{*}{\quad $\mathcal{E} >0$} &  - ;\,\,\, T-DHS & trapped region $\rightarrow$  untrapped region & decreases \\ \cline{2-4}   
 &\, {\tiny{+}}\,;\,\,\, AT-DHS & untrapped region $\rightarrow$  anti-trapped region & increases \\ \hline
\multirow{2}{*}{\quad $\mathcal{E} <0$} &  - ;\,\,\, T-DHS &  untrapped region $\rightarrow$ trapped region & decreases \\ \cline{2-4}
 & \, {\tiny{+}}\, ;\,\,\, AT-DHS & anti-trapped region $\rightarrow$ un-trapped region & increases     \\ \hline
\end{tabular}
\caption{\footnotesize{\emph{Properties of time-like Dynamical Horizon Segments.} The last column refers to the change in area along the projection of $k^a$ into the DHS.
}} 
\label{tab2}
\end{center}
\end{table}
\vskip-0.1cm

\emph{(3.a)} \emph{Space-like DHS:} In this case $f$ is positive. Therefore, $\Lie_{\uk} \theta_{(k)}$ and  $\mathcal{E}$ have opposite signs. Suppose $\mathcal{E}$ is positive. Then, $\Lie_{\uk} \theta_{(k)} <0 $, i.e., as we move along $\uk^a$ across $\DH$, the expansion $\theta_{(k)}$ becomes negative in a neighborhood of $\DH$. Therefore if $\DH$ is a T-DHS, then this motion sends one from the untrapped region to the trapped region. This is realized in the BH formation via a (Vaidya) null fluid collapse depicted in the middle panel of Fig. 1. If, on the other hand, $\DH$ is a AT-DHS, then as one moves along $\uk^a$ across $\DH$, one passes from an anti-trapped region to an untrapped region. An example is provided by the \emph{time-reversal} of the null-fluid collapse, i.e., disappearance of a white hole via emission of a null fluid. See the left panel in Fig. 2. 

\begin{figure}[]
  \begin{center}
    \includegraphics[width=0.95\columnwidth]{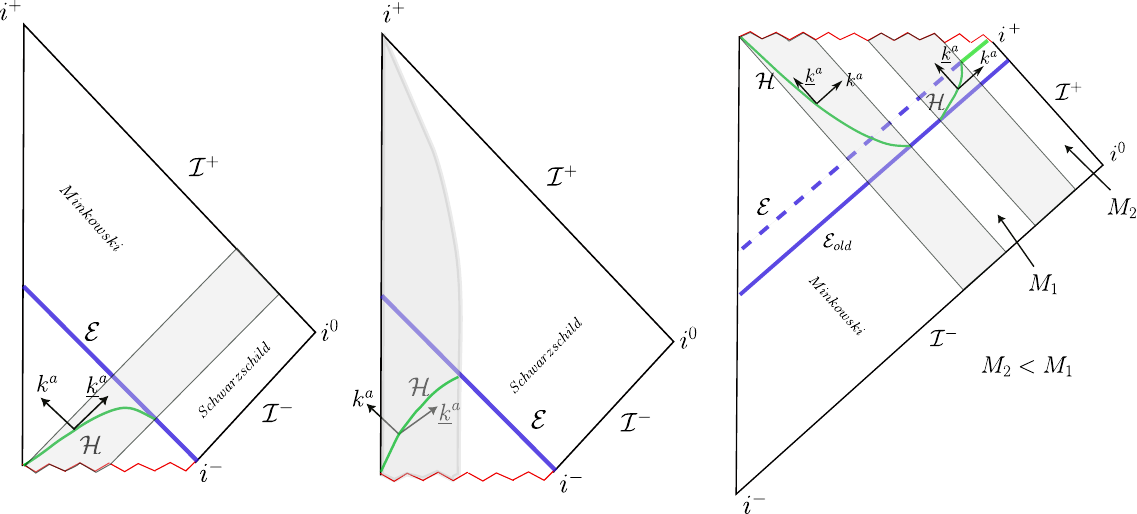}
    \end{center}
  \caption{\footnotesize{\emph{Left Panel: Disappearance of a WH via
        emissions of a null fluid.} This is a time reversal of the
      middle panel of Fig. 1. The { spherical} DHS is again space-like but now a
      AT-DHS.\,\, {\emph{Middle Panel: Disappearance of a while hole
          by emission of an OS star.}} This is a time reversal of the
      left panel of Fig. 1. DHS is again time-like but now a
      AT-DHS.\,\, {\emph{Right Panel: Two roles of infalling null
          fluids.}} An infalling null fluid with $\mathcal{E} >0$
      leads to a BH of mass $M_1$. Here we have a space-like
      T-DHS.  A second in-falling null fluid but now with
      $\mathcal{E} <0$ reduces its mass to $M_2 < M_1$. Here we have a
      time-like T-DHS. This process mimics the formation of a BH by
      Vaidya collapse and subsequent (partial) evaporation due to
      infalling negative energy flux as in the Hawking process \cite{Sawayama:2005mw,Hayward_2006}. 
      Note that now there are MTSs outside the EH $\mathcal{E}$; this is possible
      because the energy condition is violated by the second infalling null fluid.
      Finally, all these examples, QLHs are spherically symmetric. More general examples of MTSs and
      DHSs can be found in \cite{Podolsky:2009an}
      }}
\end{figure}

On the other hand, if $\mathcal{E}$ is negative, $\Lie_{\uk} \theta_{(k)} >0$. Therefore, if $\DH$ is a T-DHS, then as one moves along $\uk^a$ across $\DH$ one passes from a trapped region to an untrapped region. Similarly, if $\DH$ is an AT-DHS, then as one moves along $\uk^a$ across $\DH$ one passes from an untrapped region to an anti-trapped region. For these cases to occur, not only is the presence of matter violating the weak energy condition essential but as we will find in section \ref{s3.2}, its stress energy has to satisfy a stringent condition. These DHSs are not relevant to the BBH problem in which one works with vacuum Einstein's equations for which $\mathcal{E}$ is non-negative. Nonetheless, from a mathematical physics perspective, we have an open issue: \vskip0.1cm
\texttt{Open Issue 3 (OI-3)} Do space-like DHs with $\mathcal{E} <0$ appear in any physically interesting situations? 
\vskip0.1cm
\emph{(3.b)} \emph{Time-like DHS:} In this case $f$ is negative. Therefore, $\mathcal{E}$ and  $\Lie_{\uk} \theta_{(k)}$ have the same signs. If $\mathcal{E}$ is positive, then $\Lie_{\uk} \theta_{(k)} >0$. Consequently, if $\DH$ is a T-DHS, as one moves along $\uk^a$ across $\DH$, one moves from a trapped region to an untrapped region. This situation is realized in the BH formation by an Oppenheimer-Snyder (OS) stellar collapse depicted in the left panel of Fig.~1 \cite{Bengtsson_2013}. If $\DH$ is an AT-DHS, then as one moves across $\DH$ along $\uk^a$, one moves from an untrapped region to an anti-trapped region. An example is provided by the disappearance of a white hole via emission of an OS star. See the middle panel in Fig. 2. 

On the other hand, if $\mathcal{E}$ is negative, then $\Lie_{\uk} \theta_{(k)} <0$. Therefore, if $\DH$ is a T-DHS, then as one moves along $\uk^a$ across $\DH$ one passes from an untrapped region to a trapped region. An example is provided by the following modification of the right panel in Fig. 1: In the Vaidya segment, use a null fluid with negative infalling flux $\mathcal{E}$ so that the final IHS (which coincides with the future portion of the final EH) moves inwards. In this case the DHS within the null fluid is time-like and trapping. This situation occurs also in the semi-classical phase of BH evaporation {\cite{Sawayama:2005mw,Hayward_2006,g2022blackholeentropyloop,ashtekar2023regularblackholesloop,ashtekar2025blackholeevaporationloop}}. Similarly, if $\DH$ is an AT-DHS, then as one moves along $\uk^a$ across $\DH$ one passes from an anti-trapped region to an untrapped region. An example is provided by a space-time in which a Schwarzschild BH of mass $M_1$ is formed by a Vaidya collapse of a null fluid with $\mathcal{E} >0$ and then loses part of its mass because of infalling null fluid with $\mathcal{E} <0$ and settles to a Schwarzschild BH with mass $M_2 < M_1$. See the right panel in Fig. 2.
\vskip0.1cm

We will round up this general discussion on properties of DHs by summarizing an interesting  interplay between and DHSs $\DH$ and space-time Killing vector fields $\xi^a$. In particular, this interplay explains why DHSs cannot exist in stationary space-times and it is extremely useful in the analysis of DHSs in axisymmetric space-times. 

Let $S$ be an MTS in $\DH$ and suppose there is a space-time neighborhood $\mathcal{N}$ of $S$ in which the 4-metric $g_{ab}$ admits a Killing field $\xi^a$. Then there are strong restrictions on the behavior of $\xi^a$ at $S$. We summarize them in the order of increasing generality:\vskip0.1cm
(i) Suppose $\xi^a$ is tangential to $\DH$. Then it must be tangential to every MTS $S$ of $\DH$ in $
\mathcal{N}$ \cite{Ashtekar:2005ez}.\\
 This case is of particular interest to axisymmetric space-times.
\vskip0.1cm 
\indent (ii) Suppose $\DH$ is space-like, $\theta_{(\uk)} <0$ and $\mathcal{E}$ is nowhere vanishing on $\DH$, and the null energy condition (NEC) holds. Then $\xi^a$ cannot be everywhere transversal to any MTS $S$ \cite{Ashtekar:2005ez}.\\
Thus, (i) and (ii) consider cases in which $\xi^a$ has complementary behavior with respect to $S$. But they leave open the possibility that $\xi^a$ may be tangential to a closed subset of $S$ and transverse to its complement. This issue is taken up in the results that follow.\vskip0.1cm 
\indent (iii) Denote by $\xi^a_{\perp}$ the projection of $\xi^a$ orthogonal to $S$. Suppose that $S$ is stable along a direction parallel to $\xi^a_{\perp}$. Then $\xi^a$ is either everywhere tangential to $S$, or everywhere transversal to it \cite{Andersson:2007fh}. (For a discussion of stability of MTSs, see section \ref{subsec:stability}.)\\
If, in addition, $\mathcal{E}$ does not vanish anywhere on an MTS $S$ and the NEC holds, then, by (ii), $\xi^a$ cannot be everywhere transversal. Hence $\xi^a$ must be  everywhere tangential to $S$.
\vskip0.2cm
\emph{The remaining results do not require stability, nor the condition on $\mathcal{E}$.}\vskip0.1cm 
\indent (iv) A neighborhood $\mathcal{N}$ of an MTS $S$ of a DHS cannot not admit a time-like Killing field \cite{Mars:2003ud,ak-kvfs}.\\
In particular, then, \emph{a region $R$ of space-time in which the metric $g_{ab}$ is stationary  cannot contain a DHS.} This is precisely what one would expect given that DHSs are `dynamical'.
This result holds also for $\xi^a$ that are causal (rather than time-like) if NEC holds. 

The case in which $\xi^a$ is space-like in $\mathcal{N}$ turns out to be more involved because in this case the 3-flats spanned by $\xi^a$ and the tangent space $T_S$ at any point of $S$ can be space-like, null or time-like and one has to consider the sub-cases separately. But the final result is rather simple.
\vskip0.1cm 
\indent{(v)} The neighborhood $\mathcal{N}$ of an MTS $S$ in a DHS cannot contain a  space-like Killing field $\xi^a$ that is everywhere transverse to $S$.  If it is everywhere tangential to $\DH$, then by (i) it is tangential to every MTS $S$ in $\DH$ in this neighborhood.  If $\xi^a$ is transverse to $S$ only on an open subset $\mathcal{O}$ of $S$, then the 3-flats spanned by $\xi^a$ and $T_S$ must be null on $\mathcal{O}$ and $\mathcal{E}$ must vanish on $\mathcal{O}$ \cite{ak-kvfs}.\\
Thus, the case in which a space-like Killing field $\xi^a$ fails to be everywhere tangential to MTSs is non-generic. Furthermore, in this case, using the fact that the 3-flats are null, one can show that the projection $\t{\xi}^a$ of $\xi^a$ into the MTS $S$ is a Killing field of the intrinsic metric $\t{q}_{ab}$ of the MTS.\vskip0.1cm
In NR, one typically solves the Cauchy problem and DHSs arise as world tubes of MTSs $S$ located on Cauchy surfaces $\Sigma$, whence $S$ lie in $\Sigma$. If $\xi^a$ is tangential to $\Sigma$, then the 3-flat it spans together with the tangent space to $S$ would be space-like. Hence the second part of (v) implies that $\xi^a$ must be everywhere tangential to $S$. But this result, as well as (i) - (v) assume that $S$ is an MTS in a DHS. The last result drops this assumption and focuses just on MTSs $S$ that lie on $\Sigma$, without reference to a DHS:\\
\indent{(vi)} If an initial data set on a Cauchy slice $\Sigma$ admits a Killing vector that it not tangential to an MTS $S$ in $\Sigma$, then that MTS is unstable \cite{Booth_2024} (see also \cite{Carrasco:2009sa}.)\\
Again, this result supports the intuitive expectation that space-like Killing fields would be generically tangential to MTSs that lie on a DHS. 

In most physical applications to dynamical space-times, space-like Killing fields $\xi^a$ would be rotational. In this case, the projection $\t\xi^a$ is also a Killing field of the intrinsic metric $\t{q}_{ab}$ with closed orbits. Thus, in this case, every MTS of $\DH$ is axisymmetric with $\xi^a$ or $\t\xi^a$ as the rotational Killing field. This concludes discussion on the interplay between DHSs and isometries.  We will now make a few remarks on the entire subsection.\vskip0.1cm 
\vskip0.1cm 

\emph{Remarks:}

1. In the above discussion Einstein's equations were used \emph{only} to motivate the descriptor `null energy flux' for $\mathcal{E}$. Properties of DHSs discussed above hold in any metric theory of gravity. 

2. Like NEHSs,  DHSs are again 3-manifolds defined in their own right; the definition \emph{does not make any reference} to a foliation of space-time by (partial) Cauchy surfaces. In particular, then, if one were to draw a Cauchy surface $\Sigma$ passing through an MTS $S$ of a given DHS, that $S$ need not be the \emph{apparent horizon} in $\Sigma$ (i.e., it need not be the outermost MTS that lies in $\Sigma$). In fact this situation arises in every BBH merger! As we discuss in some detail in section \ref{s4}, immediately after the common DHS $\DH_{\rm R}$ (representing the remnant) forms, the DHSs\, $\DH_i$ ($i$=1,2)\, representing the two progenitors continue to exist. So if one were to draw a Cauchy surface $\Sigma$, it would intersect all three DHSs in MTSs. Only the MTS that lies on $\DH_{R}$ can be an apparent horizon on $\Sigma$; the MTSs that lie on \,$\DH_i$\,  would \emph{not} be. But the results on MTSs on DHSs are applicable to MTSs on $\DH_i$ as well.

3. On the other hand, since in numerical GR one solves the Cauchy problem, one is led to find MTSs on Cauchy slices which, upon evolution, yield DHS $\DH$. In these simulations, the DHSs \emph{are} closely tied to the 3+1 split used in the simulation. Mathematical issues related to this procedure of finding DHSs are discussed in section 5 of \cite{Ashtekar:2005ez}. In particular, different choices of space-time foliations can lead to distinct DHSs, all representing a given BH. While there are results that control `how many', much further work remains to understand this freedom. We will return to this issue in sections \ref{s3.4} and \ref{s7}. 

4. In the context of BHs in space-times that admit an asymptotic region (as opposed to cosmological situations)  one can distinguish between `outward pointing' and `inward pointing' null normals to any given 2-sphere $S$. If $k^a$ is `outward pointing' the MTS $S$ (on which $\theta_{(k)} =0$) is called \emph{outer} and denoted by MOTS. On the other hand a different terminology is often used while discussing  trapping horizons \cite{Hayward:1993wb}. These horizons have been called \emph{outer} if $\Lie_{\uk} \theta_{(k)} <0$ and \emph{inner} if $\Lie_{\uk} \theta_{(k)} >0$, without reference to whether $k^a$ is outward pointing or inward. These conflicting usages often leads to confusion. Similarly, trapping horizons have been called \emph{future} if $\theta_{\uk} <0$ and \emph{past} if $\theta_{\uk} >0$. But then in the BH evaporation process the past horizon so defined can lie to the future of the future horizon so defined \cite{Ashtekar_2020}, which again causes confusion. Therefore, we will avoid using the terminology `outer/inner'  or `future/past' in the context of QLHs.\vskip0.1cm

We will conclude this subsection with some heuristics on the nature of DHSs one can expect to encounter in the BBH problem in GR. Let us assume that matter, if any, is such that $\mathcal{E} >0$ on the DHS, and focus on T-DHSs since we are interested in BHs rather than white holes. Then of the 8 possibilities shown in the two Tables only 2 are viable: The first rows in each of the two tables. Let us analyze the two cases.\\
\indent (i) Time-like DHSs $\DH$. These can occur because, e.g., the progenitor BHs may be formed via an OS stellar collapse. Suppose this happens for the first BH. Immediately after the collapse ends, we would have vacuum equations and the time-like DHS of this BH would join a null IHS on which $k^a$ as the null normal and $\uk^a$ points from the untrapped into the trapped region (as in the left panel of Fig.~1). But very soon, gravitational radiation from the second BH starts falling into the trapped region of the first and the IHS becomes a DHS. But initially the flux would be weak and the process adiabatic. Hence $\uk^a$ would continue to point from the untrapped into the trapped region. Therefore it follows from the first row of Table 2 that this DHS cannot be time-like; it has to be space-like as in the first row of Table 1. With further infall of radiation its area would continue to grow as long as it remains a DHS. Thus, except for the initial formation process, the DHS of the first BH would be space-like once we enter the vacuum region.  Similarly, the common DHS representing the remnant would approach the Kerr IHS in the distant future with $k^a$ as its normal and $\uk^a$ pointing from untrapped to trapped region. Therefore this DHS will also be space-like. \\
\indent (ii) Space-like DHSs $\DH$, as e.g. with  progenitors formed by a significantly inhomogeneous stellar collapse. Then the above considerations show that, once the collapse ends and one has vacuum equations, $\DH$ would continue to be space-like with infall of radiation until it ceases to be a DHS.\vskip0.05cm 
\noindent In both cases, $\DH$ can cease to be a DHS if the intrinsic metric starts changing signature, or $\theta_{(\uk)}$ starts changing signs on an MTS, which can occur if we arrive at an MTS that fails to be strictly stable in the sense of \cite{Andersson:2007fh,Andersson:2008up}. As we discuss in section \ref{s4}, this does occur on the QLHs of progenitor BHs but only after a common DHS enveloping the two progenitors has formed. Then there is a short phase of nontrivial dynamics during which the QLHs of the progenitor BHs join on to the common QLH of the remnant which quickly becomes a space-like DHS that continues to expand. While this phenomenon is very interesting from a geometrical perspective, the short phase during which the QLHs cease to be space-like DHSs is not captured in the NR simulations because they do not probe the evolution inside the common DHS.

These considerations provide an intuitive understanding of why space-like DHSs are the most relevant ones for the BBH problem, as is evident from NR simulations \cite{PhysRevLett.123.171102,Booth:2021sow,Chen_2022}. Therefore, in our discussion of issues related to gravitational waves in section \ref{s6} we will focus on space-like DHSs.

\subsection{\emph{The second law of horizon mechanics}}
\label{s3.2}
\vskip0.1cm

Hawking's discovery of the second law of BH mechanics using event horizons was a major breakthrough in the early 1970s \cite{Hawking:1971vc}. However, that law is a \emph{qualitative} statement: it only says that the EH area cannot decrease under physically well motivated assumptions. It does not relate the increase in area with any local physical processes. Indeed, such a relation cannot exist for EHs: As the Vaidya space-time of the middle panel of Fig. 1 shows, the EH area can increase over an extended interval even when it lies in a flat region of space-time! As we will now discuss, the situation is completely different for QLHs: there is a direct and \emph{quantitative} relation between the change in the area of MTSs on a DHS and \emph{local, physical} fluxes across it.
\begin{figure}[]
\begin{center} 
    \includegraphics[width=0.6\columnwidth,angle=0]{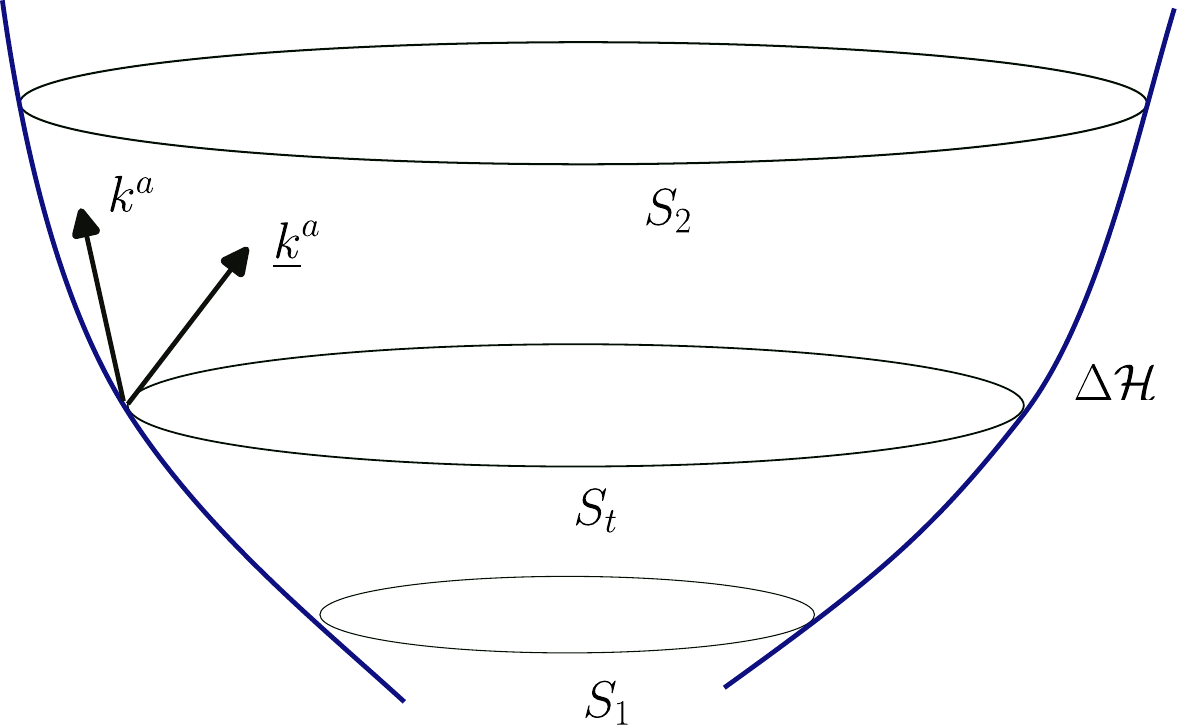}
\caption{\footnotesize{\emph{Space-like DHS:} $\Delta \DH$ is a portion of the DHS $\DH$ bounded by two cross-sections $S_1$ and $S_2$ and an intermediate cross-section $S_t$ at time $t$.}}
\end{center}
\label{fig-dh}
\end{figure}

Let $\DH$ be a dynamical horizon. Recall that $\DH$ is foliated by MTSs and that the area of these MTSs changes monotonically. We will use the areal radius $R$ of the MTSs as a coordinate on $\DH$ and focus on the portion $\Delta \DH$ of $\DH$ bounded by two MTSs, $S_1$ and $S_2$ with $R_2 >R_1$. Since $\DH$ is a space-like or time-like submanifold, the intrinsic metric $q_{ab}$ and the extrinsic curvature $K_{ab}$ of $\DH$ are subject to the four constraint equations of GR that follow directly from the geometric Gauss-Codazzi equations. For a space-like $\DH$ depicted in Fig.~3, we have
\ba \label{constraints1} H &:=& \mathcal{R} - K^2 +K^{ab}K_{ab} - 16\pi G\, T_{ab}\, \h\tau^a \h\tau^b\, \=\,0, \quad {\rm and}\nonumber\\ 
H^a &:=& D_b(K^{ab} - K q^{ab}) -8\pi G\, T_{bc}\, \h\tau^b q^{ac}\,\=\,0\, , \ea
where $D$ and $\mathcal{R}$ are the derivative operator and the scalar curvature of the 3-metric $q_{ab}$ on $\DH$;\, $K_{ab}$ is the extrinsic curvature of $\DH$;  and $\h\tau^a$, the unit future directed time-like normal to $\DH$. {We will fix the rescaling freedom in $k^a, \uk^a$ by setting $\sqrt{2}\h\tau^a\, \=\, (k^a + \uk^a)$.} 

For a time-like $\DH$, we have
\ba \label{constraints2} H &:=& -\mathcal{R} - K^2 +K^{ab}K_{ab} - 16\pi G\, T_{ab}\, \h{r}^a \h{r}^b\, \=\,0, \quad {\rm and}\nonumber\\ 
H^a &:=& D_b(K^{ab} - K q^{ab}) -8\pi G\, T_{bc}\, \h{r}^b q^{ac}\, \=\,0\, , \ea
with $\h{r}^a$ the unit space-like normal to $\DH$ { and we fix the rescaling freedom by requiring $\sqrt{2}\, \h{r}^a\,\=\, k^a - \uk^a $}. For the second law, one only needs these equations, which are again \emph{local to $\DH$}. This is why the second law for the portion $\Delta \DH$ of $\DH$, bounded by two MTSs $S_1$and $S_2$,  will refer only to a local physical process occurring \emph{on} $\Delta \DH$ (see Fig.~3). There will be no teleology. 

Physically, it is natural to expect the change in area to be related to the flux of an appropriately defined energy across $\Delta \DH$.  Now, the notion of energy is associated with a causal vector field $\xi^a$. Since $k^a$ naturally provides a causal direction on any DHS $\DH$, it is natural to set $\xi^a \, \= \, \sqrt{2}\,N\, k^a$ for some `lapse function' $N$. The choice that is well-adapted to the second law of DHSs turns out to be $N:\= \, |D R| \equiv (\pm\,\, q^{ab}\, D_a R\, D_b R)^{\f{1}{2}}$ where $\pm$ refers space-like/time-like $\DH$. (Technical simplifications occur with this choice because the intrinsic volume element of $\DH$ satisfies $N\,\rmd^3 V= \rmd R \rmd^2 {V}$ where $\rmd^2 {V}$ is the volume element on MTSs $S$. Other choices of lapse functions also give rise to interesting equalities relating geometrical quantities on $S_1$ and $S_2$ to fluxes across $\Delta \DH$. In particular, they provide an `active' version of the first law relating changes in the observables associated with the DHS to physical processes; see \cite{Ashtekar:2003hk}.) 

It turns out that the second law of DHS mechanics --that relates the change in the area of MTSs to the flux of $\xi^a$-energy-- it equivalent to the statement that the integral of a linear combination of constraints along $\xi^a$ vanishes! Let us consider first the case when $\DH$ is space-like. Then $\xi^a\, \=\,  {N}\,(\tau^a + \hat{r}^a)$. By the setting the linear combination  $\int_{\Delta \DH} N (H + H_a \h{r}^a)\, d^3 V\, \=\,0$ after much simplification one arrives at the following equivalent form of this equality \cite{Ashtekar:2003hk}:
\be \label{2ndlaw1} \hskip-1cm \f{1}{2G}\, (R_2 - R_1) \,\=\, \int_{\Delta \DH} T_{ab} \hat{\tau}^a \xi^b \rmd^3 V \, +\, \f{1}{8\pi G}\, \int_{\Delta \DH} N\, \big(\sigma_{(k)}^{ab}\, \sigma^{(k)}_{ab}\, +\,  2 \zeta_{(k)}^a\,\zeta^{(k)}_a \big)\,\rmd^3 V\,\,\,\, \ee
where, as before $\sigma^{(k)}_{ab}$ is the shear of the null normal $k^a$ and $\zeta_{(k)}^a = \t{q}^{ac} (\h{r}^b \nabla_b k_c)$. (We will see in section \ref{s3.3} that $\zeta_{(k)}^a\mid_S$ encodes the angular momentum content of any MTS $S$ in $\DH$.) Note that, since $\theta_{(k)} \,\=\,0$ on any MTS,  the left side is just the difference between the Hawking's quasi-local  mass (\ref{MH})  associated with the MTSs $S_2$ and $S_1$. The first term on the right side is just the flux of the matter contribution to the energy $E_{\Delta \DH}^{(\xi)}$ (defined by $\xi^a$) across $\Delta \DH$. It is { non-negative} if $T_{ab}$ satisfies the dominant energy condition. The second term is  manifestly positive. Furthermore, there are detailed arguments that lead one to interpret the second term as the flux of gravitational energy across $\Delta \DH$ defined by $\xi^a$ \cite{Ashtekar:2003hk}. Interestingly, while we do not have a gauge invariant notion of gravitational energy flux across a generic space-like surface in GR, a viable notion emerges if the surface happens to be a DHS because it is foliated by MTSs! Thus, the second law of DH-mechanics can be interpreted simply an energy balance law: The change in the Hawking mass in the passage from $S_1$ to $S_2$ is given by the flux of energy $E_{\Delta \DH}^{(\xi)}$ across $\Delta \DH$.  This is in sharp contrast with the second law of EH mechanics which is a qualitative statement, with mysterious teleological connotations as illustrated by the Vaidya collapse. The second law of DHS mechanics in this collapse is simply a statement that the area of the DHS increases in direct and local response to the flux of matter into the BH. \vskip0.4cm \goodbreak

\emph{Remarks}: \vskip0.05cm
1. The fact that the area of MTSs changes monotonically on any DHS $\DH$ is a rather trivial consequence of the fact that $\theta_{(k)} \, \= \,0$ and $\theta_{(\uk)}$ does not vanish anywhere on $\DH$. But constraint equations of GR provide much more:  a \emph{quantitative relation} between this growth in area and an energy flux across $\Delta \DH$. In the physically interesting case when $\DH$ is a T-DHS, this area increase occurs because of the energy flux from the untrapped region into the trapped region (as in the middle panel of Fig.~1). In the case when it is an AT-DHS, it is because of the energy flux from the anti-trapped region to the untrapped region (as in the left panel of Fig.~2).\vskip0.05cm

2. The field $\zeta_a^{(k))}$ may seem somewhat mysterious for those who are accustomed to perturbations at the EH of the Kerr solution because it vanishes on null surfaces. More generally, the flux of energy across perturbed an IHS involves only the $|\sigma_{ab}^{(k)}|^2$ term \cite{Ashtekar:2021kqj}; the appearance of an additional $|\zeta_a^{(k)}|^2$ term in (\ref{2ndlaw1}) is a genuinely non-perturbative feature of DHs. Its origin lies in the fact that $\DH$ is a space-like surface, rather than null. On a null surface, there are only two \emph{phase space} degrees of freedom per point (encapsulated in $\sigma_{ab}^{(k)}$ on a perturbed IHS, as well as at $\scrip$). On a space-like surface, on the other hand, there are four \emph{phase space} degrees of freedom and these are captured in $\sigma_{ab}^{(k)}$ and  $\zeta_a^{(k)}$.

3. The second law (\ref{2ndlaw1}) of DHSs has an interesting consequence on gravitational wave luminosity at DHSs. Let us suppose that vacuum Einstein's equations hold at $\DH$. Then, the flux of energy carried by gravitational waves across $\Delta\DH$ is given by\, $ \Delta E_{\rm GW} =\f{1}{8\pi G}\, \int_{\Delta \DH} N\, \big(\sigma_{(k)}^{ab}\, \sigma^{(k)}_{ab}\, +\,   2 \zeta_{(k)}^a\,\zeta^{(k)}_a \big)\,\rmd^3 V$\,. To compute luminosity, we need a notion of time. A natural notion at the horizon is provided by the area-radius, i.e., $c^{-1}R$, since it is monotonic and takes constant values of MTSs. (Here, for comparison with the literature we have reinstated the speed of light $c$.)
Then, using the fact that $N \rmd^3V = \rmd R \rmd^2V$ one obtains for luminosity:
\be \label{lum1} \mathfrak{L}_{\rm GW} [S] := \lim_{\Delta R \to 0}\, \f{\Delta E_{\rm GW}}{c^{-1}\Delta R}\,\, = \,\, \f{c^5}{8\pi G}\oint_S \Big(\sigma_{(k)}^{ab}\, \sigma^{(k)}_{ab}\, +\,  2 \zeta_{(k)}^a\,\zeta^{(k)}_a \Big)\,\rmd^2 V.\,\,\,\, \ee
By taking the same limit on the left hand side of Eq. (\ref{2ndlaw1}) one finds that this luminosity equals
\be \label{lum2} \mathfrak{L}_{\rm GW} [S]  = \f{c^5}{2G} \, .\ee
{ Note that(\ref{lum2}) can be regarded as a reinterpretation of the second law of DHS mechanics; it is just the infinitesimal version of (\ref{2ndlaw1}), and holds only if the MTS $S$ lies on a DHS. For example, if $S$ were an MTS in an IHS, then there would be no energy flux across $S$ whence the right side of (\ref{lum2}) would be identically zero.}

This result has two intriguing aspects. First, it says that this luminosity is the same for all MTSs of a given DHS and, furthermore, it is the same for all DHSs; it does not carry any imprint 
{{of specifics of the}  dynamical phase of the BH (so long as it is evolving, i.e. represented by a DHS rather than an IHS). { This feature seems surprising because one's first expectation would be that luminosity would be very low in the long in-spiral phase and high at merger. But luminosity is the energy radiated per unit time and on the DHS (which lies in the strong field region) the natural clock  is provided by the monotonic area of its MTSs. As seen by a far away observer, a `unit time-interval' measured by this clock in the in-spiral phase lasts \emph{much} longer than that during merger. This is why the luminosity at the DHS is the same in the inspiral phase as it is at merger.} A second intriguing aspect is that the value of the luminosity of a DHS is a fundamental constant. Now, It was suggested by Misner, Thorne and Wheeler \cite{MTW} that there may be an upper limit on the luminosity of gravitational waves of the order of $\sim\f{c^5}{G}$. This expectation is unrelated to our finding because, again, one has to keep in mind that this suggestion referred to luminosity at null infinity where an asymptotic time translation provides the notion of time. 

4. While (\ref{constraints1}) are the constraint equations of GR, in essence they are consequences of geometric Gauss and Codazzi identities. Therefore, if we simply substitute $8\pi G\,T_{ab}$ by the Einstein tensor in (\ref{2ndlaw1}), the resulting second law
\be \label{2ndlaw2} \hskip-1.8cm \f{1}{2G}\, (R_2 - R_1) \,\=\, \f{1}{8\pi G}\, \int_{\Delta \DH}\!\Big[(R_{ab} - \f{1}{2}\,R g_{ab})\, \hat{\tau}^a \xi^b \, +\, N\, \big(\sigma_{(k)}^{ab}\, \sigma^{(k)}_{ab} \,+\,2\, \zeta_{(k)}^a\,\zeta^{(k)}_a\big)\,\Big]\,\rmd^3 V\, \ee 
holds in \emph{any metric theory gravity}, providing us with a quantitative measure of area increase of MTSs on a space-like DHS in terms of `local happenings' on $\Delta \DH$. (Interestingly, in scalar-tensor theories, if $\DH$ is a DHS in the Einstein conformal frame and the conformal factor $\Omega$ relating the Einstein and Jordan frames satisfies $\Lie_{k} \Omega \,\=\,0$, then $\DH$ is also a DHS in the Jordan frame and Eq. (\ref{2ndlaw2}) holds in \emph{both} conformal frames. But physically one would be interested in the change in the area of MTSs only in the Einstein frame).  In a general metric theory, however, there is no a priori reason for $R/2G$ to be a viable quasi-local mass associated with an MTS, nor for the right hand side to be the energy flux. \vskip0.1cm
\texttt{Open Issue 4 (OI-4):} Do the integrals on the right hand side admit a physical interpretation in alternate metric theories of relativistic gravity --such as $f(R)$ and scalar-tensor-- that are have received considerable attention in the gravitational wave community?
\vskip0.2cm

Finally, let us consider the case when $\DH$ is time-like. One again obtains a quantitative relation similar to (\ref{2ndlaw1}) between the increase in the Hawking quasi-local mass as one passes from an MTS $S_1$ to an MTS $S_2$ and a flux across the portion $\Delta \DH$ of $\DH$ bounded by $H_1$ and $H_2$ (see Appendix B of \cite{Ashtekar:2003hk}). But the energy flux --e.g. $T_{ab} \h{r}^a \xi^b$-- across a time-like surface can have either sign because the normal $\hat{r}^a$ to it is space-like, while the left side is again positive. Therefore there are non-trivial constraints on the integrals that appear on the right hand side. Explicit calculations shows that they \emph{are} satisfied in the OS collapse as well as the three other examples we considered for time-like $\DH$, as indeed they must be.

\subsection{\emph{Multipole Moments of DHSs}}
\label{s3.3}
\vskip0.1cm

To begin with, let us consider axisymmetric DHSs. Since DHSs are foliated by MTSs, as we saw in section \ref{s2.2}, the rotational Killing field $\varphi^a$ is necessarily tangential to each MTS. It is natural to use MTSs for 2-spheres $S$ that feature in the expression (\ref{dimensionless1}) of multipoles of IHSs. The seed function for the shape multipoles $\texttt{I}_{\ell}$ is { again the} scalar curvature $\t\R$, and the seed 1-form $\t\omega_a$ for the spin multipoles is given by $\t\omega_a = -\t{q}_a{}^b \uk_c \nabla_b k^c$. (Note that for IHSs this definition yields the rotational one form as the one introduced in section \ref{s2.2} but on DHSs the expression features the space-time derivative operator $\nabla_a$ rather than $D_a$ that is intrinsic to the horizon.) This definition of $\t\omega_a$ is justified, e.g., by the fact that, in the standard Arnowitt-Deser-Misner phase space, the horizon contribution to the angular momentum is again given by 
\be \label {J-DH} J_{\DH}^{(\varphi)} = - \f{1}{8\pi G}\, \oint_S \t\omega_a \varphi^a\, \rmd^2 {V}\, , \ee 
for any MTS $S$. Then multipoles are defined using the same expression (\ref{dimensionless1}) as in the case of IHSs. However, except for the shape monopole and the spin dipole they now change in time, i.e., as we move from one MTS $S$ to another: we obtain a 1-parameter family of multipoles $\texttt{I}_\ell [S],\,\, \texttt{S}_\ell [S]$ that encodes the dynamics on $\DH$ in an invariant manner. For details, see \cite{Schnetter:2006yt}.

However, except for an axisymmetric collapse and a head-on collision of non-spinning BHs, the intrinsic metrics $q_{ab}$ on DHSs of physical interest can be very far from being axisymmetric. In particular, immediately after a common DHS forms in a merger, not only are the MTSs highly distorted but the spin vector can be highly dynamical. But eventually the DHS asymptotes to a Kerr IHS which \emph{is} axisymmetric. Therefore, it is of considerable interest to consider DHSs which are themselves not axisymmetric but which asymptote to an axisymmetric IHS. For simplicity, we will also assume that the DHS is space-like because that is most interesting case in BBH mergers. { And} we will focus on the geometrical multipoles $\texttt{I}_{\ell, m},\, \texttt{S}_{\ell,m}$; the mass and angular momentum multipoles can be obtained by rescaling these by the same dimensional factors as in Eq. (\ref{dimensionfull}).

The key problem is that, since metrics $\t{q}_{ab}$ on the MTSs $S$ are not axisymmetric, we do not have a direct access to Legendre polynomials $P_\ell (\zeta)$ and, furthermore, we now need full spherical harmonics $Y_{\ell,m}$, not just $P_\ell$. Not only should these weighting functions $Y_{\ell,m}$ be introduced in a covariant manner --i.e., using only the geometrical structures that are available on the DHS-- but  we also need to ensure that the $Y_{\ell,m}$'s used on different MTSs are `the same' --i.e., are time-independent-- in an appropriate sense. Otherwise the time dependence in the multipoles $\texttt{I}_{\ell, m},\, \texttt{S}_{\ell, m}$ would not just be due to dynamics of geometries of MTSs but would be contaminated by the spurious time dependence in the weighting functions $Y_{\ell,m}$ themselves. This issue is rather subtle. It was resolved in  \cite{Ashtekar:2013qta} where properties of the resulting multipoles are also discussed. We will summarize only the final result.

The key idea is to first note that in the asymptotic future the DHS tends to a Kerr IHS which is axisymmetric and then use the fact that on axisymmetric IHSs one can introduce a round metric in a natural, covariant fashion (as in section \ref{s2.4}), and consider the spherical harmonics $Y_{\ell, m}(\theta,\varphi)$ it defines. These $Y_{\ell, m}$ are defined in the distant future. In the next step, one Lie-drags them to all the  MTSs of $\DH$ along a suitable vector field $X^a$. This  vector field must have the following properties:\\
(i) It must be constructed using just the geometric structures naturally available on $\DH$;\\
(ii) The 1-parameter family of diffeomorphisms it generates must preserve the foliation of $\DH$ by MTSs;\\
(iii) It must satisfy $\Lie_X \,(R^{-2} \t\epsilon_{ab})\, \= \,0$, so that the $Y_{\ell, m}$'s induced on every MTS constitute an orthonormal set; and,\\
(iv) If the 3-metric $q_{ab}$ happens to be axisymmetric, then the diffeomorphisms generated by $X^a$ should preserve the rotational Killing field $\varphi^a$ of $q_{ab}$. This property guarantees that the multipoles constructed using the general procedure agree with those defined in the axisymmetric case in \cite{Schnetter:2006yt}, using the presence of the Killing field $\varphi^a$ .\\
It turns out that one can explicitly construct the class of vector fields $X^a$ satisfying these properties. Any two vector fields in this class are { proportional.} The weighting fields $Y_{\ell, m}$ they induce by the Lie-dragging procedure are insensitive to the rescaling freedom in $X^a$. 

With spherical harmonics on $\DH$ at hand, the geometric multipole moments are defined in terms of  $\t\R$ and $\t\omega_a$ as in Eq.(\ref{dimensionless1})
\be  \label{dimensionless2} \texttt{I}_{\ell,\,m} [S]\,  - \, i \texttt{S}_{\ell,\, m} [S]:= \f{1}{4}\, \oint_S \big[\,\t\R\, +\, 2i \,\t\epsilon^{ab} \t{D}_a \t\omega_b\,\big]\, Y_{\ell,\,m} (\theta,\varphi) \, \rmd^2 {V}\, . \ee
But now they depend on the choice of the MTS $S$, and this dependence encodes the dynamics of the horizon shape and spin structure in an invariant fashion. Also, while on IHSs the term in the square bracket could be replaced by $ -\,4 \Psi_2$, this is no longer the case on DHSs; there are additional terms involving the shear of $k^a$ and $\uk^a$ which vanish on IHSs but not on DHSs (see section \ref{s5.3}). Einstein's equations lead to balance laws:  Differences $\texttt{I}_{\ell,m}[{S}_2] - \texttt{I}_{\ell,m}[{S}_1]$ and $\texttt{S}_{\ell,m}[{S}_2] - \texttt{S}_{\ell,m}[{S}_1]$ of multipoles evaluated at two different cross-sections are expressible as fluxes across the portion $\Delta \DH$ of $\DH$ bounded by $S_1$ and $S_2$ where the integrands contain only \emph{local} geometric and matter fields. They generalize the balance law (\ref{2ndlaw1}), with multipoles replacing the Hawking mass on the left side. For details, see  \cite{Ashtekar:2013qta}.

Finally, there is an interesting interplay between symmetries and multipoles. First, if $(\DH, q_{ab})$ admits a rotational Killing field, $\varphi^a$, then of course all $\texttt{I}_{\ell,\,m} [S]$ vanish if $m\not=0$ for any MTS $S$. If $\varphi^a$ extends as a Killing field of the 4-metric to an arbitrarily small neighborhood of $\DH$, then $\Lie_\varphi\,\t\omega_a =0$ on $\DH$ whence all $\texttt{S}_{\ell,\,m} [S]$ also vanish if $m\not=0$ for any MTS $S$. Furthermore, if $\varphi^a$ happens to be hypersurface orthogonal in this neighborhood, then one can show that $\texttt{S}_{\ell,\,m} [S]$ must vanish for all $\ell, m$ \cite{ak-kvfs}! There is an intuitive expectation that in axisymmetric space-times angular momentum information is encoded in the twist of $\varphi^a$. On DHSs  this expectation is realized in a strong sense: if the Killing field is twist-free, it is not just the spin vector that vanishes on every DHS; \emph{all spin multipoles vanish}! DHSs with hypersurface orthogonal Killing field $\varphi^a$ arise in head-on collisions of on-spinning compact objects discussed in sections \ref{s4.3} and \ref{s6.1} as well as in the collapse of axisymmetric non-rotating stars.
\vskip0.1cm
\goodbreak

\emph{Remarks:}\vskip0.05cm
\noindent 1. One can show that the seed field $\t\omega_a$ that encodes the angular momentum multipoles and the 1-form $\zeta^{(k)}_a$ that enters (\ref{2ndlaw1}) have the same exterior derivative. Therefore, one can replace $\t\omega_a$ by $\zeta^{(k)}_a$ in the definition of spin multipoles. In particular, if $\zeta^{(k)}_a =0$ then all spin multipoles vanish. This is the sense in which $\zeta^{(k)}_a$ encodes the full angular momentum content of each MTS in $\DH$. 

2. Different numerical simulations of BBHs use different codes, gauge choices and coordinates. Therefore it is often difficult to compare their outputs. Multipoles provide a useful tool: DHSs obtained in two different simulations are the same if and only if the multipoles agree. In addition, the balance laws provide checks on accuracy of individual numerical simulations.\\
3. On any given DHS, the evolution of multipoles provides a gauge invariant characterization of its dynamics. We will see in section \ref{s6} that their time dependence brings out an unforeseen role of the quasi-normal frequencies of the remnant in the evolution of the \emph{horizon geometry}. On a DHS that asymptotes to an IHS, the multipoles defined on MTSs tend to those of the IHS in a precise manner \cite{Ashtekar:2013qta}. Therefore, a few cycles after the merger, the geometry of the DHS is well approximated by that of a perturbed IHS. As discussed in section \ref{s6.2}, in this approximation the time dependence of the DHS multipoles plays a key role in relating the strong-field dynamics at the horizon and wave forms at null infinity.  By contrast, in dynamical situations one cannot define multipoles on EHs, first because EHs are generally not smooth manifolds \cite{Chrusciel:1996tw,gadioux2023creasescornerscausticsproperties} and, even when they happens to be smooth, they do not admit a natural foliation.\\
4. As with IHs, the importance of multipoles of a DHS lies in the fact that they provide an invariant characterization of the dynamics of horizon geometry. And this can be achieved by distinct sets of multipoles. Indeed, another set of invariantly defined multipoles was introduced by Owen \cite{Owen:2009sb}. However, in that construction, the spherical harmonics $Y_{\ell, m}$ were introduced on each MTS separately using structures on each MTS, which of course are time dependent. While the procedure is fully covariant, these multipoles owe their time dependence not just to that of $\t\R$ and $\t{D}_{[a} \t\omega_{b]}$, but also of the weighting functions $Y_{\ell, m}$ chosen to define the multipoles. In the procedure described above, the weighting functions, being Lie-dragged along $X^a$, are the same as those used in the final equilibrium state. Therefore, these multipoles directly capture the difference between the shapes and spin structures encoded in $\t\R$ and $\t{D}_{[a} \t\omega_{b]}$ on a given MTS $S$ and those of the final equilibrium state. That is why it is these multipoles have been used in recent numerical simulations (see in particular \cite{Chen_2022}).

\subsection{DHSs: Uniqueness}
\label{s3.4}

DHSs are foliated by a 1-parameter family of MTSs. As we saw in the last two sub-sections, by following the time dependence of the geometry of these MTSs one obtains a detailed picture of BH dynamics in the strong field region of space-time. This powerful application of DHSs raises a natural question:
\begin{itemize}
\item[(a)] Is the foliation by MTSs of the 3-manifold $\DH$ representing a DHS unique? 
\end{itemize} 
There is also a broader question of uniqueness. As pointed out in remark 3 of section \ref{s3.1}, a connected trapped region which can be intuitively thought of as containing a single BH may admit several DHSs. Results summarized above hold for \emph{all of them.} It is nonetheless of interest to ask if there is a canonical choice or at least a controllable set of canonical choices. As the examples in the right Panels of Figs.~1 and 2 illustrate, a single QLH can admit several DHSs. This type of non-uniqueness is tame because these DHSs refer to different dynamical phases of the BH. But one could also have many distinct DHSs associated with the same space-time region. Therefore we are led to ask:
\begin{itemize}
\item[(b)] Are there restrictions on the distinct DHSs all of which are associated with a single dynamical phase of a BH?
\end{itemize} 
\vskip-0.15cm
We now summarize the current situation on these two uniqueness issues.\vskip0.1cm

For space-like DHSs, the answer to the first question is succinct: \emph{The foliation is unique}   \cite{Ashtekar:2005ez}. More precisely using an argument closely related to the  geometric maximum principle for minimal surface theory in Riemannian geometry \cite{andersson1997}, it was shown that if a 2-sphere $S$ is an MTS lying entirely in a space-like DHS $\DH$, then $S$ must necessarily coincide with one of the MTSs in the foliation that $\DH$ comes equipped with. The fact that $\DH$ is not null plays an important role in this analysis. Indeed, the uniqueness does not hold for IHSs --such as the EHs of Kerr BHs-- on which every 2-sphere cross-section is an MTS! 

Now, as noted in section \ref{s3.1} (and also sections\ \ref{s4.1} and \ref{s4.3} below), the QLHs associated with classical BHs are most often space-like DHSs to which this uniqueness result is directly applicable. But time-like DHSs do appear, e.g., in the OS-type stellar collapse, and during the semi-classical phase of the quantum evaporation. In the time-like case, the uniqueness result continues to hold using the time-like version of the maximum principle proved in \cite{Galloway:1999ny} (personal communication from G.~Galloway).\vskip0.1cm

For the second question, (b), \noindent let us begin with space-times that admit isometries. Bartnik and Isenberg \cite{Bartnik:2005qj} have shown that for spherically symmetric, space-like DHs in spherically symmetric space-times, one can solve the \emph{full set} of GR constraints explicitly, obtain the freely specifiable data, and prove general results on existence and uniqueness. In particular, {if a spherically symmetric space-time admits a spherically symmetric DHS, it is unique  \cite{Bartnik:2005qj}.} It should be possible to extend some of that analysis first to axisymmetric space-times, and then to general space-times, without any symmetry restrictions. Thus we are led to ask
\vskip0.1cm
\texttt{Open Issue 5A (OI-5A):} Can one extend the Bartnik-Isenberg analysis and obtain further insights into existence and uniqueness of space-like DHSs? A formulation of constraints as an evolutionary system \cite{R_cz_2015} may be particularly useful to this analysis.
\vskip0.1cm
\noindent More general QLHs in spherically symmetric space-times have been analyzed by Bengtsson and Senovilla using other techniques.  They have shown that if a spherically symmetric space-time $(M, g_{ab})$ admits trapped surfaces, then it admits a unique spherically symmetric QLH  $\mathfrak{H}$ \cite{Bengtsson_2011} (i.e., $\mathfrak{H}$ that is left invariant by the rotational Killing vectors).  

For results on axisymmetric space-times, let us first recall results from section \ref{s3.1}: If a space-time admits a rotational Killing field $\varphi^a$ then generically $\varphi^a$ is tangential to each MTS $S$ of every DHS. In the exceptional case, $\varphi^a$ could be tangential on a closed portion of MTS $S$ and transverse on its complement. But then $S$ would not be stable in any direction `in between' of the projection $\varphi^a_{\perp}$ of $\varphi^a$ orthogonal to $S$ and $k^a$ \cite{Andersson:2007fh} (and, furthermore, the 3-flat spanned by $\varphi^a$ and the tangent space $T_S$ to the MTS has to be null wherever $\varphi^a$ is transverse to $S$). It would therefore seem that in axisymmetric space-times, DHSs associated with a single dynamical phase of BHs (e.g. post-merger or pre-merger phases in BBH coalescence) would be quite restricted. Thus we are led to ask: \vskip0.05cm

\texttt{Open Issue 5B (OI-5B)} Can one further sharpen restrictions on DHSs that an axisymmetric space-time can admit? Assuming it admits a DHS, does it admit a unique axisymmetric DHS?
\vskip0.1cm
Next, let us consider DHSs in more general spacetimes, without any symmetry restrictions. These have been analyzed for DHSs that are  \emph{regular} in the sense that $\DH$ is globally achronal and also `generic' in that $\mathcal{E}$ never vanishes on $\DH$ \cite{Ashtekar:2005ez}. (A space-like $\DH$ is guaranteed to be locally achronal and the first of the two regularity conditions demands that it be \emph{globally} so, i.e., no two points on $\DH$ can be joined by a time-like curve. Similarly $\mathcal{E}$ cannot vanish identically on any MTS of the DHS but now the second condition asks that it be \emph{nowhere} vanishing. However, the results we report will probably go through it vanishes on sets of zero measure (personal communication from G.~Galloway). 

It turns out that there are strong constraints on the location of possible MTSs in the vicinity of a regular DHS. In turn, they imply that the number of regular DHSs is much smaller than what one might have expected a priori. In particular, regular DHSs cannot be so numerous as to provide a foliation of space-time by them even locally. Even in absence of the genericity condition, such a foliation cannot exist unless space-time is very special. Note that this result holds only for (regular) DHSs. There exist vacuum solutions to Einstein's equations admitting regions which can be foliated by IHSs  \cite{Pawlowski_2004}. Similarly, there exist vacuum solutions that admit regions foliated by \emph{space-like} QLHs  \cite{Senovilla_2003}. But these QLHs fail to be DHSs because $\mathcal{E}$ vanishes on them \emph{everywhere}. (As noted before, these space-times are very special in that all curvature scalars constructed from the metric and their derivatives vanish everywhere). \vskip0.1cm  

Another set of results assumes the null energy condition (NEC) that is commonly used in classical GR. Let $\DH$ be a regular DHS and $D(\DH)$ its domain of dependence. (Thus $D(\DH)$ is the set of points $p$ in the physical space-time such that every inextendible causal curve meets $\DH$). If $\t\DH$ is a  regular DHS that is contained in $D(\DH)$ but not in $\DH$, then no MTS in $\t\DH$ can lie strictly to the past of $\DH$. In effect this means that $\t\DH$ must `weave' $\DH$ so that part of it lies to the future and part to the past of $\DH$. In this case, $\t\DH$ cannot contain an MTS that lies strictly to the past of $\DH$; only a `portion' of it can. These results can be interpreted as saying that DHSs that refer to the same dynamical phase of a BH are `intertwined'  and not disparate. Note also these results only use intrinsic properties of the 3-dimensional DHSs themselves without any reference to a space-time foliation. 

But as mentioned in Remark 2 of section \ref{s3.1} (and discussed more extensively in section \ref{s4.3}), since one solves the Cauchy problem in NR, DHSs are generated by finding MTSs on a 1-parameter family of  (partial) Cauchy slices and stacking them. Such DHSs are further constrained. Let $\DH$ and $\t\DH$ be two regular DHSs obtained by using a given foliation $\Sigma_t$ of partial Cauchy slices in a space-time $(M, g_{ab})$ satisfying NEC, with $t_1 <t <t_2$. Then if for some time $t_0 \in (t_1, t_2)$, the MTSs of  $\DH$ and $\t\DH$ determined by $M_{t_0}$ agree then $\DH = \t\DH$. Thus, DHSs generated by a given foliation cannot bifurcate. We will see in section \ref{s4.3}, the DHSs \emph{can} merge. This is precisely the behavior of BHs we are led to expect from properties of EHs, but now the analysis is quasi-local. 

Another uniqueness result concerns regular DHSs $\DH$ and $\t\DH$  that asymptote to one another (i.e., their past Cauchy horizons agree to the future of a partial Cauchy surface), again in a space-time that satisfies the NEC . If  they are generated by the same foliation $\Sigma_t$, then  $\DH = \t\DH$. This result is of direct interest to the DHS representing the remnant in a BBH merger: It says that a given foliation cannot generate two different DHSs each of which can be taken to represent the remnant. 

The uniqueness results to date addressing the second point, (b), are not surprising in the light of findings of NR simulations. Rather, they serve to explain \emph{why} in any \emph{given} NR simulation one does not encounter a number of phenomena that are a priori possible --such as bifurcating DHSs, or, multiple DHSs, each claiming to represent the remnant in a BBH merger. And providing such an understanding of NR findings constitutes an integral part of the role of mathematical GR.  However, from a physical standpoint, two  important limitations remain: (i) So far, these investigations have only considered DHSs on which $\mathcal{E}$ is \emph{nowhere} zero (or can vanish only on a set occasion zero measure); and, (ii) While there are qualitative results such as `weaving' we are not aware of results providing the detailed relation between DHSs obtained using two different foliations. Also, because of the `global achronality' requirement, results to date on uniqueness refer only to space-like DHSs. Thus we are led to ask: \vskip0.1cm
\texttt{Open Issue 6 (OI-6)}: Can these limitations be overcome? In the case of space-like DHSs, we are led to ask: (i) Are there interesting uniqueness statements that can be established for DHSs where $\mathcal{E}$ vanishes on open regions of their MTSs?, and (ii) Are there results relating the DHSs of, say, the remnants that result from a suitably restricted class of foliations?  In the case of time-like DHSs, what can we say about the possible DHSs representing the same dynamical phase of the BH?

For BBH mergers, issue (i) may not be not relevant in generic situations because in simulations we are aware of $\mathcal{E}$ is always positive except at isolated points in axisymmetric situations (where the isometry forces it vanish at poles). But it is of interest from a mathematical GR perspective and also to other physical applications such as formation of BHs. The uniqueness question for time-like DHSs is also not of great interest in classical GR  but it is directly relevant to the analysis of the BH evaporation process.

\section{Dynamics of QLHs}
\label{s4}

Issues related to the formation and evolution of QLHs in dynamical processes in GR have been explored using geometric analysis and NR. In this section we will summarize the main results which will also serve to bring out the synergy between these two lines of investigation. The discussion is divided into three parts. Subsection  \ref{subsec:stability} introduces a key notion, namely, \emph{the stability operator}, which is then used in \ref{s4.2}, to discuss the time evolution of MTSs using geometric analysis. In the third part, \ref{s4.3}, we summarize illustrative results from high precision numerical simulations that have provided new insights into the strong field dynamics encountered in BBH mergers.

\subsection{The stability operator}
\label{subsec:stability} \label{s4.1}

Recall from section \ref{s2} that a QLH can be thought of as the time evolution of an MTS. We are therefore led to investigate conditions under which a marginally trapped surface evolves smoothly to form a regular world tube. This line of investigation has proven to be a very fruitful, leading to numerous rigorous mathematical results, as well as a synergistic interplay between mathematical and numerical analyses.  

As discussed in section \ref{s2.1}, an  MTS $S$ is a 2-sphere  on which the expansion $\theta_{(k)}$ of a null normal $k^a$ vanishes. In the literature this null normal is said to be pointing `outward' by convention. An important ingredient in mathematical investigations is the notion of \emph{stability} which in turn is based on the first variation of  $\theta_{(k)}$  in outward directions\, \cite{Newman:1987,Andersson:2005gq,Booth:2006bn,Andersson:2007fh,Booth:2020qhb,Booth:2021sow,Booth_2024}.  Given an abstract 2-sphere $\mathcal{S}$, one constructs a sequence of 2-spheres in the given space-time $M$ through a smooth map\,\, $\map:\,\mathcal{S}\times (-\epsilon,\epsilon) \rightarrow M$ with $\epsilon > 0$. For each $\lambda\in(-\epsilon,\epsilon)$, the function\, $\map\,(\cdot\, ,\lambda)$ maps $\mathcal{S}$ to a 2-sphere $S_\lambda$, such that $S_{\lambda=0}$ is the given MTS $S$ and for $\lambda\not=0$,\, $S_\lambda$, represents a variation of the $S$. Since $\map$ is smooth, $S_\lambda$ is a smooth 2-sphere for each $\lambda$. For a fixed point $p$ on $S$, as $\lambda$ varies\, $\map\,(p,\lambda)$ yields a curve. Let $s^a(\lambda)$ be the tangent to this curve.  With this structure at hand, one can define the first variation of geometric fields on MTSs. The general procedure is to calculate the relevant geometric quantity on each $S_\lambda$, and differentiate it w.r.t. $\lambda$ at $\lambda=0$. Each $S_\lambda$ is equipped with null normals $(k^a{(\lambda)},\,\uk^a{(\lambda)})$ depending smoothly on $\lambda$, and one can  calculate the expansion $\theta_{(k,\lambda)}$ as a function of $\lambda$. The variation of $\theta_{(k)}$ is defined as 
\begin{equation}
  \delta_{\vec{s}}\,\,\theta_{(k)} = \left.\frac{\partial\theta_{(k,\lambda)}}{\partial\lambda}\right|_{\lambda=0}\,.
\end{equation}
While this definition encompasses variations along any direction depending on the nature of $s^a$, let us first consider the case when $s^a$ is orthogonal to the MTS, unit, space-like and outward-pointing (i.e., $s_ak^a >0$). Then we can use the rescaling freedom  $(k^a, \uk^a) \to (\alpha k^a, \alpha^{-1} \uk^a)$ so that we have $s_a k^a =1$. With this choice,  variations along $\psi s^a$ for any function $\psi$ leads one to a differential operator $L_{\vec{s}}$\,:
\begin{equation}
  \delta_{\psi \vec{s}}\,\,\theta_{(k)} = L_{\vec{s}}\,\,\psi\,.
\end{equation}
It can be shown that $L_{\vec{s}}$ is an elliptic operator (which, however, is not necessarily self-adjoint on $L^2(S)$):
\begin{equation}
  \label{eq:stability-cauchy} \hskip-2cm
  L_{\vec{s}}\,\,\psi = -\t\Delta\,\psi + 2{\t\omega}^aD_a\psi + \big({\textstyle{\frac{1}{2}}}\t\R - \sigma_{ab}^{(k)}\sigma^{ab}_{(k)} - {\t\omega}_a {\t\omega}^a + D_a{\t\omega}^a - G_{ab}k^a(k^b+\uk^b) \big)\,\psi \,
\end{equation}
Here $\t\Delta$ is the Laplacian on $S$, $\t\R$ the 2-dimensional Ricci scalar of $S$, and as in {\smash{section \ref{s3},} $\t\omega_a$ is the rotational 1-form, $\sigma_{ab}^{(k)}$ is the shear of $k^a$ and $G_{ab}$ is the Einstein tensor.} { $L_{\vec{s}}$ is called  the \emph{stability operator} because the expression (\ref{eq:stability-cauchy}) of the variation of $\theta_{(k)}$ allows one to express the stability properties of the MTS { (in the direction of $\vec{s}$ chosen above)} in terms of the spectral theory of elliptic operators. Note that the operator on the right hand side depends only on fields on $S$ and does not make  \emph{direct} reference to $s^a$. However, for it to equal $L_{\vec{s}}\,\,\psi$, one has to use null normals $k^a$ and $\uk^a$ whose `boost' rescaling freedom is fixed using $s^a$. At various points in the review  --e.g. in the discussion of the passage from weakly isolated to isolated horizons in section \ref{s2.2} and in the discussion of passage to equilibrium in section \ref{s6.2.1}-- this operator is used in its own right for pre-specified $k^a$ and $\uk^a$. There, the operator  it would equal $L_{\vec{s}}$ only for unit space-like vector fields $\vec{s}$ that are orthogonal to the MTS $S$ and satisfy $k^a {s}_a =1$ but this property of the operator does not play any role in these discussions.}
 
Stability of an MTS refers to its properties under deformations, and in particular it probes whether it is possible to make it untrapped by some outward variation. An MTS $S$ is said to be \emph{stable} w.r.t. a direction $s^a$ if and only if there exists a function $\psi\geq 0$ (which is not identically zero) such that $\delta_{\psi \vec{s}}\,\,\theta_{(k)} \geq 0$.  Moreover, $S$ is said to be 
\emph{strictly stable} w.r.t. $s^a$ if the variation makes $S$ untrapped at least at some points, i.e. if $\delta_{\psi \vec{s}}\theta_{(k)}> 0$ somewhere on $S$. The stability of an MTS so defined has a counterpart in terms of the eigenvalues of $L_{\vec{s}}$.  Note first that $L_{\vec{s}}$ is not necessarily self-adjoint and may thus have complex eigenvalues. Nevertheless, one can show that the eigenvalue $\lambda_0$ with the smallest real part, referred to as the principal eigenvalue, must be real. Furthermore, $S$ is stable if $\lambda_0\geq 0$ and strictly-stable if $\lambda_0> 0$. (Detailed proofs can be found in \cite{Andersson:2007fh} where, however, the authors use the terminology \emph{stably outermost} and \emph{strictly-stably-outermost} in place of \emph{stable} and \emph{strictly stable}.)  Thus if $\lambda_0>0$ then we can find a variation that makes the MTS untrapped. Conversely, if such a variation can be found, then $\lambda_0>0$. 
\vskip0.05cm
    
\emph{Remarks:}
\vskip0.05cm

1. The variations along space-like directions $\vec{s}$ we discussed above are well-tailored to the Cauchy evolution, used in NR. Therefore, in the context of Cauchy evolution, one generally considers variations along the outward pointing normal $s^a$ to the MTS $S$, that lies in the Cauchy surface and uses terms \emph{stable} and \emph{strictly stable} without explicitly stating the direction. But one can also discuss stability w.r.t.  deformations along arbitrary directions orthogonal to $S$  \cite{Andersson:2007fh}. If NEC holds, then by the Raychaudhuri equation, every MTS $S$ is stable along $-k^a$. It turns out that if a $S$ is strictly stable
w.r.t. a space-like direction $\vec{s}$ then it is also strictly stable along $-\uk^a$, but the converse need not hold.\hskip0.5cm

2. Interestingly, the stability operator has also been used to obtain bounds on the angular momentum of an MTS \cite{Jaramillo:2011pg,Dain:2011kb,Dain:2011pi,Dain:2010qr}.  We introduced the definition of angular momentum for a DHS in Sec.~\ref{s3.3}.  Let us now consider an axisymmetric MTS $S$ with axial symmetry vector $\varphi^a$, area $A_S$ and angular momentum $J_{\DH}^{(\varphi)}$. $S$ need not lie on a DHS, but again all considerations will be entirely quasi-local; axisymmetry is only required on $S$ and not in a spacetime neighborhood of $S$.  Require now in addition that $S$ is strictly-stable, i.e. $\delta_{\vec{s}}\,\theta_{(k)} > 0$ where $s^a$ is either an outward pointing space-like vector, or points along $-\uk^a$. Then the following result holds.\\ 
\emph{Theorem 1 of \cite{Jaramillo:2011pg}:} If an MTS is axisymmetric and strictly-stable for axisymmetric variations, and if the spacetime has non-negative cosmological constant and satisfies the dominant energy condition, then $A_S\geq 8\pi G\, J_{\DH}^{(\varphi)}$. Furthermore, the equality holds if and only if $S$ has the same geometry as that of the throat of the extreme Kerr horizon.\vskip0.05cm
\noindent Thus, even in dynamical situations, the ``Kerr bound'' on angular momentum is satisfied by all MTSs satisfying the assumptions  of the theorem. \hskip0.5cm

3. For stable MTSs it is also possible to obtain bounds on the
curvature \cite{Andersson:2005me}.  The principal eigenvalue
$\lambda_0$ was shown to have a physical interpretation analogous
to a pressure difference across a soap bubble (i.e. minimal surfaces),
thus leading to a fluid interpretation of the stability operator
\cite{Jaramillo:2013rda}.  {Finally, the notion of
    stability of cross-sections of an IHS $\Delta$ is shown to be
    related to properties of $\theta_{(\uk)}$ and the sign of the
    surface gravity $\kappa_{(k)}$ \cite{Mars:2012sb}.  First,
    considering the stability operator along $-\uk^a$, it can be shown
    that all cross-section of $\Delta$ have the same principle
    eigenvalue $\lambda_0$; thus we can talk sensibly about the
    stability of $\Delta$ itself.  Roughly speaking, IHSs with
    $\theta_{(\uk)}\leq 0$ and positive surface gravity are stable.
    Moreover, if $\Delta$ admits a cross-section with
    $\theta_{(\uk)}=0$ then $\Delta$ is marginally stable,
    i.e. $\lambda_0=0$. Conversely, if $\Delta$ is marginally stable
    then there exists a cross-section $S$ of $\Delta$ such that
    $\oint_S\theta_{(\uk)} = 0$.}

\subsection{The time evolution of an MTS: Mathematical Results}
\label{s4.2}

Let us begin with a technical clarification. Smooth maps $\map:S\times (-\epsilon,\epsilon) \rightarrow M$ considered in the discussion of the stability operator can be either  embeddings or immersions. In both cases, the differential of the map is well defined. If $\map$ is an embedding, then the map\, $\map\,(\cdot\, ,\lambda)$ is one-one and the image of $S_\lambda$ is a smooth 2-sphere embedded in $M$. If on the other hand $\map$ is an immersion, then $S_\lambda$ can have self-intersections in which case $S_\lambda$ would not be a smooth sub-manifold of $M$. While most MTSs of physical interest do not have self-intersections, these unusual MTSs do arise in binary BH spacetimes very near the merger inside the common DHS representing the remnant. 

With this clarification out of the way, let us consider a portion of spacetime $(M, g_{ab})$ satisfying NEC, foliated by (partial) Cauchy surfaces $\Sigma_t$ parameterized by a time function $t \in [0,T]$. Then we have the following key result.\\
\emph{Theorem 9.2 of \cite{Andersson:2007fh}:} Let $\Sigma_{0}$ (corresponding to $t=0$) contain a smooth, immersed strictly-stable MTS $S_0$, where stability is with respect to variations $\delta_{\psi \vec{s}}$\, within $\Sigma_0$ in outward directions $s^a$ orthogonal to $S_0$. Then for some $\tau\in (0,T]$, on each $\Sigma_{t}$ for $t\in [0,\tau)$, there exist MTSs $S_t\subset \Sigma_t$ all of which are  immersed, strictly-stable, and regular (i.e. the differentials of the mappings of $S_t$ into $\Sigma_t$  are well defined). Moreover, the collection of all these $S_t$ form a smooth 3-manifold $\mathfrak{H}$ which is nowhere tangent to the (partial) Cauchy surfaces  $\Sigma_t$. \vskip0.1cm

The causal properties of $\mathfrak{H}$ can also be determined \cite{Andersson:2005gq,Andersson:2007fh}:
\begin{itemize}
\item $\mathfrak{H}$ is space-like or null everywhere.
\item If $\mathcal{E} = \sigma_{ab}^{(k)} \sigma_{(k)}^{ab} + R_{ab}k^a k^b$ is non-zero somewhere on $S_0$, then $\mathfrak{H}$ is space-like everywhere near $S_0$.
\item If $\sigma^{(k)}_{ab}$ and $R_{ab}k^ak^b$ vanish identically on $S_0$ then $\mathfrak{H}$ is null everywhere on
  $S_0$. 
\item MTSs are barriers, i.e. (local) boundaries which separate regions containing trapped and untrapped surfaces within $\Sigma_t$.
\end{itemize}

Thus, if $S_0$ is strictly stable in $\Sigma_0$, then Einstein's evolution equations guarantee that it would evolve to form a QLHS at least locally. If $\sigma^{(k)}_{ab}$ and $R_{ab}k^ak^b$ vanish identically on $S_0$, then in a neighborhood of $S_0$ the QLHS is an NEHS $\Delta$. If $\mathcal{E}$ does not vanish identically on $S_0$ \emph{and} $\theta_{(\uk)}$ is nowhere zero on $S_0$ then the QLHS is a \emph{space-like} DHS, again in a neighborhood of $S_0$. This is why space-like DHSs are found to be ubiquitous. What about the DHS in the O-S collapse which, as discussed in section \ref{s3}, is time-like? In this case neither $\mathcal{E}$ nor $\theta_{(\uk)}$ vanishes anywhere on any MTS. However, none of the MTSs on this DHS is strictly stable. That is why the theorem is not applicable to this DHS.

Several comments are in order. The proof of theorem 9.2 of \cite{Andersson:2007fh} is based on the implicit function theorem, and it relies on the invertibility of the stability operator $L_{\vec{s}}$. This is guaranteed if $\lambda_0>0$ which is of course, just the strict-stability condition. However, the operator is invertible more generally; one only needs that zero is not one of its eigenvalues. It is known that the spectrum of this operator is discrete. Therefore, generically the operator will be indeed invertible and the MTS will evolve smoothly.  This corresponds to the findings in NR (described below) where it is seen that even MTSs which fail to be strictly stable in the direction $s^a$ usually evolve smoothly. We also see this analytically in the O-S collapse where every MTS on the DHS fails to be stable. However, in these cases, several of the other results listed above do not generally hold: $\mathcal{H}$ could be time-like (as is the case for the O-S collapse), and the barrier property of an MTS will generally not hold (as in the left and the middle panels of Fig.~2). Again, this is in concordance with NR results in BBH simulations where, for example, near the merger the QLHs of progenitors cease to have definite signature. But they evolve continuously. Moreover,
by this time the progenitor QLHs are surrounded by the common QLHS which is the one that is physically relevant. And in numerical simulations, soon after its formation, the MTSs on this segment are found to be strictly-stable and the resulting DHS is found to have all the properties listed above; in particular, it is space-like.  

\subsection{The time evolution of an MTS: Numerical Results}
\label{s4.3}

There is a synergistic two-way bridge between the analytic results just discussed and findings in numerical simulations. We will now discuss on this interplay.

A critically important ingredient in numerical simulations is the
method for locating MTSs. Efficient procedures have been developed for
this purpose over the years \cite{Thornburg:2006zb}. However, the ones
that are commonly used often fail to locate the highly distorted
horizons that are found in a binary BH spacetime and new, more
accurate MTS-finders were needed to overcome these issues
\cite{Pook-Kolb:2018igu,PhysRevD.100.084044,pook_kolb_daniel_2020_3885191,Booth:2021sow,Booth:2020qhb,Cadez1974}. So
far these new methods have been applied primarily to the simplest initial data set for BBH coalescence: The one due to Brill-Lindquist that represents two non-spinning BHs at the moment of time symmetry, so that each BH has vanishing ADM 3-momentum and angular momentum \cite{Brill:1963yv}. (Upon evolution these BHs undergo a head-on collision so that the entire space-time is axisymmetric.) In this sub-section we first present this initial data and then summarize results on MTSs in two steps. The first step is `kinematical' in the sense that it explores the behavior of MTSs in the Brill-Lindquist \emph{initial data} obtained just by varying the separation between the two black holes. The second step is dynamical in that it tracks the behavior of the QLH constituted by the world tube of MTSs under \emph{time-evolution} of a given Brill-Lindquist initial datum. While this initial data set is very special, as we will see, the MTSs it accommodates can be far from being simple. Also, under time evolution the data does not remain time-symmetric. {See also
\cite{Jaramillo:2009zz} for a study with Bowen-York data.}

\begin{figure*}[h]
  \begin{subfigure}[t]{0.38\textwidth}
    \centering    
    \includegraphics[width=\textwidth]{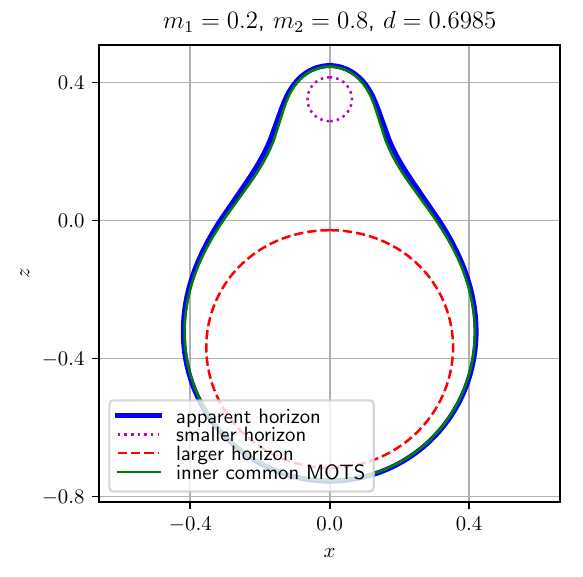}
    \caption{\footnotesize{The various MTSs shortly after the common MTS is
      formed. At this separation, the inner common MTS and the AH
      are very close to each other.}}
    \label{fig:sub1}
    \centering
  \end{subfigure}
  \hfill
  \begin{subfigure}[t]{0.38\textwidth}
    \centering
    \includegraphics[width=\textwidth]{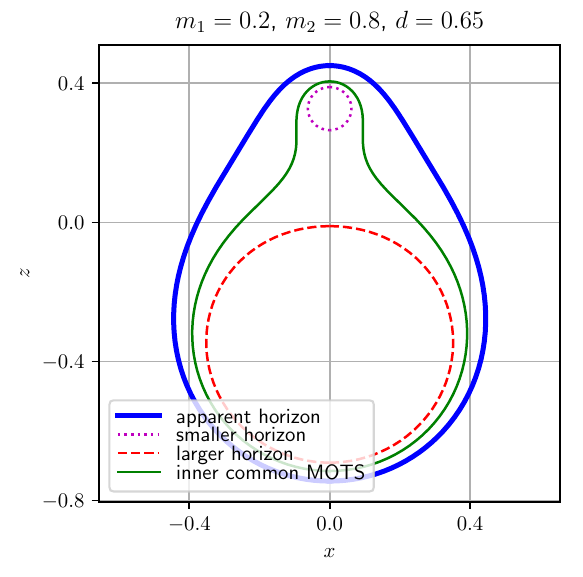}
    \caption{\footnotesize{The inner common MTS and the AH rapidly move away from
      each other as $d$ is decreased. }}
    \label{fig:sub2}
    \centering
  \end{subfigure}
  \hfill
  \begin{subfigure}[t]{0.38\textwidth}
    \centering
    \includegraphics[width=\textwidth]{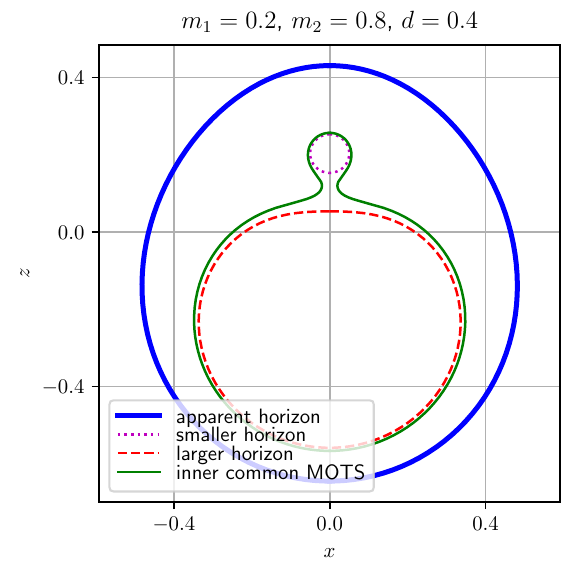}
    \caption{\footnotesize{As $d$ is further decreased, the inner common MTS
      starts getting highly distorted while the AH becomes more uniform. }}
    \label{fig:sub3}
    \centering
  \end{subfigure}
  \hfill
  \begin{subfigure}[t]{.38\textwidth}
    \centering
    \includegraphics[width=\textwidth]{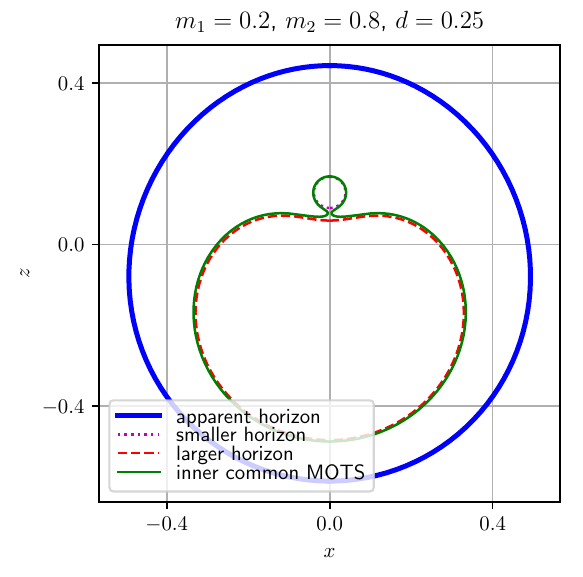}
    \caption{\footnotesize{The inner common MTS and the larger individual MTS are
      now highly distorted.}}
    \label{fig:sub4}
    \centering
  \end{subfigure} 
  \caption{\footnotesize{(Figure from
      \cite{Pook-Kolb:2018igu}) \emph{Sequence for MTS for Brill-Lindquist
        data:} In each of the plots the masses are fixed at
      $m_1 = 0.2$, $m_2 = 0.8$, and the separation $d$ becomes
      successively smaller. Each panel shows four MTSs: the two
      associated with the individual progenitor BHs (larger one in
      red, and smaller one in magenta), the outer MTS associated with
      the remnant BH (in blue and referred to as the apparent horizon
      (AH)), and the inner MTS (in green).}}
    \label{fig:brill_lindquist1}
\end{figure*}

Recall that the initial data set for GR consists of a 3-metric $\gamma_{ab}$ and the extrinsic curvature $K_{ab}$ on a Cauchy slice $\Sigma$, subject to constraint equations. In the Brill-Lindquist case, because of the assumed time-symmetry, we have $K_{ab} = 0$. Therefore the `momentum constraint' on the initial data is automatically satisfied. The 3-metric $\gamma_{ab}$ is assumed to be conformally flat: $\gamma_{ab} = \phi^4\delta_{ab}$, where $\delta_{ab}$ is a flat 3-metric. The conformal factor is determined by the `Hamiltonian constraint' and diverges at two points, called the `punctures'. If one denotes points on $\Sigma$ by the position vector $\vec{r}$ using the flat 3-metric, then $\phi$ is given by
\begin{equation}
  \label{eq:conformal_factor}
  \phi(\vec{r})  = 1 + \frac{m_1}{2r_1} + \frac{m_2}{2r_2} \,,  
\end{equation}
where $r_1$ and $r_2$ are the \emph{Euclidean} distances of the point $\vec{r}$ on $\Sigma$ from the two punctures, and $m_1,m_2$ are parameters, interpreted as the `bare masses' of the two BHs. With respect to the physical metric, $\Sigma$ has three asymptotic regions; one corresponding to $r_1 \to 0$, another to $r_2 \to 0$ and the third corresponding to $r\to \infty$. 2-spheres with large \emph{physical} radii in either of the first two asymptotic regions surround an individual BH while those in the third asymptotic region surround both BHs. As shown in \cite{Brill:1963yv}, the total ADM mass of the system (computed in the limit $r \to \infty$) is given by $M_{ADM} = m_1+m_2$.  Using the physical metric, one can also compute the ADM masses of individual BHs in the limit $r_1 \to 0$ and $r_2 \to 0$ to obtain
\begin{eqnarray}
  M^{(1)}_{\rm ADM} &=& m_1 + \frac{m_1m_2}{2d}\,,\\
  M^{(2)}_{\rm ADM} &=& m_2 + \frac{m_1m_2}{2d}\,,
\end{eqnarray}
where $d$ is the \emph{Euclidean} distance between the punctures. The difference
\begin{equation}
  M_{ADM} - M^{(1)}_{\rm ADM} - M^{(2)}_{\rm ADM} = -\frac{m_1m_2}{d}
\end{equation}
is interpreted as the binding energy between the two BHs.\,%
\footnote{\,\,\,More generally, for non-spinning BHs the difference $M_{\rm ADM} - M^{(1)}_{H} - M^{(2)}_{H}$ between the total ADM mass and the Hawking masses associated with individual horizons  can be thought of as the binding energy in a physically meaningful way \cite{Schnetter:2006yt,Krishnan:2002wxg}.}
Fig.~\ref{fig:brill_lindquist1} shows the behavior of MTSs for the Brill-Lindquist initial data as $d$ is successively reduced with $m_1$ and $m_2$ fixed. This is a `kinematical sequence' of initial data sets since evolution equations are not used in its construction. Nonetheless, it provides useful insights.  In particular, the extreme distortions of horizons along this sequence is evident; it is this feature makes the 
commonly used
horizon finders inadequate.  A close examination of this sequence of initial data sets reveals the following interesting features of the stability properties of the MTSs they contain.  When $d$ is large and the data contains only the two individual MTSs without a common horizon, the principal eigenvalues $\lambda_0$ of the stability operator for the two MTSs are positive.  Thus, they are strictly-stable as defined above.  When the common MTS first appears, it is born with $\lambda_0=0$.  When $d$ is further reduced and we have a bifurcation, the outer MTS has $\lambda_0>0$ and therefore stable, while the inner MTS is unstable with $\lambda_0<0$.

While these findings are interesting, they do not provide us direct
insights into horizon dynamics.  Indeed, since this is a kinematic
sequence, the ADM masses of individual BHs change as we change
$d$ keeping $m_1$ and $m_2$ fixed, providing concrete evidence that
the sequence cannot correspond to a dynamical evolution. To explore
the dynamics of the QLHs we need to evolve any one initial data set
using the full Einstein equations. Again, this requires new numerical
methods for locating MTSs, and the new MTS-finder has been also
extensively used in these numerical time-evolutions.  Interestingly,
{following pioneering numerical studies
\cite{Rezzolla:2010df,Jaramillo:2011rf,Jaramillo:2012rr}}, the full
evolution shows that the general picture is qualitatively similar to
that obtained by the above `kinematical' considerations
\cite{Gupta:2018znn}. When the two BHs in the binary are
initially well separated, there are two independent MTSs, one for each
BH. These two MTSs approach each other and, at a certain
point, a common MTS appears which surrounds the individual MTSs.  This
common MTS immediately bifurcates into an inner and outer MTS.  The
inner common MTS shrinks and approaches the two individual MTSs, while
the outer common MTS (which is the apparent horizon) grows and sheds
its multipole moments as time evolves and approaches an equilibrium
state. In the simple case of a head-on collision resulting from a
Brill-Lindquist datum, the equilibrium state is represented by a
Schwarzschild remnant. More generally, it would be a Kerr remnant.
\begin{figure*}[h]
    \includegraphics[width=\textwidth]{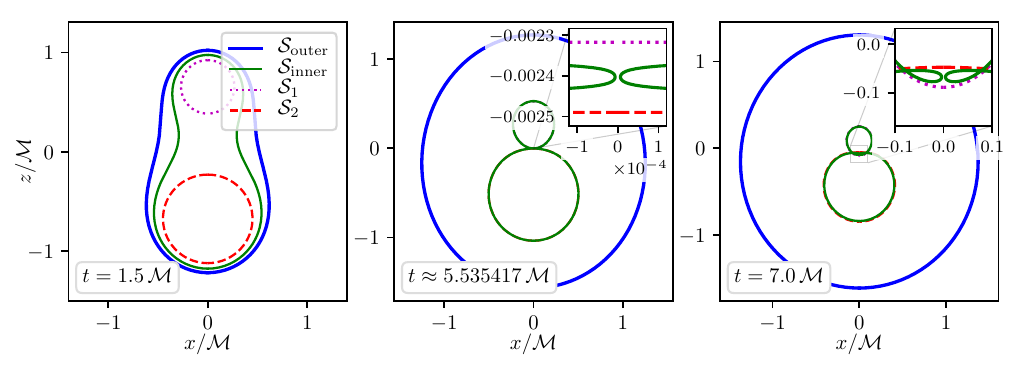}
    \caption{\footnotesize{(Figure from \cite{PhysRevLett.123.171102}
      \emph{Snapshots of the MTS structure of a simulation of a
      Brill-Lindquist initial datum:} Taken at different simulation
      times, they show the progression of the marginally trapped
      surfaces as the two BHs merge to form a final remnant
      BH.  The first panel, shortly after the start of the
      simulation, shows the two marginal surfaces for the two
      individual BHs labeled $S_1$ and $S_2$. The common
      horizon has bifurcated into an inner and outer branch labeled
      $S_{\rm inner}$ and $S_{\rm outer}$ respectively. The middle
      panel shows the situation just before $S_1$ and $S_2$ touch,
      i.e. at the time $\ttouch \approx 5.5378\Munit$ where $\Munit$ is the ADM mass. Here
      $S_{\rm inner}$ is on the verge of forming a cusp, while on the
      other hand $S_{\rm outer}$ has lost most of its distortions.
      Finally, the right panel shows the surfaces after
      $\ttouch$. Here $S_{\rm outer}$ is very close to its final
      spherically symmetric state, while $S_{\rm inner}$ has developed
      a self-intersection, and $S_1$ and $S_2$ have penetrated each
      other.}}
    \label{fig:merger1}
\end{figure*}

An illustrative result, taken from \cite{PhysRevD.100.084044}, is
shown in Fig.~\ref{fig:merger1}.  The behavior of the various MTS
under time-evolution is similar to that of the `kinematical' initial
data sequence before the merger.  It is interesting to note that the
common horizon, when it is initially born, has vanishing principal
eigenvalue.  Subsequently, the outer common horizon is strictly
stable, while the inner horizon segment has negative principle
eigenvalue.  Regarding the individual MTSs, there is however one
important difference from the kinematical case discussed
above with initial data sequences. Namely, the MTSs no longer
remain minimal surfaces under time evolution and they can, and do,
intersect each other.  The behavior of the inner MTS is
particularly interesting mathematically: it is seen to develop
self-intersections starting precisely at the time when the individual
MTSs touch (see Fig.~\ref{fig:merger1}).  Furthermore, one would
expect that at sufficiently late times all of the inner horizon
segments would have disappeared. This process has been further
investigated in
\cite{Pook-Kolb:2021gsh,Pook-Kolb:2021jpd,Booth:2021sow,Booth:2020qhb}
and the final picture is rather intricate and interesting from a
mathematical perspective.  It turns out that there is an infinite
sequence of mergers of the inner MTSs, and also bifurcations where
more inner horizon segments are formed with yet more
self-intersections and with additional negative eigenvalues. In the
end, one is left with just the outer horizon at late times 
(whose MTSs are apparent horizons). In contrast to all of the complexity associated with 
inner horizon, the outer 
horizon has an extremely simple behavior
as it reaches its equilibrium state in the distant future by shedding
its shape (and spin, when applicable) distortions.  This process of
approaching equilibrium has important applications in gravitational
wave astronomy, discussed in section \ref{s6}.

These high precision numerical results have provided much intuition about the QLH dynamics. In particular they have shown that, contrary to the earlier wide spread belief, the horizons of progenitors do not abruptly jump to the horizon of the remnant at the merger; there is a continuous QLH. The \emph{apparent horizons} do jump but that is simply an artifact of the requirement in the  definition of an apparent horizon that it be the \emph{outermost} MTS on a given Cauchy surface. While these results on the global structure of QLHs were obtained using the rather special Brill-Lindquist initial data sets, we saw that the MTSs that one encounters in these space-times are quite generic in the sense that they can be highly distorted. {Some aspects of the dynamics of MTSs described above have been shown to hold generally \cite{Andersson:2008up}.  In particular, it is shown in \cite{Andersson:2008up} that when {MTSs of the QLHs of the two progenitors} approach each other during time evolution with Einstein equations, a new common MTS appears at some time $\tau$ before the two MTSs touch.  This appears as a ``jump'' of the apparent horizon, and the new MTS is shown to be formed with vanishing principle eigenvalue.  The world-tube generated by the newly formed MTSs is tangential to the Cauchy slice at time $\tau$, whence it has an inner- and outer-branch. However, the world tube of these MTSs is not guaranteed to be a DHS { near the bifurcation.} All of these mathematical results are in accordance with the numerical results described above. Moreover the numerical results go beyond the analysis of \cite{Andersson:2008up} which does not address the issue of the ultimate fate of the inner MTSs and the case of unstable MTSs.}

We also note that a key simplification arises in this case because the stability operator has real eigenvalues for the Brill-Lindquist initial data. Even under time evolution when the data is no longer time symmetric, the spectrum of the stability operator continues to remain real. This would not be the case more generally for spinning BHs, and it is of considerable interest to carry out analysis in more general situations. In particular, we see that when the common horizon is formed, it is born with vanishing principal eigenvalue, and in fact all subsequent bifurcations are associated with the vanishing of a higher eigenvalue \cite{Pook-Kolb:2021gsh,Pook-Kolb:2021jpd,Booth:2021sow,Booth:2020qhb}.  When the spectrum is complex, then the  behavior of the eigenvalues might differ qualitatively from this picture.  Thus we have:\vskip0.1cm
\texttt{Open Issue 7 (OI-7)} Do qualitative features of the QLHSs of progenitors and the remnant, including the spectrum of the stability operator continue to hold generically beyond the head-on collision, especially even when one allows for spin? 
\vskip0.1cm

In sections~\ref{s6.1} and \ref{s6.2} we shall discuss the approach of the remnant
horizon to equilibrium. In this process, the remnant BH loses its ``hair'' by absorbing just the right amounts of radiation as the multipole moments approach their asymptotic equilibrium values.  As we shall see, the horizon fluxes in this regime are associated with the quasi-normal modes of the remnant BH.  We can also consider  the two progenitors.  At very early times the individual BHs have their equilibrium multipole moments (corresponding presumably to a Kerr or a Schwarzschild BH), and these multipole moments vary due to the tidal gravitational field produced by the companion.  These tidal deformations play an important role in GW astronomy, and will be discussed further in section~\ref{s6.3}.  Here we note that the remnant BH and the two initial BHs are connected by a sequence of MTSs. The connecting sequence of MTSs will necessarily involve the complications of the inner MTSs as discussed earlier. In particular, the ``seeds'' $\t\R$ and $\t\omega_a$  for the multipole moments of the inner MTSs will not necessarily be smooth due to the formation of cusps.  The question then arises: 
\vskip0.1cm
\texttt{Open Issue 8 (OI-8)} Can we nonetheless relate the multipole moments (and other properties) of the remnant DHS with those of the two progenitor DHSs?  
\vskip0.1cm
\noindent If this can be done, it would allow us to connect the ringdown and inspiral regimes at the QLH in a continuous manner without using the gravitational wave signal. 

\section{Null QLHs $\IH$ and Null Infinity}
\label{s5}

In this section we return to \emph{null} QLHs, $\IH$, and summarize recent results showing that null infinity, $\scri$, is also a null QLH \cite{Ashtekar2_2024}. More precisely, as summarized in section \ref{s5.1}, $\scri$ is an extremal WIHS even when there is flux of gravitational waves across { it!} Consequently,  $\scri$ inherits its geometrical structures from those on { $\IH$.} This seems very surprising at first because although $\IH$ and $\scri$ are both null, they carry very different physical connotations. $\IH$ lies in the strong field regime of the physical space-time and represents a BH boundary in equilibrium. As we saw in section \ref{s2.1}, there is no leakage of energy across $\IH$ in the sense that the Hawking quasi-local mass evaluated on any 2-sphere cross-section of $\IH$ the same. $\scri$, on the other hand, lies in the asymptotic, weak field regime and serves as the natural arena to investigate gravitational and electromagnetic radiation. There \emph{is} leakage of energy across $\scrip$. { Now the Hawking mass evaluated on a family of 2-spheres in the physical space-time tends to the Bondi energy on the cross-section of $\scrip$ they approach (see, e.g., \cite{Ashtekar4_2024}) which is \emph{is time-dependent}!}

In section \ref{s5.2} we trace the origin of this difference to a single fact: While the WIHSs $\IH$ representing BH boundaries lie in the physical space-time $(M, g_{ab})$ where Einstein's equations hold, $\scrip$ is a WIHS in its conformal conformal completion $(\hM,\hg_{ab})$ where \emph{conformal} Einstein's equations hold \cite{Ashtekar1_2024,Ashtekar2_2024}. In section \ref{s5.3} we show that this interplay extends also to certain observables, paving the way to `gravitational wave tomography'  --imaging horizon dynamics using waveforms at null infinity-- discussed in section \ref{s6.2}.

\subsection{Null Infinity $\scrip$ as an extremal WIHS}
\label{s5.1}

Recall that the surface gravity $\kappa_{(k)}$ of an NEHS $\IH$ is defined by $k^a \nabla_a k^b = \kappa_{(k)} k^b$ and $(\IH, k^a)$ is said to be \emph{extremal} if $\kappa_{(k)}\, \=\,0$. As before, let us denote by $[k^a]$  the equivalence class of null normals where any two are rescalings of one another by a \emph{positive constant}. {Now,} as we saw in section \ref{s2.2}, $(\IH, [k^a])$ is said to be a WIHS if $\Lie_{k} \omega_a\, \=\,0$, or, equivalently, if $\kappa_{(k)}$ is constant on $\IH$. Therefore, given an extremal NEHS $(\IH, k^a)$, the pair $(\IH, [k^a])$ automatically constitutes an extremal WIHS. 

Note that any NEHS can be made extremal by a suitable rescaling of its null normal; so restriction to extremality is not a constraint on NEHSs. In fact every NEHS admits an infinite number of extremal WIH structures: If $(\IH, [k^a])$ is one, so is $(\IH, [fk^a])$, where $f$ is any function on $\IH$ satisfying $\Lie_{k} f \, \= \,0$. Under $[k^a] \to [f\,k^a]$ the rotational 1-form transforms via $\omega_a \to \omega_a + D_a\ln f$. Interestingly for extremal NEHSs one can eliminate this rescaling freedom by $f$ in a natural fashion: Require that $\omega_a$ be divergence-free, i.e., satisfy $q^{ab} D_a \omega_b\, \=\,0$ on $\IH$, where $q^{ab}$ is any inverse of the intrinsic 2-metric $q_{ab}$ on $\IH$ \cite{Ashtekar:2001jb}. This requirement selects the equivalence class of null normals uniquely and we will denote this distinguished family by $[\ko^a]$. Thus, any given NEH can be endowed with the structure of \emph{a canonical} extremal WIHS $(\IH, [\ko^a])$. Note that this choice is natural because the conditions that single out $[\ko^a]$ (namely, $\kappa_{(\ko)}\, \=\,0$ and $D^a\omegao_a\, \=\, 0$) refer only to the structures that are naturally available on the NEH. As we will now show, not only is $\scrip$ a WIHS in the conformally completed space-time $(\hM, \hg_{ab})$, but its natural null normal\, $\hn^a := \h\nabla^a \Omega$\, makes it the canonical extremal WIHS as well. This fact will have interesting consequences. \vskip0.05cm
 
Let us begin by recalling the definition of asymptotic flatness at null infinity. For definiteness we will focus on $\scrip$ and, following the common usage in the literature, the null normal to $\scrip$ will be denoted by $\hn^a$ and its dual 1-form by $\h\ell_a$ (in place of $k^a$ and $\uk_a$ used so far).

\vskip0.1cm
\indent \emph{Definition 5} \cite{aa-yau}: A physical space-time $(M, g_{ab})$ is said to be asymptotically flat at future null infinity if there exists a manifold $\hM$ with boundary $\scrip$ equipped with a metric $\hg_{ab}$ such that $\hM = M\cup \scrip$ and $\h{g}_{ab} = \Omega^2 g_{ab}$ on $M$ for a nowhere vanishing function $\Omega$, and,\\
(i) $\scrip$ has topology $\mathbb{S}^2\times \mathbb{R}$ and its causal past contains a non-empty portion of $M$;\\
(ii) $\Omega \, \= \, 0$, where $\=$ now refers to equality at points of $\scrip$,\, while \,$\h\nabla_a \Omega$\, is nowhere vanishing on $\scrip$; and,\\
(iii) The physical metric $g_{ab}$ satisfies Einstein's equations $R_{ab} - \f{1}{2} R g_{ab} = 8\pi G\, T_{ab}$ where $\Omega^{-2} T_{ab}$ admits a smooth limit to $\scrip$.\vskip0.1cm
\noindent $(\hM, \hg_{ab})$ is referred to as a conformal completion of the physical space-time $(M, g_{ab})$. Condition (ii) ensures that $\scrip$ lies at infinity w.r.t. the physical metric $g_{ab}$; (i) ensures that the boundary represents \emph{future} null infinity;  and  (iii) on the fall-off rate of  the stress energy tensor is the standard one; it is satisfied by matter sources normally used in general relativity, in particular, the Maxwell field. 

Let us summarize the consequences of this definition that we will need (for details, see e.g. \cite{gerochrev,doi:10.1063/1.524467,aa-yau}). Conformal Einstein equations satisfied by $\hg_{ab}$ imply that $\scrip$ is null and, without loss of generality, one can choose\, $\Omega$\, so that { $\h\nabla_a \hn^a \,\=\,0$} where $\hn^a \,= \,\h\nabla^a \Omega$ is the null normal to $\scrip$. As is standard in the literature on null infinity, \emph{we will restrict ourselves to such divergence-free frames}.  Then conformal Einstein's equations imply that a stronger condition holds: $\h\nabla_a \hn^b\, \=\, 0$. Therefore integral curves of $\hn^a$ are null geodesics for which expansion $\h\theta_{(\hn)}$, shear $\h\sigma_{ab}^{(\hn)}$, and surface gravity $\h\kappa_{(\hn)}$ \emph{all vanish}. Finally condition (iii) together with conformal Einstein's equations implies that the Ricci tensor $\h{R}_{ab}$ of $\hg_{ab}$ { is such that} $\h{R}_a{}^b \hn^a$ is proportional to $\hn^b$. Together these properties imply that $\scrip$ is an extremal WIHS. Furthermore, the 1-form $\omega_a$ vanishes identically, whence in particular its divergence vanishes. Thus $\hn^a\, \=\, \h\nabla^a \Omega$ endows $\scrip$ with the structure of the \emph{canonical} extremal WIHS. 

This is the main result of this subsection. However, the extremal WIHS $\scrip$ also has some additional noteworthy features:\vskip0.05cm

1. $\scrip$ is a sub-manifold of the conformal completion $\hM$ (rather than $M$) and its properties also refer to the conformal metric $\hg_{ab}$ (rather than the physical metric $g_{ab}$). Therefore, to regard it a WIHS, one has to restrict oneself to a specific conformal completion with $\h{g}_{ab} = \Omega^2 g_{ab}$. If we were to change the conformal factor, $\Omega \to \Omega^\prime = \omega \Omega$ (with $\Lie_{\hn} \omega\, \=\,0$), we would obtain a physically equivalent but distinct completion $(\hM, \hg^{\,\prime}_{ab})$. In the new completion, the intrinsic metric $\h{q}^{\,\prime}_{ab}$ would be different { and,} from the QLH view point, we would obtain a WIHS with distinct geometry, even though the two are physically equivalent.

2. Any given conformal completion endows $\scrip$ with a specific null normal $\hn^a\, \= \, \h\nabla^a\Omega$, rather than an equivalence class $[\hn^a]$ one would have on a general WIHS. This small extra structure turns out to be important for certain considerations, in particular of the structure of symmetry groups, { as discussed in point 3.} 

3. Recall that boundary symmetry groups arise in general relativity as subgroups of the diffeomorphism group of the boundary that preserve the `universal' structure thereon --i.e. the structure that is common to all boundaries of the type considered. For extremal WIHSs, this structure  includes the underlying manifold $\IH$ with topology $\mathbb{S}^2 \times \mathbb{R}$ and the equivalence class $[\ko^a]$ of canonical null normals. The intrinsic (degenerate) metric $q_{ab}$ can vary from one extremal WIHS to another. However, because $\IH$ has topology $\mathbb{S}^2 \times \mathbb{R}$ and each $q_{ab}$ is degenerate in the $\mathbb{R}$ direction and satisfies $\Lie_k q_{ab} \,\=\,0$,\, $\IH$ admits a 3-parameter family of unit, round, 2-sphere metrics $\{\qo_{ab}\}$ such that the given $q_{ab}$ is conformally related to it. While the conformal factor relating the given $q_{ab}$ to one of these unit round metrics varies from one $(\IH, q_{ab})$ to another, the conformal factors relating any two unit round metrics are \emph{universal}. Therefore, the universal structure of an extremal WIHS consists of the triplet, a $\mathbb{S}^2\times \mathbb{R}$ manifold $\IH$,  an equivalence class of vector fields $[\ko^a]$ along the $\mathbb{R}$-direction and a conformal class of unit round metrics $\{\qo_{ab}\}$ satisfying $\qo_{ab} \ko^a\,\=\,0$ and $\Lie_{\ko} \qo_{ab} \,\=\,0$. The symmetry group $\G$ of extremal WIHSs is the subgroup of the diffeomorphism group of $\IH$ that preserves this structure. It is closely related to the Bondi-Metzner-Sachs group $\B$, the symmetry group of $\scrip$, in that $\B$ is a normal subgroup of $\G$ and the quotient $\D := \G/\B$ is just the 1-dimensional group of dilations that just rescales $\ko^a$ by a positive constant, leaving each $\qo_{ab}$ untouched \cite{akkl1}. Thus $\G$ is a 1-dimensional extension of the BMS group. This extension occurs simply because, whereas on a general extremal WIH we only have an equivalence class $[\ko^a]$ of canonical null normals, $\scrip$ is equipped with a single null normal $\hn^a$. For details and further discussion of the observables defined by generators of $\G$ on a general extremal WIHS and by generators of $\B$ on $\scrip$, see \cite{Ashtekar2_2024, Ashtekar3_2024}. \vskip0.1cm

\emph{Remark:} Chronologically, null infinity first arose in the Bondi-Sachs framework \cite{bondi1962mgj,sachs}, that heavily uses limits along null 3-surfaces $u={\rm const}$. Soon thereafter it was cast using the notion of `asymptotic simplicity' by  Penrose \cite{Penrose:1965am} in which it arose as end-points of null geodesics. On the other hand, the notion of QLHs makes no reference to limits along a null foliation or along null geodesics. This difference made it appear that while null infinity and horizons are both null surfaces, they are qualitatively different. It is the realization \cite{gerochrev,doi:10.1063/1.524467} that $\scri$ can be introduced  without reference to null surfaces and null geodesics (as in \emph{Definition 5}) that opened the possibility of relating it to QLHs.

\subsection{Dynamics on WIHS}
\label{s5.2}

Let us now return to the question raised in  the beginning of section \ref{s5}: Given that $\scrip$ is a WIHS, how can there be a flux of gravitational (and electromagnetic) radiation across it? What is the origin of the drastic difference in physics at the WIHSs $(\IH, [k^a], q_{ab}, D)$ in the physical space-time and the WIHS $(\scrip, \hn^a, \hq_{ab}, \hD)$ in the conformal completion thereof? As already mentioned, the origin lies in the simple fact that the metric $g_{ab}$ with respect to which $\IH$ is a WIHS satisfies Einstein's equations, while 
$\scrip$ is a WIHS w.r.t. the conformally rescaled  metric $\hg_{ab}$ that satisfies conformal Einstein's equations. To begin with, therefore, let us consider \emph{geometric} WIHSs $\IH$, defined using a general metric $\bg_{ab}$ on a manifold $\bar{M}$ which is \emph{not} subject to any field equations,  and then restrict $(\bar{M}, \bg_{ab})$ to be a physical space-time $(M, g_{ab})$, and subsequently, the conformally completed space-time $(\hM, \hg_{ab})$. 

Consider then, an extremal WIHS\, $\bIH$\, with null canonical normals $[\ko^a]$, and geometry $(\bq_{ab},\, \bD)$, so that $\bD_a \ko^b\, \= \,\bomega_a \ko^a$ and $\bomega_a$ satisfies $\Lie_{\ko} \bomega_a\, \= \,0$, $\bkappa\, \= \,\bomega_a \ko^a\, \=\,0$, and $\bD^a \bomega_a\,\=\,0$. Recall from section \ref{s2.2} that $\bomega_a$ determines only part of the intrinsic connection $\bD$ on $\bIH$ and, since $\IH$ is only a WIHS rather than an IHS, the rest of the connection $\bD$ is in general time dependent. It turns out that the time dependence 
\be \dot\bD_a \uko_b \,:\!\=\, (\Lie_{\ko} D_a - D_a \Lie_{\ko})\, \uko_b \,\ee 
of $\bD$ is completely determined by the action of the right side on any 1-form $\uko_b$ (defined intrinsically on  $\bIH$)\, satisfying $\ko^b\uko_b  =-1$. One can show that the right side is in turn given by \cite{Ashtekar:2001jb,Ashtekar2_2024} 
\be \label{Ddot1}\dot{\bD}_a \uko_b\,\, \=\,\, \bD_a\bomega_b + \bomega_a \bomega_b + {\ko}^c\, \bar{C}_{c\,{\pb{ab}}}{}^d\, \uko_d + \textstyle{\f{1}{2}}\,\big(\bar{S}_{\pb{ab}} + (\alpha\,- \textstyle{\f{1}{6}}\bar{R})\, \bq_{ab}\big)\, , \ee
where $\bar{C}_{abc}{}^d$ and $\bar{S}_{ab} = \bar{R}_{ab} - \f{1}{6} \bar{R}\, \bar{g}_{ab}$ are the Weyl and Schouten tensors of $\bg_{ab}$,\, $\alpha$ is given by the condition $R_a{}^b \ko^b \, \=\,\alpha \ko^b$ in the definition of an NEHS, and the under arrow indices that indices $a,b$ are pulled back to $\IH$. On a generic WIH, none of the terms on the right side are zero, whence \emph{$\bD$ has genuine time dependence}.  It turns out that the drastic difference in the physics of BH horizons $\IH$ and null infinity $\scrip$ can be directly traced back to Eq. (\ref{Ddot1}).

Let us first consider the case when the WIHS represents a BH boundary in the physical space-time $(M, g_{ab})$ where vacuum equations hold on $\IH$, and drop the bars on various fields. Then, the last term of (\ref{Ddot1}) in round brackets vanishes on $\IH$ and the Weyl tensor term can be written in terms of fields intrinsic to $\IH$ so that we have \cite{Ashtekar:2001jb} 
\be \label{Ddot2}\dot{D}_a  \uko_b \,\, \=\,\, \D_{(a}\t\omega_{b)} + \t\omega_a \t\omega_b - {\textstyle{\f{1}{4}}}\, \t\R\, q_{ab}\, .  \ee
Note that every field on the right side is Lie-dragged by $\ko^a$, i.e., is time independent. Therefore the time dependence of $D$ is tightly constrained: $D$ is entirely determined by the triplet $(q_{ab},\omega_a, D_a \uko_b)$ on any 2-dimensional cross-section of $\IH$. { It is illuminating to consider the characteristic Cauchy problem based on two null surfaces, one being $\IH$ and the other being a null surface $\mathcal{N}$ spanned by null geodesics emanating from a cross-section $S$ of $\IH$\, (see Fig.~\ref{fig:nearhorizon}),\, in which one has specify certain fields on $\IH$ and $\mathcal{N}$ which constitute 3-dimensional data, and certain other fields on $S$ which constitute 2-dimensional data \cite{sachs,rendall:1990reduction}. The 3-d data consists of 2 freely specifiable functions at each point of $\IH$ and $\mathcal{N}$ and represents `radiative degrees of freedom' while the 2-d data on $S$\, --often called the `corner terms'--\,  represents the `Coulombic' degrees of freedom. Thus, Eq. (\ref{Ddot2}) implies that there are no radiative modes on $\IH$; one only has `corner terms' on $S$ carrying Coulombic  information.} Put differently, although $D$ is time-dependent, its dynamical content is trivial. This is why physically $\IH$ represents a BH boundary in equilibrium.

Next, let us consider the case when the WIH $\bar\IH$ is $\scrip$. Then we have the following replacements: $(\bM, \bg_{ab})\,\to\, (\hM, \hg_{ab})$,\, $\ko^a\,\to \, \hn^a$ and $\uko_a\, \to\, \ell_a$. Now we have the  \emph{complementary situation} vis a vis Eq. (\ref{Ddot1}): the Weyl tensor $\h{C}_{abcd}{}^d$ vanishes and $\h{S}_{\pb{ab}}$ is non-zero on $\scrip$. Furthermore, the conformally invariant part of $\h{S}_{\pb{ab}}$ is precisely the Bondi news tensor  $\h{N}_{ab}$  \cite{gerochrev,aa-yau} that is the \emph{freely specifiable} datum on $\scrip$ in the characteristic initial value problem for conformal Einstein's equations \cite{Ashtekar2_2024,sachs,rendall:1990reduction}. Thus, now the time dependence in the connection $\hD$ of the WIHS $\scrip$ is non-trivial: physically it encodes the two radiative modes of the gravitational field at $\scrip$ \cite{aa-radiativemodes,aa-yau}! Consequently, although $\scrip$ is a WIHS, there can be an arbitrary flux of gravitational radiation across it! On the other hand, if $\scrip$ of a given space-time happens to be an IHS, then the full $\hD$ would be time independent. In this case the freely specifiable data on $\scrip$ trivializes: now there are no radiative modes on $\scrip$ \cite{aa-radiativemodes}. (With the standard no-incoming radiation boundary condition, then, $\scrim$ is always an IHS). It is quite striking that the difference between WIHSs and IHSs --which may at first seem insignificant at first-- leads to \emph{qualitative} changes in the physics at $\scri$. The interplay between geometry and physics is both subtle and powerful!

To summarize, while $\scrip$ is a WIHS, there is a drastic difference between the physics it captures and that captured by a BH WIHS in the physical space-time. This difference can be directly traced back to the field equations satisfied at these two types of WIHSs: the physics at BH WIHSs $\IH$ is dictated by Einstein's equations (possibly with gauge fields) that hold there, while that at $\scrip$ is dictated by the conformal Einstein's equations (possibly with zero rest mass fields that fall-off appropriately). Because of this difference, complementary terms in (\ref{Ddot2}) vanish implying that there are  local degrees of freedom on $\scrip$ but not on $\IH$. In retrospect it is but to expected that physics strongly depends on field equations; it can be encoded in geometry only after field equations are specified.
\vskip0.1cm
\goodbreak
\emph{Remark:} The importance of field equations in assigning physical interpretation to geometrical structures is further brought out by the expression (\ref{MH}) of Hawking's quasi-local mass $M_H [S]$. This expression has several attractive properties, so long as the fields in the integrand refer to the \emph{physical} space-time $(M, g_{ab})$ on which Einstein's equations hold, with matter satisfying appropriate physical conditions. Thus, for example, in asymptotically flat space-times with matter satisfying the standard fall-off conditions one can evaluate $M_H[S]$ on appropriately chosen 2-spheres $S$ and then take the limit as $S$ tends to infinity. At spatial infinity $i^\circ$ along a space-like 3-surface, $M_H[S]$ tends to the ADM energy, and in a suitable limit to a cross section $\h{S}$ of $\scrip$ it tends to the  Bondi-energy. (In both cases the limiting procedure selects an asymptotic rest frame. See, e.g., \cite{Ashtekar4_2024}.) If the matter satisfies the dominant energy condition, then these energies would necessarily be positive. But of course these calculations use Einstein's equations satisfied by $g_{ab}$. 

Now, since $\scrip$ is a WIHS and since $M_H$ gives a physically viable quasi-local mass of MTSs, one may be tempted to calculate $M_H[\h{S}]$ using the rescaled metric $\h{g}_{ab}$ for the fields in the integrand of (\ref{MH}). Then the answer would be $\h{R}/2G$, where $\h{R}$ is the areal radius of $\scrip$ defined by $\hg_{ab}$. But this value has \emph{no bearing whatsoever} on the physical mass or energy associated with $\h{S}$, or indeed on any physical observable associated with the given space-time;  it is not even conformally invariant! Thus, { for an MTS $S$, $M_H [S]$ is a physically viable quasi-local mass only when $S$ lies} in the physical space-time satisfying Einstein's equations. 

\subsection{Interplay between observables}  
\label{s5.3}

To conclude this section, we will discuss an unforeseen interplay between horizon multipoles\, --QLH observables that characterize its geometrical structure-- \,and certain observables defined at null infinity that capture the shape of the asymptotic `mass aspect' and its dual. In the first step, we will consider space-times $(M, g_{ab})$ that are asymptotically flat, admit an IHS in its interior that meets $\scrip$ at $i^+$, and in which there is no radiation at $\scrip$ (so that it is also an IHS, but in $(\hM, \hg_{ab})$). In the second step we will consider dynamical situations: general asymptotically flat space-times that admit a DHS $\DH$ that asymptotes to an IHS meeting $\scrip$ at $i^+$, and in which there is gravitational radiation at $\scrip$ as well as $\DH$. For simplicity we will ignore matter fields on the $\DH$ as well as $\scrip$ .

In this analysis, the observables of interest at $\scrip$ are the BMS supermomenta and their duals --also called the NUT \cite{newman1963empty} or `magnetic' supermomenta-- both of which are charges associated with BMS supertranslations, evaluated at cross-sections $\h{S}$ of $\scrip$ (see e.g., \cite{aa-asNUT,Penrose:1985jw,aa-yau}). Being observables, they are conformally invariant. But one needs to choose a conformal frame to write their explicit expressions since individual fields that appear in their integrands are not conformally invariant; only the integrals over $\h{S}$ are. Fortunately, there is a natural choice. Let us begin with the case when there is no radiation at $\scrip$. In this case, there is unique Bondi conformal frame (i.e. one in which the metric $\hq_{ab}$ is a unit 2-sphere metric) in which the conserved Bondi-Sachs 4-momentum is purely time-like (i.e., one that represents the asymptotic rest frame). In this Bondi-frame a complete set of supermomenta $\mathtt{Q}_{\ell, m} [\hat{S}]$ and dual supermomenta $\mathtt{Q}^\star_{\ell,m} [\hat{S}]$ can be obtained by using the spherical harmonics $\mathring{Y}_{\ell,m}$ (of the round $\hq_{ab}$) to represent the generators of supertranslations. Following Bondi and Sachs \cite{bondi1962mgj,sachs} the expressions of these observables at $\scrip$ can be arrived at by the following limiting procedure. Consider a null surface $\mathcal{N}$ in the physical space-time (on which a retarded time coordinate $u$ is constant) foliated by 2-spheres $S$ labelled by the values of their areal radius $R_\circ$ such that in the limit $R_\circ\to \infty$, we have $S \to \hat{S}$ on $\scrip$. Then, in terms of fields defined in the physical space-time, $\mathtt{Q}_{\ell, m} $ and  $\mathtt{Q}^\star_{\ell,m}$ are given by
\ba \hskip-2.5cm (\mathtt{Q}_{\ell, m}\, - \, i\, \mathtt{Q}^\star_{\ell,m})\,[\h{S}]\,  &=& - \, \lim_{R_\circ \to \infty}\, \f{R_\circ}{8\pi G}\,\oint_{R=R_\circ} \!\!\! \Big[(C_{abcd}\,+ i\, {}^\star C_{abcd})\, n^a \ell^b n^c \ell^d\,\Big]\, Y_{\ell, m} \, \rmd^2 V \\
&=& -\,\lim_{R_\circ \to \infty}\, \f{R_\circ}{8\pi G}\,\oint_{R=R_\circ} \!\!\! \big[\Psi_2 \big]\, Y_{\ell, m} \, \rmd^2 V \label{bms1} 
\ea 
where $Y_{\ell,m}$ in the integrands are functions on the 2-spheres $R=R_\circ$ that smoothly tend to the spherical harmonics $\mathring{Y}_{\ell,m} (\theta,\varphi)$ on $\h{S}$ in the limit $R_\circ \to \infty$. The $[\Psi_2] := [(C_{abcd}\,+ i\, {}^\star C_{abcd})\, n^a \ell^b n^c \ell^d]$ that appears in the second step can be rewritten as the component 
\be \label{Psi2} \Psi_2 = C_{abcd} \ell^a m^b \bar{m}^c n^d\ee
of the Weyl tensor in a Newman Penrose tetrad adapted to $\scrip$. Since the Bondi news $\h{N}_{ab}$ vanishes on $\scrip$, the values of $\mathtt{Q}_{\ell, m}$ and  $\mathtt{Q}^\star_{\ell,m}$ are independent of the choice of $\h{S}$ on which they are evaluated. (There is some arbitrariness in the choice of the overall numerical factors; with the choice made in (\ref{bms1}) $\mathtt{Q}_{0,0}$ is only proportional to the Bondi-Sachs energy since $Y_{0,0} \hn^a$ does not represent a unit time translation.) 

Note that these observables at $\scrip$ have exactly the same functional form as the multipoles on $\IH$ defined in Eq. (\ref{dimensionless1-psi}). Since the supermomenta have dimensions of mass while the shape and spin multipoles are dimensionless, let us rescale the multipoles by multiplying them by the Hawking mass of $\IH$. Then, we have
\ba 
\hskip-1.5cm 
M_H\, \big(\texttt{I}_{\ell, m}\, - \, i\, \texttt{S}_{\ell,m}\big)  &=& - \f{R_\IH}{2\, G}\,\oint_{S} \Big[(C_{abcd}\,+i\, {}^\star C_{abcd})\, k^a \uk^b k^c \uk^d\,\Big]\, Y_{\ell, m} \, \rmd^2 V\\
&=& - \f{R_\IH}{2\, G}\,\oint_{S} \big[\Psi_2]\, Y_{\ell, m} \, \rmd^2 V \label{relation1} 
\ea 
where $R_\IH$ is the area radius and $S$ is any cross-section of $\IH$. It is both interesting and surprising is that the functional forms of these two infinite sets of observables (\ref{bms1}) and (\ref{relation1}) is the same. Of course the \emph{numerical values} of these observables would be different on $\scrip$ from what they are at $\IH$ since the 2-sphere integrals are evaluated on two distinct null surfaces, the first in the asymptotic region and the second in the strong field region. Note also that the physical meaning of supermomenta at $\scrip$ and multipoles of $\IH$ is quite different. Nonetheless, as we will see in section \ref{s6.2} the close similarity in the functional forms opens the door to gravitational wave tomography. 

In presence of radiation, the integrands of supermomenta acquire additional terms, while the form of the seed function for multipoles on DHSs remains the same: {\smash{$\f{1}{4}\mathcal{R}\,+\, i \t\epsilon^{ab}\t{D}_a \omega_b$}} (see Eq. (\ref{dimensionless2})). Therefore a priori one would not expect the simple relation between supermomenta and multipoles to extend. But it turns out that it does! This is a rather surprising fact and its ramifications are not yet to be fully understood. 

Consider then general space-times $(M, g_{ab})$ with gravitational radiation that admit a $\scrip$ and a DHS $\DH$ in the distant future that asymptotes to an axisymmetric IHS meeting $\scrip$ at $i^+$. Now it is natural to select the conformal frame at $\scrip$ that is adapted to $i^+$. One can again uniquely single out such a frame: A Bondi conformal frame for which the limit of the Bondi 4-momentum to $i^+$ is purely time-like (i.e., one that represents the BH rest frame in the asymptotic future). At $\scrip$ let us use the spherical harmonics defined by the round metric $\h{q}_{ab}$ of this Bondi frame and on $\DH$ the spherical harmonics that descend from the IHS to which the $\DH$ asymptotes in the distant future as in section \ref{s3.3}. Now expressions of supermomenta at $\scrip$ are more complicated because of the presence of gravitational radiation and they are no longer conserved.  In place of (\ref{bms1}) we have \cite{aa-asNUT,Penrose:1985jw,aa-yau}: 
\be \label{bms2} 
\hskip-2cm 
(\mathtt{Q}_{\ell, m}\, - \, i\, \mathtt{Q}^\star_{\ell,m})\,[\h{S}]  =  - \, \lim_{R_\circ \to \infty}\, \f{R_\circ}{8\pi G}\, \oint_{R=R_\circ}\!\!\Big[ \Psi_2\, + \,\sigma_{ab}^{(n)}\, (\sigma^{ab}_{(\ell)} + i\, {}^\star\sigma^{ab}_{(\ell)} )\Big]\, Y_{\ell, m} \, \rmd^2 V\ee
where ${}^\star\sigma^{ab}_{(\ell)} = \t\epsilon^{ac} \tq^{bd} \sigma_{cd}^{(\ell)}$, with $\t\epsilon^{ac}$  and $\tq^{bd}$ the area 2-form and the intrinsic metric on the 2-spheres ${R=R_\circ}$. Next, let us turn to the multipoles on $\DH$. A highly non-trivial fact is that because of presence of radiation, the relation between the fields $\R,\, \t\epsilon^{ab} D_a \t\omega_b$ and the real and imaginary parts of $\Psi_2$  on $\DH$ also changes \emph{precisely} such that the expression in the square bracket in the integrand of Eq. (\ref{bms2}) continues to equal $\f{1}{4}\big[ \R + 2i \t\epsilon^{ab} D_a \t\omega_b\big]$ on $\DH$! Therefore, rescaling the DHS multipoles by the Hawking mass $M_H$ we have: 
\be \label{relation2}\hskip-1.5cm 
M_H\, \big(\texttt{I}_{\ell, m}\, - \, i\, \texttt{S}_{\ell,m}\big)  = - \f{R_\IH}{2\, G}\, \oint_{S}\!\! \Big[ \Psi_2\, + \,\sigma_{ab}^{(k)}\, (\sigma^{ab}_{(\uk)} + i\, {}^\star\sigma^{ab}_{(\uk)} )\Big]\, Y_{\ell, m} \, \rmd^2 V \, .\ee 
Again, the \emph{numerical values} of these observables would be different on $\scrip$ from what they are at $\DH$; what is interesting and surprising is that the functional form of the charges (\ref{bms2}) defined intrinsically at $\scrip$ is the same as that of multipoles (\ref{relation2}) defined on $\DH$ using entirely different considerations. This property has an interesting consequence. We saw in section \ref{s3.3} that if there is a hypersurface orthogonal rotational Killing field $\varphi^a$ in an arbitrarily small neighborhood of a DHS $\DH$, then all spin-moments $\texttt{S}_{\ell,m}[S]$ vanish identically on every MTS $S$. The analogous result extends to observables at $\scrip$: If the physical space-time metric admits a hypersurface orthogonal rotational Killing field $\varphi^a$ in an arbitrarily small neighborhood of $\scrip$, then  $\mathtt{Q}^\star_{\ell,m}$ vanish for all $\ell,m$.

\vskip0.1cm
\texttt{Open Issue 9 (OI-9)} The BMS supermomenta are charges associated with  BMS supertranslations on a phase space constructed at $\scrip$ \cite{Ashtekar3_2024}. Is there a similar interpretation for shape multipoles on the DHS, given that the two expressions have the same functional form? 
\goodbreak

\section{Horizon Dynamics and Gravitational Wave Astronomy} 
\label{s6} 

One of the most important applications of quasi-local horizons is to NR, especially in simulations of BBH mergers.  The coordinate and gauge conditions used in these simulations are typically chosen for reasons of numerical efficiency and convenience, and are generally not the same as those assumed in analytical studies. Since waveforms are constructed by combining results from both, analytical approximations and NR, it is important to ensure gauge invariance of final results. QLHs are ideally suited for this purpose and are now commonly used in calculations of masses and the spins of progenitors as well as the remnant  { \cite{Dreyer:2002mx,Schnetter:2006yt,Iozzo_2021}}. 

But it is the issues related to horizon dynamics  --in particular, calculations of fluxes across DHSs and details of the approach to equilibrium-- that bring out the full utility of this framework. In this section we shall report on recent results on these aspects of the horizon dynamics. As we saw in section \ref{s5}, there is a close relation between geometrical structures at null infinity $\scrip$ and WIHSs $\IH$. Results summarized in this section  extend this relation to \emph{dynamics} at $\scrip$ and at \emph{DHSs} $\DH$. These results also bring out a synergy between mathematical GR and numerical methods. Since we are primarily interested in BBH systems, in this section we will assume vacuum Einstein's equations and, as is customary in the gravitational wave literature, we will generally set $G=1$.

\subsection{The approach to equilibrium}
\label{s6.1}

This section is divided into two parts. In the first we recall results from studies of the post-merger waveforms that bring out the important role played by quasi-normal modes (QNMs) of the remnant in the late time dynamics at $\scrip$. In the second part we will summarize three numerical studies that show that, rather surprisingly, QNMs also play an important role in the process by which the DHS of the remnant reaches equilibrium in full GR. \vskip0.2cm 

\centerline{\it Quasi-normal modes}
\label{s6.1.1}
\vskip0.2cm

Ground based detectors such as LIGO and Virgo observe gravitational wave strain $h(t)$, which is a linear combination of the two polarizations of the gravitational wave signal $h_+$ and $h_\times$ (see e.g. \cite{Jaranowski_Krolak_2009}). It is customary to break up the GW signal from a binary system of coalescing compact objects into three distinct regimes based on the signal morphology, namely the inspiral, merger, and ringdown. In the inspiral regime $h(t)$ is a ``chirp'' with slowly growing amplitude and frequency.  The merger signal has the morphology of a high amplitude burst and is expected to have non-perturbative features. The merger time is often characterized in terms of the peak of the strain signal or alternatively in terms of the peak luminosity (which is proportional to a 2-sphere integral of $\dot{h}_+^2 + \dot{h}_\times^2$). Soon after this we transition into the ringdown regime where $h(t)$ is expected to be a superposition of damped sinusoids, determined by the QNMs of the remnant BH. The precise distinction between these three regimes is not sharp. However, it was suggested in \cite{Kamaretsos:2011um} that a ringdown regime can be identified in the post-merger waveform approximately $10M$ after the peak in the luminosity of the dominant (quadrupolar) angular mode, where $M$ is the total mass of the binary. On the observational side, the analysis of gravitational wave data from the first observed binary BH merger event GW150914 \cite{TheLIGOScientific:2016src} reaffirms that a ringdown regime can be tentatively identified approximately $10M$ after the peak { (which corresponds to $\lesssim 1$ cycle in the post merger wave form!)}. QNMs have attracted a great deal of attention in the recent gravitational wave literature.  The detection of these signals and using them to test the Kerr nature of the remnant BH is an important goal of gravitational wave astronomy. Detailed reviews of results that are heavily used in the gravitational wave community can be found in e.g. \cite{Kokkotas:1999bd,Berti_2009}. 
\footnote{From a mathematical perspective, there have been  significant recent advances \cite{gajic2024quasinormalmodeskerrspacetimes,stucker2024quasinormalmodeskerrblack} that put the notion of QNMs on a rigorous footing. The two works use different methods, each with its own merits. The first uses asymptotically hyperboloidal 3-surfaces and spaces of functions that have finite Sobolev norm in bounded regions and are, furthermore, `Gevrey-regular' at null infinity (i.e., roughly speaking, lie `in-between analytical and smooth' categories at $\scrip$). The second combines the method of complex scaling near infinity with microlocal methods near the BH horizon.}
The QNMs for a Kerr BH of mass $M$ and angular momentum $J = Ma$ are characterized by a spectrum of complex frequencies
labeled by three integers $(\ell,m,n)$:\, $\omega_{\ell m n}(M,a)$,\,
($\ell \geq 2, -\ell\leq m \leq \ell, n\geq 0$). Here $(\ell,m)$ are the usual angular
quantum numbers and the index $n$ is the so-called overtone index with
the lowest mode $n=0$ referred to as the fundamental mode, and the
higher modes are referred to as the overtones. The real part of\, $\omega_{\ell m n}(M,a)$\,
is the oscillation frequency while the imaginary part yields
the (inverse) damping time.  For a Kerr BH,
it is seen that as we increase $n$ while keeping $(\ell, m)$ fixed,
the real parts of $\omega_{\ell mn}$ are relatively unchanged, while
the imaginary parts increase, whence modes decay faster. 

The analysis of the GW merger event GW150914 mentioned above \cite{TheLIGOScientific:2016src}, shows again that the observed ringdown signal is consistent with the $\ell=m=2$ QNM (which is expected to be the dominant mode).  Since the overtones decay faster, it was generally believed that they would not be important for the post-merger signal. However, more recently, a detailed analysis of numerical relativity waveforms suggested that essentially the entire post-merger regime can be modeled using linear combinations of QNMs and their overtones \cite{Giesler:2019uxc}. Similarly, a closer look at gravitational wave data from GW150914 revealed again that the post-merger signal may be better modeled using the overtones \cite{Isi:2019aib}. This is potentially important for astrophysical applications since it allows a more accurate estimation of the remnant mass and spin, and is also of importance in tests of general relativity using BH no-hair theorem and the area increase law \cite{Isi:2019aib,Isi:2020tac}. There have been several follow-ups to these claims on the data analysis side (see e.g. \cite{Baibhav:2023clw,Cotesta:2022pci,Isi:2023nif,Isi:2022mhy}) that have led to an ongoing debate on whether this role played by overtones is a reflection of a genuine feature of Einstein dynamics or simply an artifact of using a large number of fitting parameters.
\vskip0.2cm 

\centerline{\it DHSs: Results from NR simulations}
\label{s6.1.2}
\vskip0.2cm

The counterpart to the gravitational wave signals at $\scrip$ at QLHs are the fluxes across the DHS $\DH$ of the remnant. We will now summarize results on these fluxes and their connection to QNMs from three different numerical studies. The first work in this direction \cite{Gupta:2018znn} studied properties of the remnant formed by the coalescence of two equal mass, non-spinning BHs and investigated its approach to equilibrium. It was found that the dynamical horizon multipole moments decay rapidly to their final Kerr values, as expected. In addition, the decay of these multipole moments turned out to have several interesting features, illustrated in Fig.~\ref{fig:mass_moments}. Two of these features are especially noteworthy. First, there are oscillations and perhaps surprisingly, the oscillation frequencies turns out to be given by the QNM frequencies of the remnant BH. The second feature is the qualitative change in the decay rate starting at $t\sim 27M$, i.e. about $10M$ after the merger, and thus consistent with the suggestion of \cite{Kamaretsos:2011um} on when the QNM regime sets in at at $\scrip$.  A more detailed understanding of this behavior requires an investigation with higher accuracy simulations and a more detailed modeling of the horizon dynamics. The formalism summarized in the sections \ref{s3} and \ref{s4} is ideally suited for this purpose. See also \cite{Prasad:2020xgr,Prasad:2021mrj} for related studies in the inspiral regime.\vskip0.2cm

\begin{figure}
    \begin{center}
    \includegraphics[angle=-0,width=0.5\columnwidth]{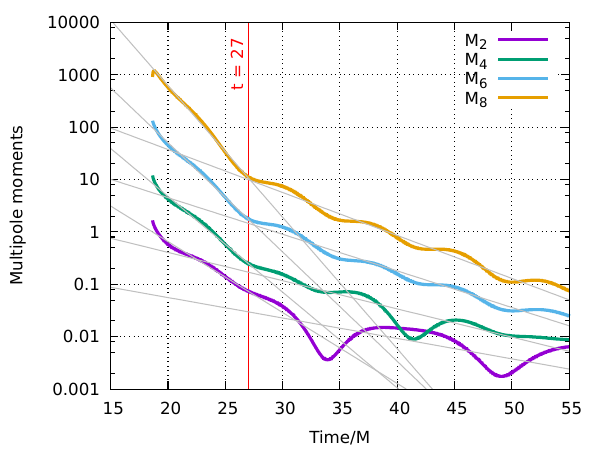}
    \caption{\footnotesize{(Figure from
            \cite{Gupta:2018znn})\emph{The behavior of the mass
            moments of Eq.(\ref{dimensionfull}):} 
           $\texttt{M}_{\ell,0} \equiv \texttt{M}_{\ell}$ for $\ell=2,4,6,8$
          are shown on a logarithmic scale for the remnant black hole
          for the merger of two equal mass non-spinning black
          holes. The x-axis is the time parameter used in the
          simulation in units of the total mass.  In the simulation
          time, the common horizon is formed at $\sim 17M$ and the
          multipole moments are calculated from this time onwards. The
          decay of the multipoles is seen to be exponential, and
          furthermore: i) there are oscillations with frequencies
          corresponding to the real parts of QNM frequencies, and ii)
          there is a qualitative change in the decay rate starting
          from $\sim 27M$, i.e. about $10M$ after the merger.}}
      \label{fig:mass_moments}
  \end{center}
\end{figure}

The second set of simulations summarized below is based on ideas discussed in section \ref{s4.3}, and makes a crucial use of the high accuracy apparent horizon finder
\cite{Pook-Kolb:2018igu,PhysRevLett.123.171102,PhysRevD.100.084044}. 
In this discussion, we will use the notation that is common in the NR literature. Thus, we will now denote the null normals to MTSs by $(\ell^a,\, n^a)$ (in place of $(k^a,\, \uk^a)$). The ingoing and outgoing directions are known unambiguously in BBH mergers and, as is common in the literature, $\ell^a$ will be the outgoing null normal and $n^a$ the ingoing.  In the context of a numerical simulation which locates an MTS $S$ numerically on a Cauchy slice $\Sigma$, common choices are
\begin{equation} \label{elln}
  \ell^a = \frac{1}{\sqrt{2}}(\tau^a + r^a)\,,\quad{\rm and}\quad n^a = \frac{1}{\sqrt{2}}(\tau^a - r^a)\,.  
\end{equation}
Here $\tau^a$ is the unit time-like normal to $\Sigma$, and $r^a$ is the unit space-like normal to $S$ on $\Sigma$. (In section \ref{s3.2}, $(k^a,\, \uk^a)$ were defined using normals $\h\tau^a$ and $\h{r}^a$ adapted to the DH $\DH$, rather than normals $\tau^a$ and $r^a$, adapted to $\Sigma$. Nonetheless, $(\ell^a,\, n^a)$ defined in (\ref{elln}) are  proportional to $(k^a,\, \uk^a)$ of section \ref{s3.2} and both are normalized; $k^a \uk_a =-1$ and $\ell^a n_a =-1$.) The basic quantities of interest are the multipole moments, discussed above, and also the shear $\sigma_{ab}^{(\ell)}$ of the null normal $\ell^a$. (Properties of $\zeta_a^{(\ell)}$  that accompanies $\sigma_{ab}^{(\ell)}$ in the expression (\ref{2ndlaw1}) of the energy flux have not been investigated in these numerical simulations because its amplitude is much smaller.)

As discussed in Sec.~\ref{s3}, $\sigma_{ab}^{(\ell)}$ appears in the area increase law and as such can be viewed as the infalling gravitational radiation into the horizon.  Following the Newman-Penrose notation, let us introduce a complex dyad $(m^a,\bar{m}^a)$ that is tangent to each MTS and satisfies $m^a m_a =0$ and $m^a\bar{m}_a = 1$. Then, information in $\sigma_{ab}^{(\ell)}$ is fully encoded in a complex scalar $\sigma$: 
\begin{equation}
  \label{eq:shear}
  \sigma = { -}\, m^am^b\sigma_{ab}^{(\ell)}\,.
\end{equation}
For simplicity, we shall drop the superscript ${}^{(\ell)}$ on $\sigma$; in this section, by `shear' we shall always mean the shear of $\ell^a$. Because $\sigma$ is obtained by contracting $\sigma_{ab}^{(\ell)}$ twice with $m^a$, it is said to carry spin-weight 2: Under a dyad rotation $m^a \to e^{i\alpha} m^a$, we have $\sigma \to e^{2i\alpha}\, \sigma$. The vectors $(\ell^a,\,n^a,\, m^a,\,\bar{m}^a)$ constitute a Newman-Penrose null tetrad on the QLH. (Thus, as is common in the literature, $(\ell^a,\,n^a\,,m^a)$ denote null vectors, while $(\ell,\,m,\,n)$ refer to the labels of QNMs; the distinction will be clear from the context.) 

It is convenient to decompose $\sigma$ on any given MTS $S$ into angular modes $\sigma_{\ell m}$ based on spherical harmonics ${}_2Y_{\ell m}$ with spin-weight $2$:
\begin{equation} \label{eq:shear-modes}
  \sigma = \sum_{\ell=2}^\infty\sum_{m=-\ell}^{\ell}\sigma_{\ell m}\,\,{}_2Y_{\ell m}(\theta,\varphi)\,.
\end{equation}
At each time step $t$ of the numerical simulation, we decompose $\sigma(t)$ on the MTS $S(t)$ as above and thus obtain the time series $\sigma_{\ell m}(t)$ for each $(\ell,m)$ that is considered. For simulations that refer to axisymmetric space-times, we only need the $m=0$ mode and it is then unnecessary to indicate the mode index $m$. In this case the shear modes will be simply denoted $\sigma_\ell$ (which is moreover real valued in this case).  Finally, as noted in section \ref{s2.4}, 
{ the spherical harmonics $Y_{\ell,0} (\theta,\varphi)$ can be defined unambiguously  on axisymmetric MTSs.} Therefore one does not have to construct a dynamical vector field $X^a$ (discussed in Sec.~\ref{s3.3}) that is needed on non-axisymmetric DHSs to transport them from asymptotic future.
  
With this notation at hand, we now summarize the main findings reported in the second set of simulations \cite{Mourier:2020mwa,Forteza:2021wfq}, which was for head-on collisions of non-spinning BHs, focusing on a specific mass ratio value $1.6$; results for other mass ratios share the same qualitative features.  Here the focus will be on the outermost horizon of the remnant BH since various inner horizon segments discussed in Sec.~\ref{s4.2} are not relevant for studies of approach to equilibrium. The accuracy of the simulation and the horizon finder allows one to reliably extract high angular modes, all the way up to $\ell = 12$. As noted in section \ref{s4.3}, the final remnant in this merger is a Schwarzschild BH. Furthermore, since the progenitors have neither spin nor orbital angular momentum in this case, one would expect all angular momentum multipoles to vanish also on the final DHS of the remnant. This is indeed the case because the space-time admits a hypersurface orthogonal rotational Killing field  \cite{ak-kvfs} (see section \ref{s3.3}). 
\begin{figure}[]
  \centering
    \includegraphics[width=0.4\columnwidth]{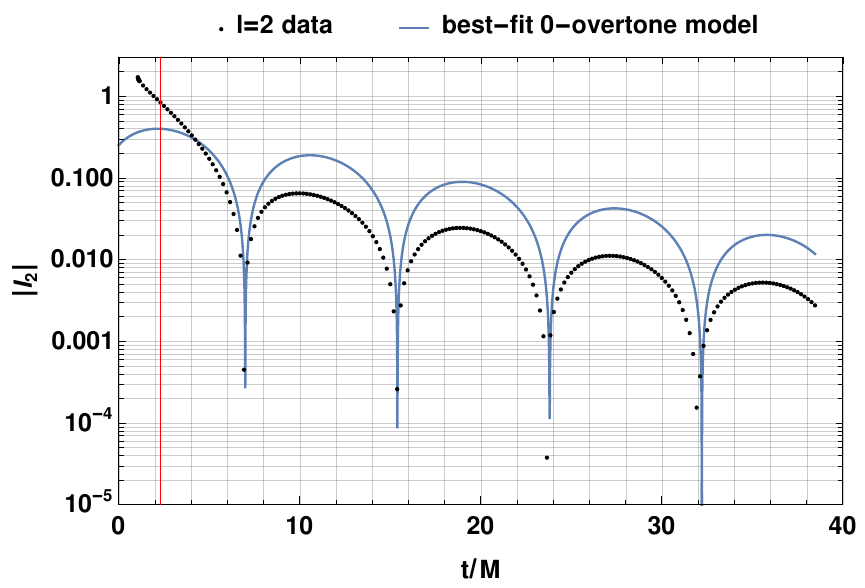}
  \includegraphics[width=0.4\columnwidth]{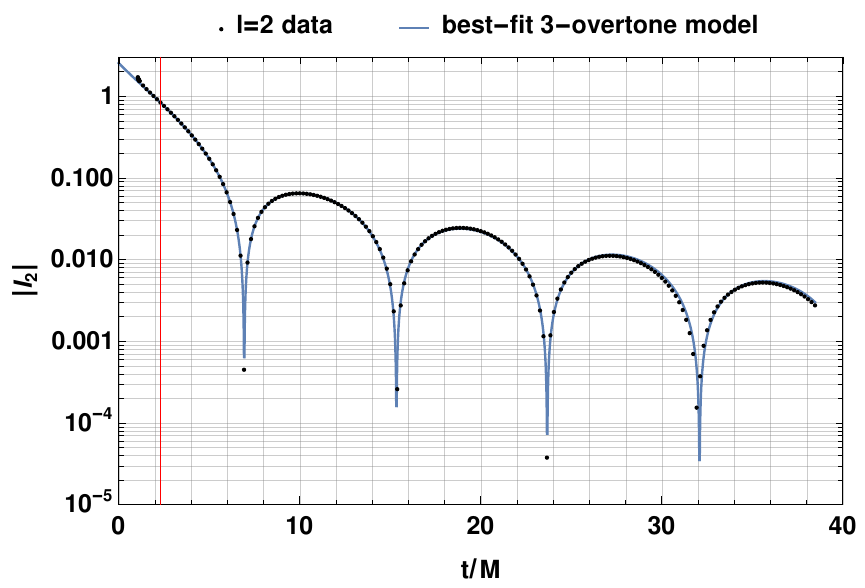}\\
  \includegraphics[width=0.4\columnwidth]{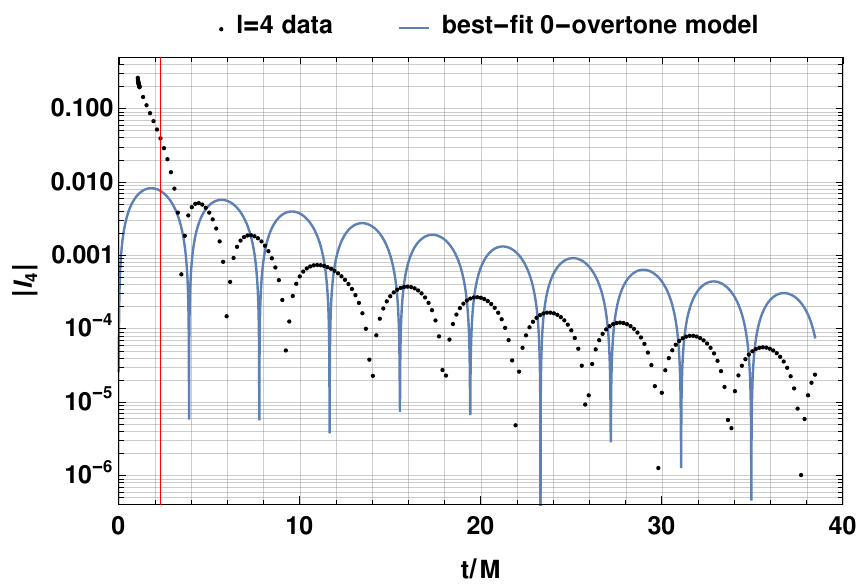}
  \includegraphics[width=0.4\columnwidth]{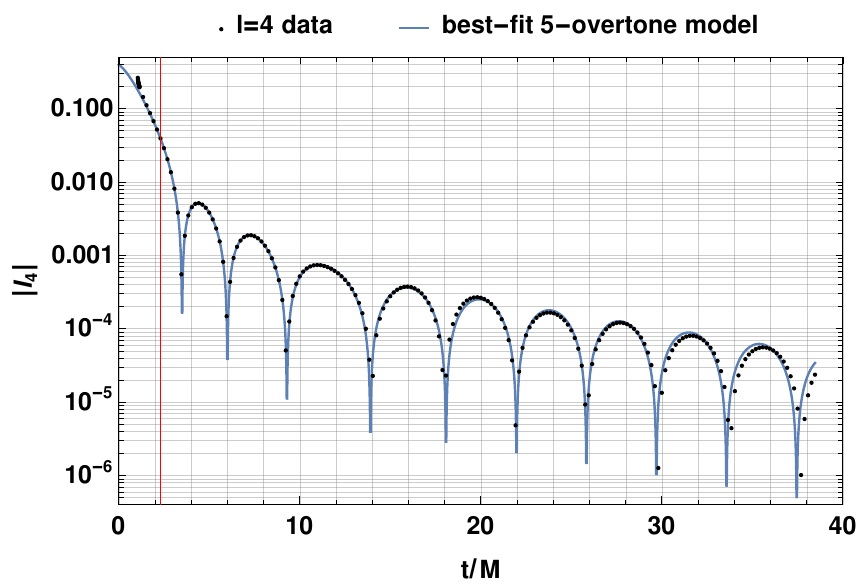}\\
  \caption{\footnotesize{(Figure from
        \cite{Mourier:2020mwa}) \emph{Results for the head-on
        collision of non-spinning black holes with mass ratio $1.6$:}
      These plots show the behavior of the moments $\mathtt{I}_2$ and
      $\mathtt{I}_4$ for the remnant black hole.  The top-row shows
      the time evolution of the $\ell=2$ mass quadrupole moment
      $\mathtt{I}_2$.  The plots shows the multipole moment as a
      function of time (black dots) and also a best-fit model using
      combinations of quasi-normal modes.  The first fit on the left
      uses only the fundamental mode, while the fit shown on the right
      includes modes up to the third overtone.  The second-row is
      similar, but for the $\ell=4$ multipole; in this case the right
      panel shows the best-fit with five overtones.  In these plots,
      the time-axis refers to simulation start time starting with
      time-symmetric initial data. The common horizon is formed
      shortly after the simulation begins, and the data points (black
      dots) begin from this point onwards.  The solid lines are the
      best-fit solution with the QNM model with or without overtones.
      Only data to the right of the vertical red line is used to
      create the best-fit solution. Note that the model with the
      overtone provides a good fit to the data even to the left of
      this red-line while the model with just the fundamental mode
      does not.  }}
  \label{fig:l2-overtonefits-multipole}
\end{figure}

The numerical results of interest are then the shape multipole moments $\texttt{I}_{\ell}(t)$ defined in Sec.~\ref{s3.3} and the modes of the complex shear $\sigma_\ell(t)$ defined in Eq.~(\ref{eq:shear-modes}), as functions of the simulation time $t$. The question is whether one can model this time evolution using combinations of QNMs on $\DH$ in the \emph{strong field} regime, just as one can model waveforms at $\scrip$ in the \emph{weak field} regime.  Furthermore, since the final remnant is a Schwarzschild BH, all multipole moments beyond $\ell\geq 1$ decay to zero.  It turns out that the time dependence of multipoles is remarkably similar to that of the post-merger gravitational wave signal at $\scrip$. Starting from the peak of the gravitational wave signal, we typically see a rapid decay followed by damped sinusoidal oscillations given by the QNMs of the remnant BH. Analogously, starting from the time when the common { apparent horizon (AH)} is formed, the higher multipole moments decay rapidly and exhibit oscillations.  Thus, as a working hypothesis, we shall associate the merger time seen in the GW signal at $\scrip$ with the time when the common AH is formed. 

Fig.~\ref{fig:l2-overtonefits-multipole} shows the quadrupole moments $\mathtt{I}_2$ and $\mathtt{I}_4$ for the common horizon as functions of time. A closer look shows that
the dynamics of the multipole moments again shows distinct regimes
with different damping rates: an earlier regime with rapid decay,
followed by damped oscillations.  In the model, this decay
is related to the quasi-normal-modes and one can find fits, two of which are shown in Fig.~\ref{fig:l2-overtonefits-multipole}. The left panel includes just the fundamental $\ell=2$ mode while the right panel uses a combination of 4 modes: the fundamental and first three
overtones.  It is clear that one obtains a much better agreement when
the overtones are considered.  A second example, shown in the second row of the same figure,
is for one of the higher mass-multipole moments, with
$\ell=4$. In this case, we need at least 5 overtones in order to
obtain a suitable fit.  It is found heuristically that
$\mathtt{I}_\ell(t)$ requires $\ell+1$ overtones; see
\cite{Mourier:2020mwa} for details.  We would like to emphasize that  while these results are suggestive, they should not be taken to be definitive proof that the overtones are indeed present in the data, since the results are obtained by fitting different models to the data and not arrived at from a first principle reasoning.
\begin{figure*}[h]
  \centering    
  \includegraphics[width=0.4\textwidth]{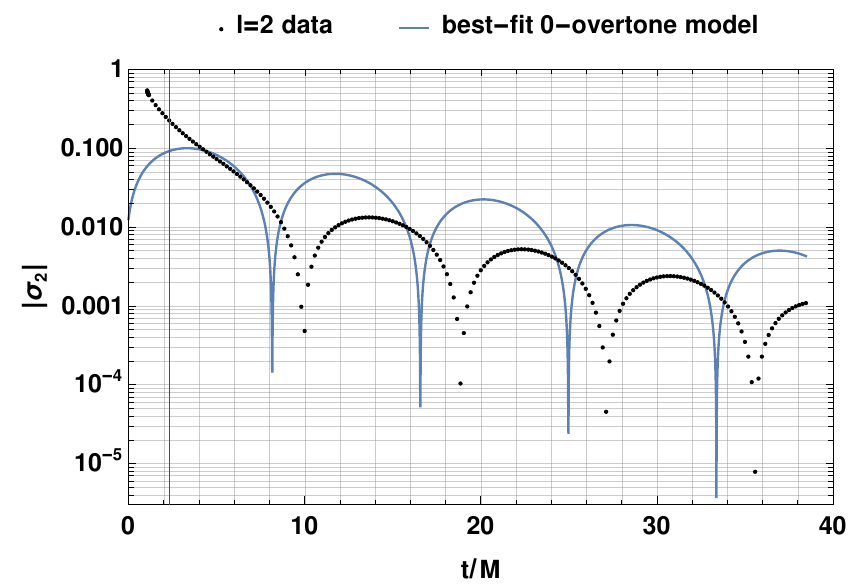}
  \includegraphics[width=0.4\textwidth]{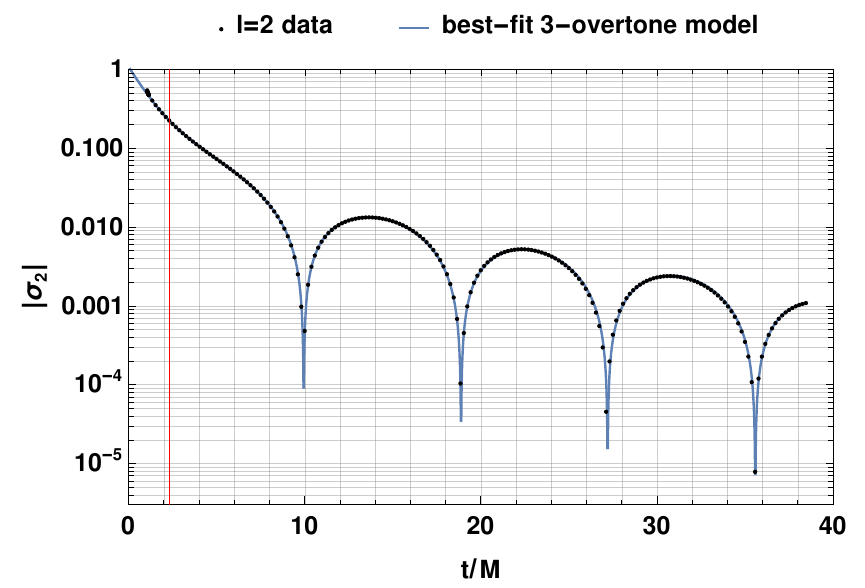}
  \includegraphics[width=0.4\textwidth]{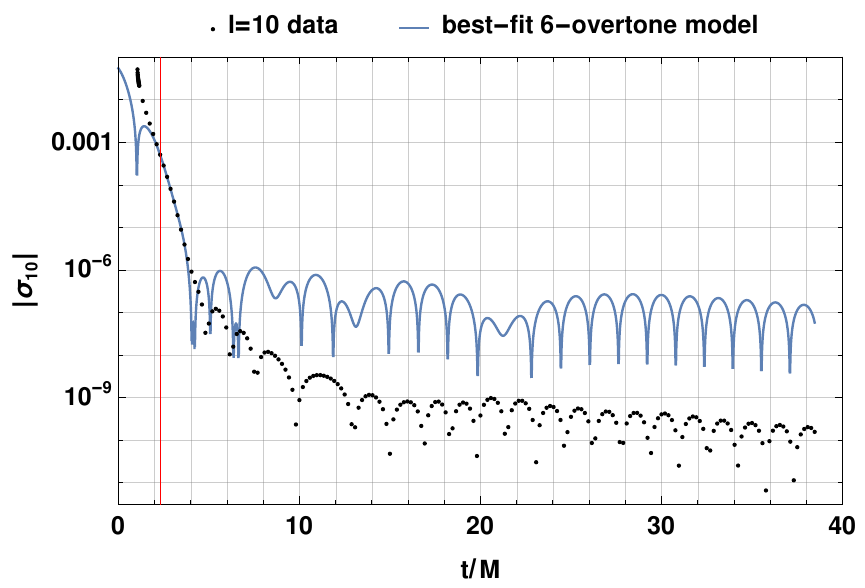}
  \includegraphics[width=0.4\textwidth]{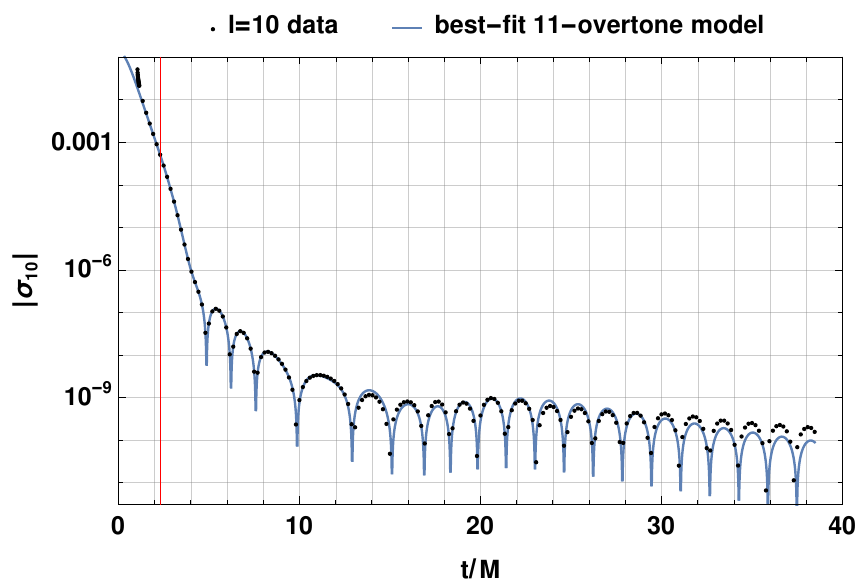}
  \caption{%
    {\footnotesize{(Figure from \cite{Mourier:2020mwa}) \emph{The
          behavior of the various shear modes $\sigma_\ell$:} These
        modes are shown for the same numerical simulation as in
        Fig.~\ref{fig:l2-overtonefits-multipole}. Again, the black
        dots are the numerical data, while the solid blue curves are
        the best fit models. The top-row shows the shear mode
        $\sigma_{2}$ and its fits with a QNM expansion including
        either just the fundamental mode (top-left) and with three
        overtones (top right).  The bottom row is for the $\ell=10$
        mode $\sigma_{10}$, and the best-fits with either 6 overtones
        (bottom left) or 11 overtones (bottom right). The time-axis,
        the vertical red line have the same meaning as in
        Fig.~\ref{fig:l2-overtonefits-multipole}} }}
    \label{fig:shear_overtone}
\end{figure*}
Let us now turn to the shear $\sigma$, which is more directly related to the
infalling gravitational radiation. $\sigma$ can also be expanded as a sum of quasi-normal modes of the remnant BH, including also
the overtones. As for the multipoles, one finds empirically
that for a given mode $\sigma_\ell$, an adequate fit is again provided by a
QNM expansion including $\ell+1$ overtones. Two examples are shown in
Fig.~\ref{fig:shear_overtone}, first for the dominant $\ell=2$ mode which is
well described by a fit including 3 overtones, and then for the $\ell=10$
mode, which requires 11 overtones.

Based on the fits for multipole moments and shear shown in the two figures (and more detailed results from \cite{Mourier:2020mwa}), one can conclude that the distinct regimes in the dynamics of these quantities on $\DH$ can be potentially explained in terms of quasi-normal-modes and the overtones. With overtones one is able to model the steep decay rate at early times, even just after the common horizon is formed: The QNM fit can describe the dynamics at the remnant DHS
throughout its history, even right at its formation which, one might have thought, is a non-perturbative phenomenon. However, we emphasize again that this is only a suggestive evidence. \vskip0.2cm

A key simplification in this analysis of approach to equilibrium came from the fact that, thanks to the restriction to non spinning BHs and head-on collisions, the full space-time metric is axisymmetric.  In particular, as explained in the beginning of section \ref{s3.3}, it is much easier to define the required spherical harmonics. The third set of NR results \cite{Chen_2022} refers to a non-axisymmetric situation where initially the two equal mass BHs are placed on a ``quasi-circular'' orbit following the procedure detailed in \cite{Buonanno:2010yk} to reduce the initial eccentricity. Specifically, the simulation used in \cite{Chen_2022} is labeled as SXS:BBH:0389 in the SXS catalog \cite{Boyle:2019kee}. While the head-on collision simulation reported above started with the simple Brill-Lindquist construction, the initial data here is based on the Extended Conformal Thin Sandwich formulation of the initial value problem \cite{Pfeiffer:2002iy,York:1998hy}.  The spins of the two progenitor BHs, as measured by the QLH angular momentum formula (\ref{J-DH}), is essentially zero numerically.  The system completes $~\sim 18$ orbits before merger.  The remnant, once it reaches equilibrium, is a Kerr BH with dimensionless spin $J_f/M_f^2 \sim 0.68644$ (with $J_f$ and $M_f$ being respectively the final angular momentum and mass of the remnant BH using the quasi-local measures for the DHS). 

Since space-time is no longer axisymmetric in this case, the issue of finding the appropriate spherical harmonics is more complicated. Nonetheless, at a late time, say $t_f$, the DHS of the remnant is sufficiently close to Kerr IH so that it is straightforward to find an axial symmetry vector field $\varphi^a$ and to introduce preferred spherical coordinates $(\theta,\,\varphi)$ and the corresponding spherical harmonics on the marginally trapped surface at $t_f$. As explained in the main part of section \ref{s3.3}, this structure can be Lie-dragged \emph{backward} in time by the appropriately constructed vector field $X^a$ on $\DH$, which takes into account the angular velocity of the horizon. The end result is a well-defined `kinematic' set of angular coordinates and associated spherical harmonics on all MTSs of $\DH$. They can be used to compute multipoles on each MTS, whose time dependence encodes the horizon dynamics in a gauge invariant fashion. This general procedure was implemented in \cite{Chen_2022}. 
\footnote{The time dependence of QNM modes refers to the affine parameter of the time translation Killing field of the Kerr remnant, while that of horizon multipoles refers to the coordinate $v$ whose level surfaces are the MTSs. For a meaningful comparison, the moments are modified in  \cite{Chen_2022} via\,\,{\smash{$\mathtt{I}_{\ell m} \rightarrow \mathtt{I}_{\ell m}e^{-im\Omega t}$}},\,\, taking into account the angular velocity $\Omega$ of the final Kerr remnant. It is not entirely clear to us that this is the optimal procedure for the required comparison.} 

It was shown that the time evolution for the $\ell=m=2$ mass quadrupole moment $I_{22}(t)$ can again be modeled as a combination of QNMs, and these turn out to be strongly correlated with the corresponding modes of the gravitational wave signal.  Moreover, again it turns out that $I_{22}(t)$ can be well modeled using overtones starting already from the formation of the common DHS, where, as remarked earlier, a priori one would think that non-linearities of full general relativity would be important. Thus, the situation is qualitatively similar to that we summarized for the head-on collision. 

But now there is an added complication, a phenomenon known as mode-mixing: in order to model $I_{22}(t)$, it is not sufficient to include just the $\ell=2$ QNMs; it is necessary to include also the $\ell=3, m=2$ QNM as well. This phenomenon occurs because perturbations of Kerr, and thus the QNMs, are most simply expressed with (spin-weighted) \emph{spheroidal} harmonics, while the waveforms at $\scrip$ and the horizon multipole moments use the usual spin-weighted spherical harmonics.
In \cite{Chen_2022}, the fitting with QNMs, including overtones was based on the ansatz 
\begin{equation}
  \label{eq:ansatz-modemix}
  I_{22}(t;t_0) = C_{320}e^{-i\omega_{320}(t-t_0)} + \sum_{n=0}^NC_{22n}e^{-i\omega_{22n}(t-t_0)}\,.
\end{equation}
Here $t_0$ can be considered as a start time for the ringdown regime, so that the model is valid for $t\geq t_0$. Moreover $t_0$ itself can be, in principle, any time after the formation of the common horizon. The complex QNM frequencies $\omega_{320}$ and $\omega_{22n}$ are fixed by the mass and spin of the remnant, while $C_{320},C_{22n}$ are free model parameters.  
It is shown that $N=3$ (i.e. 3 overtones) suffices to model
$I_{22}(t)$ starting from early times soon after the merger; including still more overtones only leads to marginal
gains.  Fig.~\ref{fig:chen_i22} shows the improvements of the fit as we
increase $N$.  The values of the mismatch improve and become flatter
as a function of $t_0$ as $N$ is increased from $0$ to $3$.  Similar results hold also
for other multipole moments as well.
\begin{figure*}[h]
  \centering    
  \includegraphics[width=0.8\textwidth]{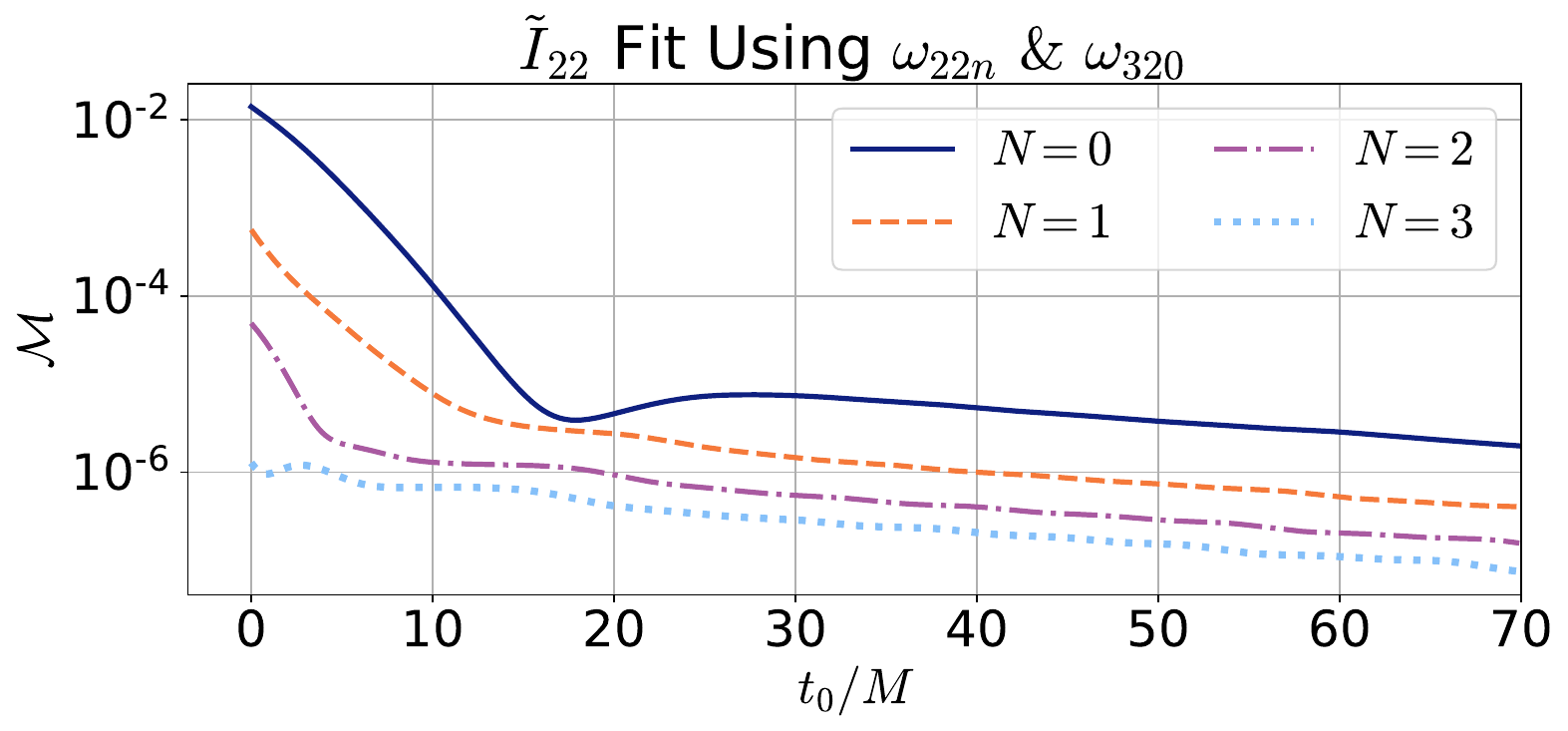}
  \caption{%
    {\footnotesize{(Figure from \cite{Chen_2022}) Improvement of the
    goodness of fit for the best model of
    Eq.~(\ref{eq:ansatz-modemix}) for different values of $N$ and the
    start time $t_0$; the time axis is labeled such that the 
    remnant horizon is formed at $t=0$.  The measure of goodness of fit uses
    the normalized $L^2$ inner product between the model and the
    numerical data over the relevant time interval (starting from
    $t_0$).  Maximizing the inner product over $(C_{320},C_{22n})$
    yields the best \emph{match} $m$ for each value of $t_0$ and for
    each $N$; the mismatch $\mathcal{M}$ shown in the plot is
    $\mathcal{M} = 1-m$. It is seen that the mismatch decreases
    as $N$ is increased and also becomes flatter over the range of
    $t_0$ (the improvement beyond $N=3$ is marginal). 
    }} }
    \label{fig:chen_i22}
  \end{figure*}

To summarize, in this subsection we summarized illustrative results from NR investigations on how the DHS $\DH$ of the remnant approaches equilibrium. They show that, in a BBH merger, the dynamics of fields on $\DH$ can be well modeled as a linear combination of quasi-normal-modes associated with the remnant \cite{Mourier:2020mwa,Forteza:2021wfq,Chen_2022}. By including overtones, one can obtain a good fit all the way to the merger, i.e., up to the event when the common horizon is first detected. While this is a striking observation, as we have emphasized, it is an empirical result based on standard fitting procedures. At this stage it is not clear at  why the quasi-normal-modes should work so well, nor how one can determine the specific quasi-normal-modes one needs to use to model a particular multipole or shear mode. As of now there is no systematic reasoning based on analytic considerations showing that linear, QNM approximation suffices to capture the physics of full GR starting from the merger. In section \ref{s6.2} we will discuss certain checks that can be performed to test how close to the merger the QNM approximation can suffice.\vskip0.1cm
\goodbreak

\emph{Remarks :}\vskip0.1cm
1. The issue of stability of QNMs has drawn considerable attention in the mathematical GR community. It has been argued that the overtones might be unstable under very small perturbations of the effective potential outside the BH
\cite{Jaramillo:2020tuu,Destounis:2021lum,Jaramillo:2022kuv,Boyanov:2022ark,Cardoso:2024mrw}. However, the kind of perturbations considered in these studies are special and it seems unlikely that they can arise in realistic astrophysical situations.  As mentioned earlier, there is recent work aimed at putting QNMs on a mathematically rigorous footing \cite{gajic2024quasinormalmodeskerrspacetimes,stucker2024quasinormalmodeskerrblack} that also addresses the issue of stability. Therefore, it is likely that our understanding of the issue of stability and more generally, mathematical control on QNMs, would significantly improve in the near future.\vskip0.05cm
2. It would be of interest to relate the multipole moments and fluxes in the ringdown regime with properties of the progenitor BHs.  In this way, one could \emph{predict} the amplitudes $C_{320}$ and $C_{22n}$ of Eq. (\ref{eq:ansatz-modemix}) from the progenitor properties, assuming the overtones are present in the physical predictions of full nonlinear GR.  This is evidently related to Open Issue OI-8 of section \ref{s4.3}, and the related task of { extending the dynamical vector field  $X^a$ (that transports the $Y_{\ell, m}$'s)} from the remnant DHS back to the DHSs associated with the progenitors. 
Furthermore, since the behavior of say $I_{22}(t)$ correlates very well with the phase of the post-merger GW signal \cite{Chen_2022}, a prediction of these amplitudes would enable us to also predict the GW signal in the ringdown regime.  Previously, the amplitudes of the QNMs in the postmerger GW signal at $\scrip$ have been related directly to the inspiral regime { using phenomenological considerations} \cite{Kamaretsos:2011um,Borhanian:2019kxt}.  Since the QLHs provide a link between the progenitor and remnant BHs, the same could be achieved { from first principles} using the QLHs as intermediaries. This provides a potential avenue to exploit the mathematical framework of QLHs in gravitational waveform modeling to address the question of which QNMs should be included and what their amplitudes should be.  \vskip0.05cm
3. There are also recent studies that point to the importance of going beyond linear  approximation \cite{khera2024quadraticmodecouplingsrotating}. A closer investigation of the $\ell=2$ component of the shear discussed above reveals the presence of yet more interesting features \cite{Khera:2023oyf}. As in Sec.~\ref{s4.3} one considers axisymmetric head-on collisions of non-spinning BHs of different mass ratios, with/without an initial boost and fits the exact NR shear on $\DH$ with  \emph{quadratic} combinations of QNMs including overtones, i.e. one includes possible sums of the complex QNM frequencies and \emph{product of the amplitudes}. A closer examination of the $\ell=2$ shear data $\sigma_2$ uncovers the fact that the inclusion of the quadratic combination $(2,0)\times(2,0)$ of modes improves the fit considerably.  Similar results hold for $\sigma_4$ and $\sigma_6$ shear components. This is closely analogous to the results of \cite{Mitman:2022qdl,Cheung:2023vki} where quadratic ringdown modes are found to provide a better fit to the waveform at $\scrip$. Thus, again the dynamics of the remnant $\DH$ appears to be closely correlated to that at $\scrip$. { Since fits improve in both regimes  by including non-linearities of GR, this analysis implies that the fits that use only linear combinations of QNMs modes may not have fundamental physically significance.}

\subsection{Gravitational Wave Tomography}
\label{s6.2}

In section \ref{s6.1} we discussed three sets of NR simulations of BBH mergers that showed a striking similarity between the post merger dynamics at null infinity and at the DHS of the remnant. In this section we will present an approach to probe this relation systematically using the observables introduced in section \ref{s5.3}: the supermomentum charges $\mathtt{Q}_{\ell,\,m}$ and their duals $\mathtt{Q}^\star_{\ell,\,m,}$ at $\scrip$, and the multipoles $\texttt{I}_{\ell,\,m}$ and $\texttt{S}_{\ell,\, m}$ of the remnant DHS $\DH$ \cite{aa-banff,aa-nk}. 

The overall situation can be summarized as follows. On physical grounds one expects the space-time metric $g_{ab}$ to settle down to a stationary state in the distant future, represented by a Kerr metric $g_{ab}^{\rm kerr}$, and the DHS $\DH$ to approach the EH of $g_{ab}^{\rm kerr}$. Therefore, there is a tame basic assumption underlying all investigations of approach to equilibrium : \emph{$\DH$ becomes asymptotically null in the distant future.} This expectation is borne out in all NR simulations to date. Now, for BBH mergers, in the foreseeable future we will be able to investigate space-time dynamics in quantitative detail only through numerical solutions. Therefore, the analysis is tailored to NR rather than mathematical GR (but allows for continued improvements in the accuracy of these simulations). Thus, one makes certain assumptions that refers to the set of NR simulations of interest. \vskip0.15cm
\noindent \emph{Assumption 1 on NR Precision:} The BBH space-time $(M, g_{ab})$ admits hyperboloidal slices $\Sigma_1$ and $\Sigma_0$ 
such that in the space-time region $R_1$ to the future of $\Sigma_1$,\, $g_{ab}$ is indistinguishable from a perturbed Kerr metric $g_{ab}^{\rm kerr} + h_{ab}$,\, and, in the region $R_0$ to the future of $\Sigma_0$\,  $g_{ab}$ is indistinguishable from $g_{ab}^{\rm kerr}$, \emph{both within the accuracy of the given set of NR simulations}. (See Fig.~\ref{fig:tomography}.)\vskip0.15cm

Note that one has direct access only to the physical metric $g_{ab}$ (through NR) and knows only that $g_{ab}$ is well approximated by \emph{some} perturbed Kerr metric. 
A priori one does not know what $g_{ab}^{\rm kerr}$ is. 
Nonetheless, equations governing dynamics of physical fields at $\DH$ are such one can locate a portion $\DH_0$ of $\DH$ in the distant future that can be identified with a portion $\IHo$ of the EH/IH of the metric $g_{ab}^{\rm kerr}$ near $i^+$. A portion of $\DH$ immediately to the past of $\DH_0$ can then be identified as a perturbed IHS, i.e, $\IHo$ perturbed by $h_{ab}$. Then, the dynamics of multipole moments $\texttt{I}_{\ell,\,m}$ and $\texttt{S}_{\ell,\, m}$ of this perturbed $\IHo$ are shown to be directly correlated to the rates of change of $\texttt{Q}_{\ell,\,m}$ and $\mathtt{Q}^\star_{\ell,\,m,}$ at $\scrip$. \vskip0.15cm

Given this setting, the framework provides:
\vskip0.1cm
\indent (i) \emph{Benchmarks for validity of the perturbative regime:} By comparing the rates of change of DHS-multipoles defined using the exact metric $g_{ab}$ with those induced by $g_{ab}^{\rm kerr} + h_{ab}$, one can identify regions in the post-merger space-time where linear perturbations around the Kerr remnant suffice, and regions where they do not; and,\vskip0.05cm

(ii) \emph{Gravitational wave tomography:} Knowing just the waveform at $\scrip$, one can create a movie showing the evolution of the shape and spin structure of $\DH$. This is striking because a causal signal cannot propagate either from $\DH$ to $\scrip$ or vice-versa. Nonetheless, thanks to Einstein's equations, there are definite correlations between the two.
\vskip0.1cm

This sub-section is divided into two parts. The first focuses just on $\DH$ and shows that in the distant future it can be regarded as a perturbed IHS within NR accuracy. In the second part we consider the space-time region $R_1$ to the future of the hyperboloid slice $\Sigma_1$ in which $g_{ab}$ is well approximated by $g_{ab}^{\rm kerr} + h_{ab}$ and discuss correlations between observables on the inner boundary $\DH_1$ of  $R_1$ and the outer boundary $\scrip_1$. \bigskip 
\begin{figure}[]
\begin{center} 
  \vskip-1cm
  \includegraphics[width=0.6\columnwidth,angle=0]{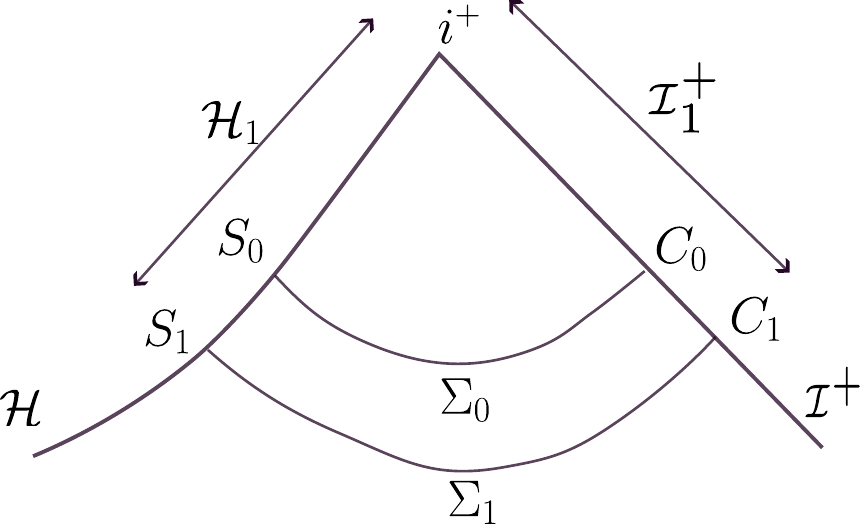}
\caption{\footnotesize{\emph{Post-merger space-time.}\, Hyperboloidal slices $\Sigma_0$ and $\Sigma_1$ join MTSs $S_0$ and $S_1$ on the DHS $\DH$ of the remnant to cross-sections $C_0$ and $C_1$ of $\scrip$. It is assumed that, \emph{within a given NR accuracy}, in the portion $R_1$ of space-time to the future of $\Sigma_1$, the physical metric $g_{ab}$ is indistinguishable from a perturbed Kerr metric $g_{ab}^{\rm kerr}+ h_{ab}$, and in the portion $R_0$ to the future of $\Sigma_0$,\, $g_{ab}$ is indistinguishable from  $g_{ab}^{\rm kerr}$. The portion $\DH_1$ of the DHS $\DH$ to the future of $S_1$ is indistinguishable from a perturbed IHs within NR accuracy. Time dependence of the multipole moments on $\DH_1$ is directly correlated with the time dependence of charges on the portion $\scrip_1$ of $\scrip$ to the future of the cross-section $C_1$. Within NR accuracy, the multipoles and charges are time independent to the future of $S_0$ and $C_0$, being equal to those of the Kerr remnant.
}} 
\label{fig:tomography}
\end{center}
\end{figure}

\vskip0.2cm
\centerline{\it Introducing a fiducial IHS structure on the DHS of the remnant}
\label{s6.2.1}
\vskip0.2cm

Consider then just the DHS $\DH$ of the remnant, formed as the result of a BBH merger. Let us label the MTSs of $\DH$ by $v= {\rm const}$ and label points of each MTS $S$ by $(\theta,\, \varphi)$ (that are dragged from distant future by the transport vector field $X^a$ of section \ref{s3.3}). Then, the intrinsic metric on $\DH$ is given by 
\be \label{DHq} q_{ab}\, \= \,\t{q}_{ab} + 4b^2 D_a v  D_b v \ee
 where as before $\t{q}_{ab}$ denotes the 2-metric on each MTS { and $(2b)^{-1}$ is the norm of 
$D_a v$ defined by $q_{ab}$.} Denote by $V^a$ the vector field that is orthogonal to the MTSs that preserves the foliation by MTSs and for which $v$ is an affine parameter:
\be \quad V^a D_a v\, \=\, 1, \quad {\rm so\,\,\, that} \quad  q_{ab} V^a V^b\,\=\,  4b^2, \,\quad {\rm whence}\quad V^a\, \=\, 2b\, \h{r}^a\,, \ee
where, as in section \ref{s3.2}, $\h{r}^a$ is the unit normal to MTSs within $\DH$. Since by assumption $\DH$ becomes asymptotically null in the distant future, $V^a$ becomes null in the limit, whence $b \to 0$ and $q_{ab} \to \t{q}_{ab}$, i.e., $q_{ab}$  becomes degenerate.  The two null normals to the MTSs that have well-defined limits in the asymptotic future are given by \cite{Ashtekar:2003hk,Ashtekar:2013qta}
\be \label{V} \hskip-0.9cm\lb^a\, =\,  b (\h\tau^a + \h{r}^a),\quad {\rm{and}}\quad  \bar{n}^a\, =\, \f{1}{2b}  (\h\tau^a - \h{r}^a),\quad {\rm so\,\,\, that} \quad V^a = \bar\ell^a - 2 b^2 \bar{n}^a\, , \ee
where, as before, $\h{\tau}^a$ is the unit (time-like) normal to $\DH$. Thus, in the asymptotic future, $V^a$ tends to that null normal $\lb^a$ to MTSs whose expansion vanishes. The function $b$ provides a measure of the difference between $\DH$  and its final equilibrium state \cite{Ashtekar:2003hk,Booth:2003ji}. Interestingly, it turns out that in the exact equations governing the dynamics on $\DH$, time derivatives of fields of interest have terms that can be grouped into those which are of order $O(b^0), O(b)$ and $O(b^2)$; there are no higher order terms! As we  explain below, if one ignores terms of $O(b)$, the DHS becomes a non-expanding horizon segment (NEHS), and there is a unique rescaling of its null normal that makes it a IHS which can be identified as the event horizon of $g_{ab}^{\rm kerr}$ of \emph{Assumption on 1}. If one keeps $O(b)$-terms but ignores $O(b^2)$ terms, one is in the linear regime in which $\DH$ can be regarded as a perturbed IHS (corresponding to $g_{ab}^{\rm kerr} + h_{ab}$). If NR simulations become sufficiently accurate to correctly monitor terms of $O(b^2)$, they would capture the full non-linear dynamics at the DHS $\DH$.

Let us explore this structure. Denote by $S_0$ the MTS to the future of which terms of  $O(b)\,  \o=\,  0$, where  $\,\o=\,$ stands for ``equals, in the numerical precision of the set of simulations under consideration".  Denote by $S_1$ the MTS (to the past of $S_0$) such that all terms $O(b^2)\,  \o=\,  0$ to the future of $S_1$. (See Fig.~\ref{fig:tomography}.) Denote by $\DH_1$ the portion of the DH to the future of $S_1$, and by $H_0$ the portion to the future of $S_0$ (so  $\DH_0$ is contained in $\DH_1$).  As noted in section \ref{s6.1}, in NR simulations, typically  $S_1$ corresponds to $\sim 10 M$ after the merger but the analysis we summarize below does \emph{not} assume this. It allows $S_1$ to occur anywhere after the common DHS is formed. This strategy leads to checks on how close to the merger one can push $S_1$, i.e., on the viability of the perturbative regime in the strong field region represented by $\DH$. 

The basic fields on the DHS $\DH$ are the 2-metric $\tq_{ab}$ on the MTSs, the rotational 1-form $\bomega_a$, the surface gravity $\kappa_{{}_{V}}$, and the shear $\sigma^{(\lb)}_{ab}$ of the null normal $\lb^a$:
\ba \label{DHfields}\bomega_a &=& - \t{q}_{a}^{b}\, \nb_c\, \nabla_b \lb^c = \t{q}_{a}^{b}\, \nb_c\, \nabla_b V^c
\qquad  \kappa_{{}_{V}} \,=\, - V^a \nb_c \nabla_a \lb^c = - V^a \nb_c \nabla_a V^c\nonumber\\
\sigma_{ab}^{(\lb)}\, &=&\, {\rm TF}\, \tq_a^c\,\tq_b^d\,\nabla_c \lb_d\, , \ea
where \,\,TF\,\, stands for ``trace-free part of". On the region $\IH_1$ of $\IH$ where terms of $O(b^2)$ are NR-negligible, Eqs. (\ref{DHq}), (\ref{V}) and (\ref{DHfields}) imply
\be \label{DHevo} V_a V^a \, \o=\, 0, \qquad q_{ab} \, \o= \, \tq_{ab} \qquad 
\Lie_V \tq_{ab}\, \o= \, 2 \sigma^{(\lb)}_{ab}\,. \ee 
Furthermore, the infinitesimal version of the balance law (\ref{2ndlaw1}), written in terms of $\lb^a$ (rather than $k^a$), implies that $\sigma_{ab}^{(\lb)}\,=\, O(b)$.  Therefore, on the future portion $\DH_0$ of $\DH$ where terms of $O(b)$ are NR-negligible, we have $\Lie_V \tq_{ab}\, \o= \, 0$. \emph{Thus, within the presumed NR accuracy, $(\DH_0,\,V^a, q_{ab} )$ is an NEHS.} Now, on a generic NEHS, one can introduce the structure of an IHS by appropriately rescaling its null normal and, furthermore, the IHS structure is unique, up to a \emph{constant} rescaling of the null normal \cite{Ashtekar:2001jb}. This small freedom can be eliminated by fixing the value of its surface gravity. 

One can find the desired IH structure on the part of $\DH$ to the \emph{future of} $S_0$ in two steps. First, one rescales $V^a$ by a positive function $\bar{f}$ so that the surface gravity of $\bar{V}^a := \bar{f} V^a$ is a constant $\kappa_\circ$, equal to the surface gravity of the Kerr remnant. This rescaling transforms the NEHS to a WIHS with desired surface-gravity. As noted in section \ref{s2.2}, this rescaling can always be performed but is not unique. But if the WIH is generic, a further rescaling of $\bar{V}^a$ picks out a unique null normal 
$\ello^a =f \bar{V}^a$ that transforms the WIHS to an IHS. For this, $f$ has to satisfy: 
\be \label{toIH} 
\mathcal{M}\, f\, := \,\,\big[\tD^2 + 2 \bomega^a \tD_a + \bomega_a \bomega_a - \textstyle{\f{1}{2}} \tilde{\mathcal{R}}\big]\, f\,\, \o=\,\, \kappa_\circ\, \Theta_{(\nb)}\,. \ee
Here, as before, $\tq_{ab}$ and $\tD$ are the metric and the derivative operators on MTSs and $\Theta_{(\nb)}$ is the expansion of the null, inward pointing normal $\nb^a$ to the MTSs. The operator $\mathcal{M}$ is the adjoint of the stability operator discussed in section \ref{s4.1} (since the term $|\sigma_{(\lb)}|^2$ is of $O(b^2)$ and $G_{ab} =0$). \emph{Assumption 1} implies that the operator $\mathcal{M}$ is invertible (i.e., we are in the generic case). Therefore the required rescaling function $f$ is given by ${\mathcal{M}}^{-1} \big(\kappa_\circ\, \Theta_{(\nb)}\big)$.  To summarize, by rescaling of the vector field $\bar{V}^a$ on the DHS by the solution $f$ to (\ref{toIH}), the portion $\DH_0$ of $\DH$ can be endowed with the structure of an IHS $\IHo$. The construction refers only to the fields available on the physical DHS and the given NR accuracy. Yet it deposits on $\DH_0$ the structure  $(\IHo,\, \ello)$ of the EH/IHS of $g_{ab}^{\rm kerr}$ within NR accuracy.

On the portion of $\DH_1$ to the past of the MTS $S_0$, one needs to keep terms of $O(b)$ because only terms $O(b^2)$ are negligible within NR accuracy on this portion. As noted above, these $O(b)$ terms represent first order perturbations. The simplest strategy is to extend the background IH structure $(\IHo, \ello^a)$ to all of $\DH_1$,  using again $f$ given by (\ref{toIH}), and regard the $O(b)$ terms as making it a \emph{perturbed} IHS \cite{Ashtekar:2021kqj}. The degenerate metric $\tq^{\circ}_{ab}$  and the rotational 1-form $\omega_a^{\circ}$ of the \emph{background IHS} are obtained by Lie-dragging $\tq_{ab}$ and $\bomega_a^{(\ello)}$ along $V^a$ from the future portion $\DH_0$ of $\DH_1$ to all of $\DH_1$. Consequently, the multipoles of the extended background IHS $\IHo$ are all time independent, and are the same as those of the Kerr remnant.

Note that equations $q_{ab} V^aV^b\,\, \,\o=\,\, q_{ab} \ello^a \ello^b\,\, \o= \,\,0$\, and\, $q_{ab} \,\o=\, \tq_{ab}$\, still hold on all of $\DH_1$. But now the fields $\t{q}_{ab}$ and $\bar\omega_a^{(\ello)}$ are \emph{time-dependent} on the portion of $\DH_1$ that lies to the past of $S_0$ because of $O(b)$ terms. (Hence they differ from $\tq^{\circ}_{ab}$ and $\bomega^{\circ}_{a}$ that are time independent everywhere on $\DH_1$, and refer to the background IHS $\IHo$). This time dependence is dictated by the first order perturbation encoded in $\sigma^{(\ello)}_{ab}$: 
\be \Lie_{\ello} \tq_{ab} \,\o=\, 2 \sigma_{ab}^{(\ello)},\qquad {\rm and} \qquad \Lie_{\ello}  \bar\omega_a^{(\ello)}\, \o=\, - \tilde{D}_{\circ}^b \sigma^{(\ello)}_{ab}\, , \ee
where $O(b^2)$ terms are dropped. In this approximation, the area of MTSs does not change in time since $\sigma_{ab}^{(\ello)}$ is traceless. But the scalar curvature $\tq_{ab}$ and the curl of $\bar\omega_a^{(\ello)}$ are time-\emph{dependent}. Since these fields serve as `seeds' in the definition of shape and spin multipoles $\mathtt{I}_{\ell,m}$ and $\mathtt{S}_{\ell,m}$, these moments are also time-dependent, just as one would expect given that we have a \emph{perturbed} IHS to the past of $S_0$. In the next subsection we will see that the time-rate of change of these $\mathtt{I}_{\ell,m}$ and $\mathtt{S}_{\ell,m}$ is directly correlated with that of the BMS charges $\mathtt{Q}_{\ell,m}$ and $\mathtt{Q}^\star_{\ell,m}$, providing us with the gravitational wave tomography. 

Finally, note that in this analysis we have ignored terms of $O(b^2)$. Therefore these multipoles of the perturbed IHS $\IHo$ will differ from those on the DHS $\DH_1$ obtained from full $\t{q}_{ab}$ and $\bomega_a$ \emph{without} discarding the $O(b^2)$ terms. Now, within the accuracy of  NR simulations, one would likely find that terms $O(b^2)$ are not negligible as one slides to the past along $\DH$ and moves close to the merger. Then, in that regime the time dependence of multipoles of the perturbed IHS would fail to capture the dynamics of the exact DHS multipoles. This failure would provide a benchmark for the cut-off, beyond which the horizon dynamics cannot be well-approximated by the linear theory. This potential failure of the linear regime can be detected numerically simply by computing $b^2$ or, equivalently, by monitoring the area of the MTSs of $\DH$, starting from the merger. In the regime in which $b^2$, or, the change in area of MTSs is not negligible numerically, space-time geometry at the horizon cannot be approximated by a perturbed Kerr metric and non-linearities of full GR come into play already within the accuracy of NR simulations. 

\vskip0.2cm
\centerline{\it Correlations between observables on $\DH_1$ and $\scrip_1$.}
\label{s6.2.2}
\vskip0.2cm
So far the focus was only on the DHS $\DH$. Let us now consider the space-time region $R_1$ to the future of $\Sigma_1$ in which one has $g_{ab} \,\o=\, g_{ab}^{\rm kerr} + h_{ab}$ for some linearized perturbation $h_{ab}$ on a background $g_{ab}^{\rm kerr}$ characterizing the remnant. As we just discussed, within any given NR accuracy, one can obtain the multipoles $\mathtt{I}_{\ell,m}$ and $\mathtt{S}_{\ell,m}$ of the background Kerr metric, as well as those of the perturbed Kerr metric using just the intrinsic geometry of the \emph{physical} DHS, without having to determine $g_{ab}^{\rm kerr}$ explicitly. At the outer boundary $\scrip_1$, one can invariantly drag the charges $\mathtt{Q}_{\ell,m}$ and $\mathtt{Q}^\star_{\ell,m}$ of the Kerr remnant from distant future (i.e., $i^+$) to any cross-section of $\scrip_1$. These would be time-independent. NR also provides the exact charges on $\scrip$ which, by \emph{Assumption 1}, agree with the time independent, background charges to the future of the cross-section $C_0$, and with charges of the perturbed Kerr metric to the future of $C_1$. Since these boundary observables are insensitive to diffeomorphisms in the \emph{interior} of the region $R_1$ (i.e., that are identity on the two boundaries), the final results on tomography are also insensitive to the gauge freedom in the choice the Kerr background $g_{ab}^{\rm kerr}$ in the interior. 

Time dependence of the multipoles of $\DH_1$ and of charges of at $\scrip_1$ is governed { by}  $h_{ab}$. Hence, to relate the two one needs to know the relation between $h_{ab}\mid_{\DH_1}$ and $h_{ab}\mid_{\scrip_1}$, both of which are determined by the Cauchy data on the hyperboloid $\Sigma_1$. Now, there is a strong indication from NR simulations that this data is quite special: Within NR accuracy, it is represented by a linear combination of quasi-normal modes (QNMs) of the remnant. Thus, we are led to make a second assumption: 
\vskip0.15cm
\noindent \emph{Assumption 2 on NR Precision:} \emph{Within the accuracy of the given set of NR simulations,} in the region $R_1$ the perturbation $h_{ab}$ is indistinguishable from some linear combination of QNMs of the remnant, i.e.,  of $g_{ab}^{\rm kerr}$.
\vskip0.1cm

\noindent While this assumption permeates the literature on waveforms and tests of general relativity, it is yet to be justified from fundamental considerations. Indeed, from a mathematical GR perspective, the mechanism that propels the perturbation $h_{ab}$ to this form at late times is still  a mystery. We will comment on this issue in Remark 1 below. For now we note that, given this assumption, there is a direct path to relate $h_{ab}\mid_{\DH_1}$ and $h_{ab}\mid_{\scrip_1}$ because there is sizable literature on relating the amplitudes of QNMs at $\scrip$ with those at $\IHo$  using hyperboloidal slices (see, e.g.,  \cite{Berti_2009,Ansorg_2016,Ripley_2022,khera2024quadraticmodecouplingsrotating,Frauendiener:2022bkj}). This relation then determines the correlations between observables at $\scrip_1$ and $\DH_1$, leading to tomography. 

We will discuss this step in detail using the special case in which the remnant is a Schwarzschild BH where the argument can be given succinctly. As discussed in section \ref{s4.3}, a head-on collision of non-spinning BHs results in a Schwarzschild remnant, and it can arise even in a merger of spinning BHs if the initial data is fine tuned astutely \cite{Haley}. The basic argument for a general Kerr remnant is the same but it is obscured by certain technical complications that are irrelevant for the underlying conceptual structure. From any perturbations $h_{ab}$ on a Schwarzschild background $g_{ab}^{\rm sch}$, one can construct the even and odd parity gauge invariant functions\, $\Psi^{(\pm)}_{\ell,m}$\, \'a la Cunningham-Price and Zerelli-Moncrief.  One sets  
\be \Psi_{\ell,m}\,=\, \Psi^{(+)}_{\ell, m}\, +\,i\, \Psi^{(-)}_{\ell, m}, \quad {\rm and} \quad 
f(\ell) = (\ell-1)\ell (\ell+1) (\ell+2)\, . \ee
Then, interestingly, the time evolution of perturbed multipoles on $\DH_1$ has been shown to be given by
\be \label{dotmultipoles} 2\,GM\, \big(\dot{\mathtt{I}}_{\ell,m} + i\, \dot{\mathtt{S}}_{\ell,m}\big)\, = \, f(\ell)\, \dot{\Psi}\mid_{\DH_1} (\vo)\, ,\ee
while those of the charges at $\scrip_1$ was shown to be given by
\be \label{dotcharges}\big(\dot{\mathtt{Q}}_{\ell,m} + i\, \dot{\mathtt{Q}^\star}_{\ell,m}\big)(u)\, =\, f(\ell)\, \dot{\Psi}\mid_{\scrip_1} (u). \ee
Note that the \emph{same} gauge invariant variable appears on the right hand sides of the two equations! But it is evaluated at $\DH_1$ in (\ref{dotmultipoles}) and at $\scrip_1$ in (\ref{dotcharges}). Now, given the waveform $h_{ab}$ that is a linear combination of quasi-normal modes, for any $\ell,m$ we have
\be \hskip-1.5cm \Psi_{\ell,m}\mid_{\DH_1} (u) = \sum_n\, A_{n,\ell,m}\, e^{iu\,\omega_{n,\ell,m}} \quad {\rm and}\quad 
\Psi_{\ell,m}\mid_{\scrip_1} (\vo) = \sum_n \t{A}_{n,\ell,m}\, e^{i \vo\,\omega_{n,\ell,m}} \ee
for some constants $A_{n,\ell,m}$ and $\t{A}_{n,\ell,m}$ representing the amplitudes. (Here $u$ is the affine parameter of the time translation Killing field at $\scrip_1$ and $\vo$ that of $\lo^a$ at $\DH_1$.)
Furthermore, there are scripts available that provide the constants $C_{n,\ell,m}$ relating the amplitude at $\scrip_1$ to that at $\DH_1$. Therefore, given the waveform at $\scrip_1$, Eqs. (\ref{dotmultipoles})  and (\ref{dotcharges})
determine the time-dependence of multipoles of $\DH_1$, and provide a movie of the evolution of the horizon geometry. As already remarked, this is possible even though no causal signal can pass between $\DH_1$ and $\scrip$ because Einstein's equations provide the constants $C_{n,\ell,m}$ that are necessary to calculate the amplitudes at the horizon from those at $\scrip$.

To summarize, while it is widely expected that, in a BBH merger, the space-time metric $g_{ab}$ in the distant future would be well-approximated by a perturbed Kerr metric $g_{ab}^{\rm kerr} + h_{ab}$, it is difficult to make this statement precise especially in the strong field region near the horizon. Equations governing fields on the DHS $\DH$ of the remnant facilitate this task because they provide a convenient function $b$ to discuss passage to equilibrium. Within NR precision, one can calculate the rate of change of multipoles of $\DH$ that characterizes the dynamical phase of the exact DHS as well as the rate of change of multipoles induced by terms $O(b)$ that represent the dynamics governed by just by the first order perturbation. Their disagreement near merger will signal the breakdown of the linear approximation at the horizon already within the accuracy of NR simulations. Identical considerations hold for the time dependence of charges at $\scrip$. In the domain of validity of the linear approximation, Einstein's equations imply that  there is  strong correlation between the time dependence of charges $({\mathtt{Q}}_{\ell,m},\, {\mathtt{Q}^\star}_{\ell,m})$ at $\scrip_1$ and multipoles $({\mathtt{I}}_{\ell,m},\,\, {\mathtt{S}}_{\ell,m})$ of $\DH_1$. This surprising correlation enables one to determine the evolution of shape and spin structure of $\DH_1$ knowing only the waveform at $\scrip_1$. \vskip0.1cm
 \goodbreak
 
\emph{Remarks}
\vskip0.1cm

1. In the discussion of tomography, one needs to set up a correspondence between the retarded time $u$ at $\scrip_1$ and the affine parameter  $\vo$ of $\ello^a$ at $\DH_1$. In the case when the remnant is a Schwarzschild BH, one can fix $u$ and `slide' the coordinate $\vo$ to obtain the best match between $(\dot{\mathtt{I}}_{\ell,m}\,+\,i\, {\mathtt{S}}_{\ell,m})$ and $ (\dot{\mathtt{Q}}_{\ell,m}\,+\,i\, {\mathtt{Q}^\star}_{\ell,m})$, for say $\ell= m=2$, and then deduce the time dependence of other multipoles knowing only the waveform at $\scrip_1$. In the case when the remnant is a Kerr BH, one has to take into account the angular velocity of horizon as well because the null generator $\ello^a$ of $\IHo$ would be a linear combination of the asymptotic time translation and rotation \cite{Chen_2022}. \vskip0.05cm
2. As discussed in section \ref{s6.1}, there is considerable recent literature showing that the waveform at $\scrip$ obtained by solving full Einstein's equations using NR can be approximated extremely well by suitable linear combinations of quasi-normal modes, especially when overtones are included. And the same holds for the time dependence of multipoles of the DHS $\DH$. But there is a controversy as to whether there is genuine physics in these findings or if they are simply artifacts of fitting the NR results using the freedom in the choice of amplitudes of a large number of QNMs, including overtones. The framework summarized in this subsection provides appropriate tools to analyze this issue. First, as discussed at the end of the first part, difference between exact GR and first order perturbation theory is captured in the terms of $O(b^2)$ in dynamical equations at $\DH$. Therefore, one can simply check if these terms are genuinely negligible within the accuracy of the set of NR simulations used. For example, the area of MTSs is constant in time in the  first order perturbation theory. Therefore if a simulation were to detect time dependence in the area and also find that the evolution of the exact multipoles can be recovered by a linear combination of QNMs, one would conclude that the match is only a fit, devoid of a deeper physical significance. 

Secondly, so far matching has been done \emph{independently} for the waveform at $\scrip$, and for the time dependence of multipoles at $\DH$. As discussed in the second part of this subsection, once a match is made at $\scrip$, then the candidate linear combination of QNMs is determined. Then one has to use the \emph{same} linear combination at $\DH$ using the relation between the amplitude of individual QNMs at $\scrip$ and $\IHo$, provided by linearized Einstein equations. This calculation would provide an important restriction on the space-time region in which $g_{ab}$ is indistinguishable from $g_{ab}^{\rm kerr}+ h_{ab}$ within the accuracy of the given simulation. Finally, it may well happen that the linear approximation is valid in the asymptotic regime $\scrip$ over a certain range of retarded time $u$, but not on the corresponding range of $v_{\circ}$ at $\DH$. This would signal that the domain of validity of the QNM approximation in the strong field regime is more restricted than that at in the weak field regime. Such checks will sharpen our understanding thereby providing clear cut guidance to address the current controversy. Note, however, that this analysis is carried out \emph{assuming} GR. Therefore, it would not be directly useful to discussions  of tests of GR using QNMs.
\vskip0.05cm
3. Within the approximation we worked, $\scrip_1$ and $\DH_1$ are null surfaces. Therefore one might think of solving a characteristic initial value problem for linearized fields $h_{ab}$ by specifying data on these two intersecting null surfaces and evolving backward in time. If so, one would be able to specify data freely on the two null surfaces. How can there be correlations between observables on $\scrip_1$ and $\DH_1$ then?  Indeed, if one were to consider a \emph{finite} portion of $\DH_1$, say between $S_1$ and $S_0$, and replace $\scrip_1$ with a null hypersurface $\mathcal{N}$ generated by past directed null geodesics emanating from $S_0$ (as in Fig.~\ref{fig:nearhorizon}), then the argument would hold and one would be able to specify data freely on that portion of $\DH_1$ and on $\mathcal{N}$. However, it does not hold with $DH_1$ and $\scrip_1$ because of the singularity at $i^+$. Indeed, in the mathematical GR literature such correlations have been found without recourse to any approximations: In the case of Reissner-Nordstrom BHs one can use them to distinguish between extremal and non-extremal BHs knowing only the wave forms of perturbations at 
$\scrip$ \cite{sa-banff,aretakis2024observationalsignatureextremalblack}. The correlations we summarized in this section are much more detailed and also of more direct physical interest. But  they were obtained under approximations tailored to NR, while the mathematical results on Reissner-Nordstrom BHs are exact.
\vskip0.05cm
4. One knows from general considerations involving Green's functions that there are three types of solutions $h_{ab}$ to linearized Einstein's equations on a Kerr background: Linear combinations of the `direct part' (consisting of normal modes that are oscillatory), quasi-normal modes (that are damped sinusoidal), and tails (that have a power-law falloff)  \cite{Leaver1985,Leaver1986JMP,Leaver1986PRD,Kokkotas:1999bd,Berti_2009}. However, as remarked earlier, in the NR simulations of BBH mergers, one finds only linear combinations of QNMs at late times. This finding is in complete agreement with the initial intuition of S. Chandrasekhar about the similarity between the ringing of a bell and BH perturbation theory. But it also implies that, somehow, the BBH evolution funnels dynamics inducing very special initial data on an hyperboloid slice --such as $\Sigma_1$ of Fig.~\ref{fig:tomography}-- soon after the merger.  Why are direct part and tails effectively absent in these data? A recent investigation tailored to quasi-local horizons suggests that the absence of the direct part may be intimately tied with the `no incoming radiation' condition at $\scrim$ \cite{metidieri2025blackholetomographyunveiling}. This appears to be a promising direction to pursue because it suggests that the origin of this feature lies in a \emph{very} global issue; perhaps this is why the puzzle has persisted so long. A detailed investigation of the viability of this suggestion would be of considerable interest. 

The absence of tails in NR simulations is perhaps even more puzzling because  mathematical investigations have established that in the approach to $i^+$ either along $\DH$ or $\scrip$, only the tail terms survive asymptotically. The general belief is that one does not find tails in the NR simulations because tails become relevant only at \emph{very} late times when their amplitude exceeds that of the exponentially damped QNMs. While this is likely to be the case, the puzzle still remains.  General initial data on a hyperboloid slice such as $\Sigma_1$, close to the merger, would be a linear combination of QNMs and tail terms (ignoring the direct part). If their amplitudes were comparable, then the tail terms would in fact dominate soon after the merger. Somehow the amplitudes of tails and QNMs seem to become comparable only at a very late time --on a hyperboloid slice such as $\Sigma_0$-- so that, if one were to evolve backward in time starting from $\Sigma_0$ to $\Sigma_1$, the QNM amplitude would grow exponentially, dominating the amplitude of the tail terms that would have only a power law growth. Thus, we are led to an interesting puzzle:
\vskip0.1cm
\texttt{Open Issue 10 (OI-10)} Why are the amplitudes comparable only at a late time such as $\Sigma_0$ rather than at earlier time $\Sigma_1$ soon after the merger? Is this feature also tied to a very global boundary condition such as `no incoming radiation' at $\scrim$?
\vskip0.1cm

\subsection{Tidally Perturbed Quasi-Local-Horizons and Love numbers}
\label{s6.3}

The deformation of a star placed in an external field depends quantitatively on the composition of the star. In Newtonian gravity, investigations on this subject go back to the work of Love \cite{AEHLove} in 1909, who investigated the effect of tides on the shape of Earth, and Chandrasekhar \cite{10.1093/mnras/93.6.462,1933MNRAS..93..449C} in the 1930s who investigated the properties of a tidally distorted star with a polytropic Equation of State (EoS) and a binary companion. Like planets and stars, the shape of a BH is modified by an external tidal field.  This is evident in numerical simulations of BBH mergers where we see explicitly the time-varying coordinate shape of the apparent horizon  during the inspiral and merger. As discussed in sections \ref{s2.4} and \ref{s3.3}, shape multipole moments provide an invariant characterization of such deformations. Therefore, multipoles have been used to quantify the tidal deformations of the horizon geometry due to the binary companion. When the progenitors are well separated, the effect of the binary companion on a given BH can be encoded in a \emph{static} (or slowly time-varying) external tidal field at that BH. Therefore the perturbed multipoles are also time-independent. This discussion will complement that of the last two sub-sections, where the focus was on time dependence of perturbations.

Let us begin by illustrating the underlying ideas using tidally distorted stars. The subject is of considerable interest in its own right to gravitational wave astronomy for the following reason. While most of the detections of merging compact objects have turned out to be binary BH systems, there \emph{are} some events with double neutron stars (NSs), and potentially NS-BH binaries as well. A merger of two neutron stars was first observed in 2017 and the event, referred to as GW170817, was observed simultaneously by electromagnetic observatories as well \cite{LIGOScientific:2017vwq,LIGOScientific:2017ync,LIGOScientific:2017zic}. An analysis of the GW data from this event led to constraints on neutron star radii, as well as the nuclear EoS at high densities that are not accessible in any ground based collider experiment (see e.g. \cite{LIGOScientific:2018cki,De:2018uhw,Capano:2019eae}). These constraints rely critically on the tidal deformabilty of neutron stars, and furthermore, these calculations potentially allow us to
distinguish between neutron stars and (low mass) BHs.

The most widely used formalism to discuss tidal deformabilitesdeformability in general relativity \cite{Hinderer_2008,Flanagan_2008} can be briefly summarized as follows. Consider a static, spherically symmetric star immersed in an external tidal field. In the most common application, this external field is due to a binary companion which is sufficiently far away so that its time dependence can be ignored. The equilibrium spherically symmetric solution is obtained by solving the Tolman-Oppenheimer-Volkov (TOV) equations with a particular EoS (for example, polytropic models are considered in \cite{Hinderer_2008,Flanagan_2008}).  Static quadrupolar perturbations to this solution are constructed using the formalism developed in \cite{1967ApJ...149..591T}. The exterior solution obtained in this manner is then matched to an exterior asymptotic expansion in the weak-field regime as follows. 

Let $(t,x^i)$ be an asymptotic coordinate system, as in \cite{Thorne:1984mz,PhysRevD.58.124031}, with $x^i$ (nearly) Cartesian coordinates.  In the weak field Newtonian limit, the quantity $-(1+g_{tt})/2$ would be the Newtonian potential, $g_{tt}$ being the time-time metric component.  For the unperturbed star, which is assumed to be spherically symmetric, it is just $-M/r$ with $M$ the mass and $r=\sqrt{\sum_i (x_i)^2}$. If the star is placed in an external static tidal field described by a constant symmetric tracefree tensor $\mathcal{E}_{ij}$, we have the following asymptotic expansion:
    \begin{eqnarray}
      \label{eq:gtt-expansion}
      -\frac{(1+g_{tt})}{2} &=& -\frac{M}{r} - \frac{3Q_{ij}}{2r^3}\left(n_in_j-\frac{1}{3}\delta_{ij} \right) + \mathcal{O}(r^{-4}) \nonumber \\&+& \frac{1}{2}\mathcal{E}_{ij}x^ix^j + \mathcal{O}(r^3)\, ,
    \end{eqnarray}
where $n^i := \f{x^i}{r}$ and $Q_{ij}$ is the quadrupole moment induced by the external field $\mathcal{E}_{ij}$. The coordinates are assumed to be `mass-centered' so that the dipole moment vanishes, and we have organized the  expansion so that the first row contains terms that decay at infinity and the second row those that diverge at infinity due to the presence of an external field. It is also assumed here that one is sufficiently far away from the star so that this asymptotic expansion is meaningful. The perturbed TOV solution mentioned in the previous paragraph is matched  with Eq.~(\ref{eq:gtt-expansion}) to yield $Q_{ij}$ and $\mathcal{E}_{ij}$.  Then the induced quadrupole moment $Q_{ij}$ turns out to be related linearly to the external tidal field at leading order:
    \begin{equation}
      \label{eq:lambda-def}
      Q_{ij} = -\lambda\mathcal{E}_{ij}\,.
    \end{equation}
The constant $\lambda$ can be regarded as the tidal Love number (although the general convention is to consider a dimensionless version thereof), and it depends on the EoS. In a binary system where the two companions have Love numbers $(\lambda_1,\lambda_2)$, the emitted GW signal contains an imprint of a particular combination of these Love numbers and can thus, in principle, be measured. The analogs of these Love numbers vanish for BHs, whence they provide a way to distinguish low-mass BHs from neutron stars \cite{Brown:2021seh}. 

Let us now return to BHs. Then Eq.~(\ref{eq:gtt-expansion}) no longer applies in a straightforward manner since one does not expect the Cartesian coordinates $x^i$ to extend to the horizon, or to the center of the back hole. In general, it is not clear if there is a simple analog of the Newtonian potential in the strong field region, nor how the external tidal field $\mathcal{E}_{ij}$ should be defined.  Effective Field Theory approaches do offer an alternative route (see e.g. \cite{Damour:2009wj,Ivanov:2022hlo,Bini:2014zxa,Goldberger:2004jt}), but one can also attempt a first-principle calculation of the tidally distorted spacetime geometry. Specifically, since $Q_{ij}$ in Eq.~(\ref{eq:gtt-expansion}) is read off from the symptomatic expansion, it represents the `field quadrupole moment' while, as we saw in section \ref{s3.4}, the (leading order) deformation of the horizon itself is enclosed in its shape/mass quadrupole moment which leaves its imprints in the space-time geometry in a large neighborhood of the horizon. Therefore, it is more natural to discuss horizon deformability using the horizon multipoles. Is there then a well-defined notion of a `horizon Love number'?

Such Love numbers have been previously discussed in the literature as ``surficial'' Love numbers \cite{Damour:2009va,Landry:2014jka}.  The framework of QLHs allows for a reformulation and generalization of this concept wherein the horizon geometry is taken to be the starting point.  As we shall see, the local spacetime geometry can be reconstructed from the horizon geometry { as a Taylor series}, and the formalism naturally includes also Love numbers of the ``magnetic'' type arising from the angular momentum multipole moments (previous results on surficial Love numbers 
are restricted to surficial mass multipole moments). This issue has been investigated by modeling a BH in a static external field as non-extremal IHS $\IH$ with perturbed, but time independent, horizon multipole moments \cite{RibesMetidieri:2024tpk}. Thus, in this set up one ignores infalling fluxes across $\Delta$ thereby focusing only  on the `Coulombic aspects' of the gravitational field near $\Delta$. Therefore, the perturbed $\IH$ is again an IHS with the same mass, and one can choose the gauge freedom so that the perturbed IHS $\IH$ is the same as the unperturbed IHS as a manifold, but now with perturbed fields. The unperturbed IHS can be taken to be that of a Kerr space-time but for simplicity of presentation we will assume it to be a Schwarzschild IHS, so that its only non-vanishing multipole is the mass monopole $\mathtt{M}_0$.

In the space of solutions to vacuum Einstein's equations, the perturbation moves one to a neighboring isolated horizon $\IH$ with the same mass $\mathtt{M}_0$ but with perturbed higher multipole moments $\mathtt{M}_\ell^\prime$ (with $\ell\geq 2$) and $\mathtt{J}_\ell^\prime$ (with $\ell \geq 1$). ($\mathtt{M}_1$ and $\mathtt{J}_0$ vanish even after perturbation because they vanish on every IHS \cite{Ashtekar:2004gp}.) The perturbed source multipole moments $\mathtt{M}_\ell^\prime$ and $\mathtt{J}_\ell^\prime$ are assumed to be small and to depend linearly on a smallness parameter which, in the case of a binary companion, would be determined at leading order by the mass of the companion and the distance to it.  

Next, recall from section \ref{s2.4} that multipole moments of an IHS $\IH$ are determined by the moments of the Newman-Penrose Weyl tensor component $\Psi_2$. Thus, tidal perturbations of, say, a Schwarzschild BH can be conveniently represented by adding  to the spherically symmetric $\Psi_2$ of the background $\IH$, higher $(\ell\geq 2)$ multipole moments. Can this be done leaving other Weyl components unperturbed? The answer turns out to be in the negative. 

It is easiest to explain this feature using the Weyl tensor components in a convenient Newman-Penrose null tetrad. One can construct it on $\IH$ as follows. The first tetrad vector is the null normal $\ell^a$ to $\IH$. To introduce the other three, it is convenient to introduce a foliation of $\IH$ by 2-spheres. Such a foliation can be introduced in a covariant manner using properties of the rotational 1-form $\t\omega_a$ of the perturbed IHS \cite{RibesMetidieri:2024tpk}. With this foliation at hand, one chooses the ingoing, past-directed null vector $-n^a$ orthogonal to the foliation and satisfying $\ell \cdot n = -1$, and a complex null dyad $(m^a,\bar{m}^a)$ on $\IH$, with $m^a$ tangent to the leaves of this foliation and satisfying $m\cdot\bar{m}=1$. This covariant construction provides a suitable null tetrad $(\ell^a,\,n^a,\,m^a,\,\bar{m}^a)$ on $\IH$. In the Newman-Penrose notation, the components of the Weyl tensor are given by:
\begin{eqnarray} 
\Psi_0 &=& C_{abcd}\ell^am^b\ell^cm^d\,,\qquad
   \Psi_1 = C_{abcd}\ell^am^b\ell^cn^d\,,\\
   \Psi_2 &=& C_{abcd} \ell^a m^b \bar{m}^c n^d\,,\\
   \Psi_3 &=& C_{abcd}\ell^an^b\bar{m}^cn^d\,,\qquad
   \Psi_4 = C_{abcd}\bar{m}^an^b\bar{m}^cn^d\,.
 \end{eqnarray}
\begin{figure}
  \begin{center} \vskip-2cm
    \includegraphics[width=0.6\columnwidth]{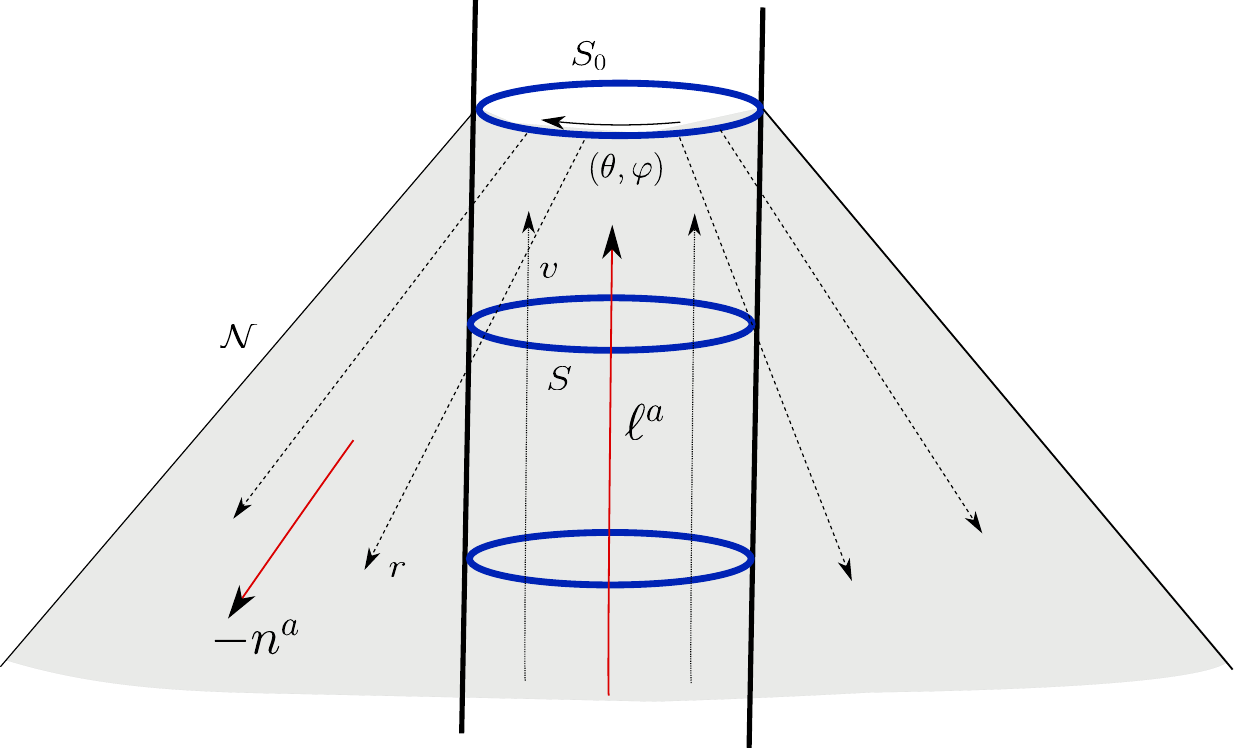}
    \caption{\footnotesize{\emph{Setting for the characteristic initial value problem with interesting null surfaces:}} The past directed null vector field $-n^a$ orthogonal to the preferred family of cross-sections is extended to a neighborhood of $\IH$ using geodesic equations. The neighborhood is large because these geodesics don't cross in the perturbed space-time. Coordinates $(v,\theta,\varphi)$ at $\IH$ and the vector fields $\ell^a, \, m^a, \bar{m}^a$ on $\IH$ are parallel transported to this neighborhood along $-n^a$. Together with the  affine parameter $r$ along the geodesics, we have coordinates $(v,r,\theta,\varphi)$ and a null tetrad  $(\ell^a,\, n^a,\, m^a, \bar{m}^a)$ in this neighborhood. The characteristic initial value problem for constructing the near horizon geometry uses two null surfaces: the perturbed IHS $\IH$  and $\mathcal{N}$, generated by $-n^a$, starting from a cross-section $S_0$ in the preferred foliation of $\IH$. The spacetime metric is constructed starting with suitable data on $\Delta$, $\mathcal{N}$ and $S_0$.}
    \label{fig:nearhorizon}
  \end{center}
\end{figure}

On a general NEHS, $\Psi_0$ and $\Psi_1$ vanish identically but the remaining six components of the Weyl tensor do not \cite{Ashtekar:2000hw}. For the specific class of Kerr-Newman horizons, it turns out that one can choose a null tetrad in the full space-time such that we have in addition: $\Psi_3=\Psi_4=0$, so that  $\Psi_2$ is the only non-vanishing Weyl tensor component. (More generally such spacetimes are said to be of `type D' in the Petrov classification). However, for a \emph{generic} quadrupolar perturbation of $\Psi_2$, vacuum equations and Bianchi identities imply that $\Psi_3$ and $\Psi_4$ would be necessarily non-zero (and the space-time would no longer be type D \cite{Lewandowski:2000nh,Dobkowski-Rylko:2018usb}). In particular then, although our unperturbed metric in a neighborhood of $\IH$ is Schwarzschild, after a generic quadrupolar perturbation of $\Psi_2$, $\IH$ ceases to be an IHS of a Kerr space-time.

The next step is to move away from $\IH$ and construct the perturbed  spacetime geometry in a (large) neighborhood of the horizon by using the characteristic initial-value formulation of general relativity
(see e.g. \cite{Sachs:1962zzb,Friedrich:1983vi,rendall,Chrusciel:2012ap}) in which one of the two intersecting null surfaces is the IHS $\IH$ \cite{Lewandowski:1999zs,Krishnan:2012bt,Lewandowski:2018khe,Scholtz:2017ttf,Flandera:2024awl,RibesMetidieri:2024tpk}.
This procedure is analogous to the Bondi-Sachs construction near $\scrip$ \cite{sachs,bondi1962mgj} and yields a Newman-Penrose tetrad $(\ell^a,\, n^a,\, m^a\, \bar{m}^a)$ and coordinates $(v,r,\theta,\varphi)$ in a large neighborhood of the $\IH$. Here $v$ is is an extension of the affine parameter along $\ell^a$ at the $\IH$ (such that constant $v$ lines join $\IH$ and $\scrim$); $r$ is a null coordinate directed towards $\scrim$. See Fig.~\ref{fig:nearhorizon} for details.

For the characteristic initial value problem, we need a second null surface $\mathcal{N}$ that intersects $\IH$ in a cross section $S_0$ of the foliation. If $v\, \=\, v_0$ on $S_0$, we can take $\mathcal{N}$ to be the null surface $v=v_0$ in the neighborhood under consideration (see Fig.~\ref{fig:nearhorizon}). The initial data consists of $\Psi_0$ on $\IH$ (which vanishes identically),\, $\Psi_4$ on $\mathcal{N}$ and $\Psi_2$ on their intersection $S_0$. Because the perturbation caused by the  distant companion BH is assumed to be time-independent, one seeks linearized solutions on the Schwarzschild background for which $\partial/\partial v$ is a Killing field. Therefore, the required $\Psi_4$ on $\mathcal{N}$ is determined by the $\Psi_4$ of the perturbed metric, evaluated on $S_0$. Thus, the task reduces to specifying perturbed $\Psi_2$ and $\Psi_4$ on $S_0$. Let us begin with $\Psi_2$ and write it as 
\be \Psi_2 = \Psi_2^{\rm (sch)} + \h\Psi_2 \, .\ee
Because of time-independence of the perturbed solution, all three fields in the equation are $v$ independent and the restrictions $\Psi_2^{\IH}$ and $\h\Psi_2^{\IH}$ of $\Psi_2$ and $\h\Psi_2$ to $\IH$ depends only on angles $(\theta,\, \varphi)$. We will illustrate the main results using the simplest case: axisymmetric, purely mass-quadrupolar perturbation 
\be \label{quadrupole} \widehat{\Psi}_2^{\Delta} = -\frac{K}{MR^4}\,\,\mathtt{M}_2\,\,Y_{20}(\theta)\, ,\ee 
where $K$ is a dimensionless constant, $\mathtt{M}_2$ captures the perturbation parameter, $R$ is the  area-radius and $M$ the mass of the unperturbed horizon.  In an astrophysical scenario with a binary companion, $\mathtt{M}_2$ will be determined by $R$ and the distance to the companion and its mass. 

Starting with this perturbed data at the horizon, one can integrate the linearized Einstein field equations outwards assuming time-independence. This procedure has been carried out in \cite{RibesMetidieri:2024tpk} (in a somewhat more general context that allows the background to be a Kerr BH with small angular momentum). Here we summarize the main results, focusing mostly on the bits relevant for gravitational wave astronomy. In the solution to the linearized Einstein's equations, for large $r$ the Weyl tensor component $\Psi_2$ turns out to have the asymptotic the form 
\begin{equation}
  \label{eq:psi2result}
  \Psi_2 = -\frac{M}{r^3} + C^{(\infty)}\, Y_{20}(\theta,\varphi)\,.
\end{equation}
Here $C^{(\infty)}$ is a non-vanishing constant, so that $\Psi_2$ tends to a constant when $r\rightarrow\infty$ in any given direction.  Thus, as anticipated, the perturbed spacetime is  not asymptotically flat at $r\rightarrow\infty$.

This result can be interpreted using the discussion following Eq.~(\ref{eq:gtt-expansion}) on tidal deformations of regular stars. The analog of $g_{tt}$ of Eq.~(\ref{eq:gtt-expansion}) is now $g_{vv}$ since $\partial/\partial v$ is the Killing vector.  As shown in \cite{RibesMetidieri:2024tpk}, by solving for linearized Einstein's equations, the perturbed metric can be shown to have the form
\begin{equation} \label{eq:gvv-expansion}
  -\frac{(1+g_{vv})}{2} = -\frac{M}{r} + \widehat{U} \qquad \textrm{with} \qquad  \hat{\Psi}_2 = \frac{1}{2}\,\partial_r^2\,\widehat{U}\,  \, ,
\end{equation}
and $\widehat{U}$ is explicitly known.  This equation provides the
link between the Newtonian potential\, $-\f{(1 + g_{tt})}{2}$ \,of
Eq.~(\ref{eq:gtt-expansion}) used in the analysis of tidal
deformability of stars and the Weyl component $\Psi_2$ of the
perturbed BH space-time. A comparison of the right had sides of (\ref{eq:gtt-expansion}) and  (\ref{eq:gvv-expansion}) suggests that $\Psi_2$ should have the form %
\begin{equation}
  \Psi_2 = -\frac{M}{r^3}\, +\, k\,\frac{Q_2\, Y_{20}}{r^5}\,\,\, +\,\mathcal{O}\left(\frac{1}{r^6}\right)\, +\, C^{(\infty)}\, Y_{20} +\mathcal{O}(r)\,,
\end{equation}
where $Q_2 Y_{20}$ is the \emph{field} quadrupole moment, i.e. the quadrupole moment deduced from asymptotic properties of $\Psi_2$,\, and $k$ is a dimensionless constant.

Comparing with the actual form (\ref{eq:psi2result}) of $\Psi_2$, it is evident that this quadrupolar term is absent so that $Q_2 =0$  even though we began by introducing a quadrupolar deformation (\ref{quadrupole}) of the IHS geometry. Thus, while the  horizon quadrupole moment, encoded in $\mathtt{M}_2$,  is not zero, it is quite striking that Einstein's  equations imply that the quadrupole moment $Q_2$ of the field, evaluated at a large distance from the horizon vanishes.  A similar phenomenon was observed in \cite{Gurlebeck:2015xpa} in the context of the Schwarzschild BHs distorted by static, axisymmetric external sources: Under the assumption that the full space-time is asymptotically flat, all \emph{field multipoles} were shown to vanish in spite of the tidal distortions produced by external sources.  Furthermore, this was a non-perturbative statement obtained using properties of Weyl metrics. However, because of reliance on Weyl metrics, this analysis encompassed only axisymmetric situations. The method summarized above is free from this restriction, has been shown to extend to Kerr BHs with small angular momentum, and sheds light on how perturbations of the IHS geometry affect space-time metric in a large neighborhood thereof.

The fact that the field quadrupole moment vanishes (even though the horizon quadrupole does not) has an important consequence which we now discuss. Note first that the asymptotic value $C^{(\infty)}$ is related to the external field that causes the perturbation.  Moreover, since one is  working within linear perturbation theory and the only small parameter in the problem is $\mathtt{M}_2$, both $C^{(\infty)}$ and $Q_2$ are linearly related to each other (since they are both linearly determined by $\mathtt{M}_2$).  Therefore, one  can introduce a constant $\lambda_2^{(\infty)}$ and write this linear relation as
\begin{equation}
  Q_2 =  \lambda_2^{(\infty)}\,\, C^{(\infty)}\,,
  \end{equation}
so that the external field can be said to induce a non-vanishing field multipole moment with $\lambda_2^{(\infty)}$as the (field) Love number.  The vanishing of $Q_2$ implies that the Love number $\lambda_2^{(\infty)}$ vanishes for a BH.  The same conclusion holds also for the other multipoles as long as the BH is slowly spinning \cite{RibesMetidieri:2024tpk} (and, furthermore, there does not seem to be any obvious obstacle for using the method for larger spins).

The fact that BHs have vanishing Love numbers is already known
in the literature (see e.g. \cite{Poisson:2021yau,Binnington:2009bb,LeTiec:2020bos,LeTiec:2020spy,Charalambous:2021mea,Pani:2015hfa,Pani:2018inf,Damour:2009vw,Damour:2009wj,Gurlebeck:2015xpa}) but the method we summarized has some qualitatively new elements. First, it begins with deformations induced on the geometry of the horizon of a BH by its companion in the binary and these can be cast in invariant terms, thanks to the availability of the horizon multipoles. Second, using the characteristic initial value problem and field equations, it  provides a detailed link between source/horizon multipole moments and space-time geometry far away from the horizon.
Third, the method provides  a natural avenue to define Love numbers for the magnetic (i.e. spin) source multipole moments. Finally, using the same logic that led us to the field Love numbers, it is natural to define the source/horizon Love number $\lambda_2^{(\Delta)}$ via:
\begin{equation}
  \mathtt{M}_2 = \lambda_2^\Delta\, C^{(\infty)}\,.
\end{equation}
and this Love number is \emph{non-zero}. 

Remark: There is an extensive literature on extreme mass ratio systems
where a small compact object orbits around a massive BH (see
e.g. \cite{Hartle:1974gy,Hartle:1973zz,OSullivan:2014ywd,OSullivan:2015lni}). In
this work small amounts of infalling radiation at the horizon are
allowed, a phenomenon referred to as ``tidal heating''.  It should be
possible to extend the above analysis to incorporate tidal heating by
considering a perturbed isolated horizon as in section \ref{s6.2}.

\vskip0.1cm
\texttt{Open Issue 11 (OI-11)} In an extreme mass ratio system, the
small compact object ``maps'' the spacetime around the massive black
hole and the emitted GW signal can potentially be used to test the
Kerr nature of the massive BH by measuring its field multipole
moments \cite{Ryan:1995wh,Ryan:1997hg}.  As the discussion in this
sub-section shows, in the perturbative regime, the spacetime geometry
is determined by the source multipole moments.  Therefore we are led to
ask: Can the spacetime mapping problem be reformulated in terms of the
source multipole
moments? 

\section{Discussion}
\label{s7}

{\smash{S. Chandrasekhar's celebrated monograph on black holes \cite{Chandrasekhar:1985kt} begins with a preamble:}}
\begin{quote}
``In my entire scientific life, extending over forty-five years, the most shattering experience
has been the realization that an exact solution of Einstein's equations of general relativity, discovered by the New Zealand mathematician Roy Kerr, provides the absolute exact representation of untold numbers of massive black holes that populate the universe.''
\end{quote}
It then goes on to develop in detail perturbation theory around Kerr BHs, which continues to provide the underlying framework for much of the work at the interface of GWs and astrophysics even today. As we discussed in section \ref{s6}, in this analysis it suffices to represent the BH by the event horizon (EH) of a Kerr background, or a perturbed version thereof. Therefore, a great deal of research on astrophysical applications of BHs is focused on perturbations of the Kerr solution.

However, as we discussed in sections \ref{s3} and \ref{s4}, this description cannot hope to capture regimes in which the physics is dominated by dynamical processes close to BHs, triggered by the full nonlinearities of GR. And this is precisely the regime that can provide us with insights into the nature of gravity in extreme conditions. Indeed, so far the only tests of the fully non-linear content of GR have come from GW observations of mergers of compact bodies, most of them BHs. NR simulations have established unambiguously that space-time geometry close to the merger is \emph{very} different from that of a perturbed Kerr metric. In these simulations, the notion of an EH is totally inadequate to characterize BHs because the EH can only be constructed as an after thought, once the simulation is completed and one knows the metric all the way to the infinite future. Therefore they use quasi local horizons (QLHs) instead of EHs, modeling BHs by dynamical horizon segments (DHSs), and by perturbed isolated horizon segments (IHSs) for progenitors when they are far away, as well as for the remnant when it is close to reaching its final equilibrium state.

Of course there are numerous issues in fundamental physics, unrelated to mergers, in which BHs play a key and fascinating role. These were first investigated in the 1970s and 1980s using EHs because, in space-times that admit a future asymptotic boundary $\scrip$, they provide a sharp, coordinate invariant characterization of a BH. Indeed, since EH is the boundary of the region from which causal signal cannot propagate to $\scrip$, it succinctly captures the intuitive idea that \emph{black holes are black}. Furthermore, fundamental properties of EHs, enshrined in the laws of EH mechanics, led to unforeseen connections with thermodynamics that, in turn, opened new vistas on the quantum nature of relativistic gravity. Perhaps the most striking among these results is the area theorem: under physical conditions that are realized in classical GR, the area of an EH cannot decrease. 

However, as pointed out in sections \ref{s1}- \ref{s3}, EHs also suffer from serious drawbacks especially because of their { teleological nature. This limitation} makes them unsuitable in the analysis of certain fundamental conceptual issues as well. As noted in section \ref{s1}, these limitations have been pointed out in the literature in the context of cosmic censorship in classical GR and in investigations of the end point of BH evaporation in quantum gravity. In both these cases, one cannot make a priori assumptions on the nature of space-time geometry all the way to the infinite future since the primary task is to \emph{determine} what happens in the distant future! Furthermore, in quantum gravity  BHs are no longer black because of Hawking radiation, and the raison d'\^etre of EHs --capturing the idea that back holes are black-- simply disappears. More concretely, if singularities of classical BHs are resolved due to quantum effects, the resulting quantum extension of space-time would not have EHs either!

As we discussed in sections \ref{s1} - \ref{s3}, QLHSs are free of teleological features; their properties are dictated by physics in their immediate vicinity. As a consequence:\\
$-$\, There are no QLHSs (with $S^2\times R$ topology) in flat regions of space-times;}\\ 
$-$\, Their definitions and properties make no reference to the existence of $\scrip$; and, \\
$-$\, They have been used in the analysis of conceptual issues mentioned above, particularly \indent in the discussion of the BH evaporation process. \\
Also, the laws of BH mechanics extend to QLHs. Moreover, not only is the area monotonic on DHSs, but the Gauss-Codazzi equations provide a detailed balance law that relates the change in area with ``local happenings" at the DHS in any metric theory of gravity. For general relativity, these equations imply that the change in the area is governed by the flux of (an invariantly defined) energy across the DHS. Such a relation cannot hold for EHs since they can grow in flat regions of space-time where there are no fluxes.

Consequently, QLHs have drawn a great deal of attention over the last two decades. As noted in section \ref{s1} there are several detailed reviews that cover earlier phases of this research. Therefore, in this article, we focused on the more recent advances --and on tools used in these investigations-- covering mathematical GR, NR and GW astrophysics. (QLHs also have important applications to quantum gravity. We chose not to include them to keep the review focused. As mentioned in section \ref{s1}, these applications are discussed in recent review articles.)
\vskip0.1cm
On mathematical GR side, we discussed\,  \emph{(i)} The laws of BH mechanics for isolated and dynamical BHs; (sections \ref{s2.3} and \ref{s3.2})\, \emph{(ii)} Multipole moments that invariantly characterize the shapes and spin structures of the QLHs representing them (sections \ref{s2.4} and \ref{s3.3});\, \emph{(iii)}\, An interplay between space-time isometries and DHSs that bears out physical expectations, thereby strengthening the case for the notion of  DHSs; (section \ref{s3.1})\, \emph{(iv)}\, The stability operator for marginally trapped surfaces (MTSs), insights it has yielded on horizon dynamics, and its various applications (section \ref{s4.1} and \ref{s4.2}); \,  \emph{(v)}\, The surprising fact that null infinity $\scrip$  is in fact a weakly isolated horizon (WIH) but in the conformally completed (rather than physical) space-time (section \ref{s5}).   \vskip0.1cm
On the numerical side, we presented examples of new insights that have emerged from simulations:\, \emph{(i)}\, Subtleties in the process by which MTSs of progenitor BHs merge, which is far more sophisticated than the ``pair of pants" picture normally invoked to describe it. Yet, { in the axisymmetric situations that have been analyzed in detail,} the QLHs of progenitors and that of the remnant are joined continuously, without any abrupt jumps that were expected from over-reliance on apparent horizons associated with Cauchy slices (section \ref{s4.3});\, \emph{(ii)} How a perturbed BH (such as the remnant formed in a BBH merger) approaches equilibrium (sections \ref{s6.1} and \ref{s6.2}); and\, \emph{(iii)}\, The time dependence of waveform at $\scrip$,\, as well as of multipoles and shear at the DHS can be calculated using NR, without any reference to (the waveform at infinity, or to the) perturbation theory. Yet, it is is governed by the quasi-normal frequencies of the remnant, roughly $10\,M$ after the merger (section \ref{s6.1}), which corresponds just to $\lesssim 1$ cycle in wave forms observed in BBH events! \vskip0.1cm
Finally, for gravitational wave astronomy, QLHs have been used to bring out: \,\emph{(i)}\, The interplay
between horizon dynamics and GW observations in the ring-down regime (\ref{s6.1} and \ref{s6.2}), and\, 
\emph{(ii)}\, The fact that vanishing of the standard Love numbers for BHs does not imply that their horizons remain undistorted by tidal deformations. For, the standard Love numbers refer to the field multipoles while the tidal distortions refer to horizon (or source) multipoles and  linearized Einstein's equations imply that even though the Horizon Love numbers are non-zero, the standard Love numbers vanish (section \ref{s6.3}). \vskip0.1cm
This discussion brought out some surprising synergies: (i) Results from mathematical GR and NR complement as well as re-enforce one another, thereby providing a deeper understanding of non-linearities of GR (section \ref{s4}); (ii) WIHSs and null infinity share a number of common geometrical features and the same basic dynamical equation, and yet carry \emph{very} different physics because complementary terms in the dynamical equation vanish in the two cases (section \ref{s5}); and,\, (iii) Although gravitational wave tomography and tidal distortions seem completely unrelated at first, they both stem from the theory of multipole moments of perturbed IHSs; it is just that in the first case perturbations are time dependent (namely, QNMs) while in the second, they are time independent (section \ref{s6.2} and \ref{s6.3}). \vskip0.1cm

Throughout the review we listed some interesting open issues in the hope that they will stimulate further research in all three communities. We will conclude with a discussion of a deeper, foundational open issue; indeed, an elephant in the room! The question is: 
\begin{itemize} 
\item What is the best way to characterize the BH boundary in fully dynamical situations? Or, put more simply, \emph{what exactly is a dynamical BH?}
\end{itemize}
To bring out the subtleties involved, let us begin with the simplest situation, and then add generality step by step. If one were assured that a BH is truly isolated in a space-time that admits $\scrip$, then one could characterize its boundary by the EH, which would also be a Killing horizon for a time translation Killing field. But even in this simplest case, the description would be adequate only if one were to ignore quantum effects. For, the BH would become dynamical due to quantum radiation and, as mentioned above, we would no longer be able to say if the EH exists, or where it lies, until we obtain a space-time description of the evaporation process to the infinite future. 

To streamline the discussion, then, let us ignore quantum effects and ask for the best description of the BH boundary within classical gravity. In dynamical situations, there is no time-translation Killing field, whence we cannot use Killing horizons to find BH boundaries. {Furthermore, while EHs do exist, as the middle panel of Fig.~1 shows, even in simplest situation such as the Vaidya null fluid collapse (or collapse of certain Vlasov fluids \cite{kehle2024extremalblackholeformation}) it is physically inappropriate to use them as depicting BH boundaries because they have segments (with $S^2\times R$ topology) that lie in and even expand in flat regions of space-time. Moreover, the analysis  \cite{Booth_2010} of more complicated spherically symmetric situations depicted in the right Panel of Fig.~1 shows that in non-dynamical regions of space-time where an EH segment can be regarded as a BH boundary, it is extremely close to an IHS which can be located quasi-locally (in contradistinction to the EH segment in question).}

Such examples and our previous discussion suggests that QLHs may provide a better depiction of the BH boundary especially { in situations that cannot be adequately handled by perturbative treatments}. This line of inquiry has been pursued since the 1990s \cite{Hayward:1993wb,Hayward:1994yy,Eardley:1997hk}, and using increasingly sophisticated ideas and tools in recent years \cite{BenDov:2006vw,Bengtsson:2008jr,Bengtsson:2010tj}. One begins with the notion of a \emph{trapped region} as the connected subset of space-time through each point of which there passes a trapped 2-sphere (i.e., 2-spheres for which the expansion of both null normals is negative). The first idea was to represent a BH as an inextendible, connected trapped region $\mathcal{T}$ and take its boundary $\partial\mathcal{T}$ to represent the surface of a BH. However, it turns out that $\partial\mathcal{T}$ does not have properties one would expect a physically acceptable BH boundary to have \cite{Bengtsson:2008jr,Bengtsson:2010tj}. In particular, it has non-local features reminiscent of EHs; for example it can penetrate flat regions of space-time.  Furthermore, this can occur already in spherically symmetric space-times.  It is rather astonishing that the notion of a \emph{dynamical} BH is so subtle already in the spherically symmetric case!

This realization led to a more sophisticated notion of the \emph{core} $\mathcal{C}$ of a trapped region $\mathcal{T}$ to characterize a BH \cite{Bengtsson_2011}: In essence, $\mathcal{C}$ is a minimal closed subset that has to be removed from space-time to get rid of all trapped surfaces contained in $\mathcal{T}$. With this rather subtle notion, one can establish an interesting relation between cores and QLHSs in spherically symmetric space-times. Let us consider spherical cores and spherical QLHS, i.e, 4-d regions $\mathcal{C}_{(s)}$ and 3-d surfaces $\mathfrak{H}_{(s)}$ that are left invariant by the action of rotations. With these notions at hand, one can show that if the space-time admits trapped surfaces, then it admits a unique spherically symmetric core $\mathcal{C}_{(s)}$ (namely, $(r < 2m(v,r))$ in the standard $(r, v, \theta,\varphi)$  chart), and a unique spherically symmetric QLH $\mathfrak{H}_{(s)}$ (given by $r=2m(v,r)$). Furthermore, the QLH  $\mathfrak{H}_{(s)}$ is in fact the boundary of the core $\mathcal{C}_{(s)}$. The QLHs of Figs.~1 and 2 --unions of DHSs and IHSs-- are these unique spherically symmetric QLHs and they constitute the boundaries of the unique spherical cores in these space-times \cite{Bengtsson_2011,senovilla2012stabilityoperatormotscore}. To summarize, restriction to spherical structures enables one to single out, in a single stroke, a preferred core representing the BH, and a preferred  QLHS which, furthermore, serves as the BH boundary: $\partial \mathcal{C}_{(s)}= \mathfrak{H}_{(s)}$. There is a suggestion \cite{senovilla2012stabilityoperatormotscore} that, by using properties of the stability operator, it may be possible to extend the argument to general (i.e., non-spherical) space-times to select a preferred core and a preferred QLH, such that one again has  $\partial \mathcal{C}= \mathfrak{H}$.  Thus, we are led to a key open issue:
\vskip0.1cm
\texttt{Open Issue 12 (OI-12)} Can this line of reasoning be completed? If so, central foundational questions, \emph{``What is a BH?"} and \emph{``What is its boundary?'',} would be answered in a single stroke. One could first attempt to extend the arguments in \cite{Bengtsson_2011,senovilla2012stabilityoperatormotscore} to axisymmetric space-times, following a parallel line of reasoning.
\vskip0.1cm

From foundational viewpoint, an affirmative answer would be a most satisfying outcome.  From a `practical' viewpoint however, this answer would not be directly useful, for example in the analysis of the BBH problem. For, as we recalled in section \ref{s3.4}, there can be many DHSs associated with any given evolutionary phase of the progenitors and the remnant. These DHSs `interweave' one another, and to select the preferred one, one would have to first isolate that core $\mathcal{C}_\circ$ whose boundary would single out the preferred DHS $\DH_\circ$. Then $\DH_\circ$ could then be taken to represent the surface of the BH. In practice this would be very difficult because the definition of a core is subtle and it is not clear how NR simulations would find the preferred $\mathcal{C}_\circ$. Therefore, at present perhaps the best strategy would be to continue with the current practice of finding DHSs $\DH$ by first locating MTSs on Cauchy slices used in any given NR evolution, and stacking them. Of course if two simulations use quite different choices of Cauchy slices, they may end up with different DHSs, $\DH_1$ and $\DH_2$ describing the same dynamical phase of a BH. The key point is that the detailed equations discussed in section \ref{s3} will apply to \emph{both} $\DH_1$ and $\DH_2$.\, Thus, for example, the balance law (\ref{2ndlaw1}) can be used to check the accuracy in each of these simulations. Similarly, tomography and the checks it provides on the validity of the perturbative regime, discussed in section \ref{s6.2}, will also be available both for $\DH_1$ and $\DH_2$. One just has to be self-consistent, i.e, while extracting physics from a given NR simulation, one has to keep in mind that results refer to the ``black hole surface" $\DH$ determined by that simulation. 

Indeed, one could even adopt an alternate viewpoint to the strategy suggested by cores, outlined above, and say that in the fully dynamical regime in absence of symmetries, the notion of the ``surface of the black hole" is inherently fuzzy: Any of the interweaving DHSs can be taken to represent the surface, each providing a self-consistent and  quantitative description of the dynamical process. From this perspective, the main remaining task is to quantify and better understand the relation between results that emerge from the use two different DHSs in the same region of a given space-time. Since all these DHSs would be on the same physical footing, there could be an underlying group of transformations that relates these results. Unforeseen interesting structures may well emerge. As the BH approaches an equilibrium stage, these DHSs would all asymptote to the same IHS. Then, physical properties --in particular, the multipoles of the remnant-- would be independent of which DHS one uses. Indeed, this is the underlying viewpoint behind the procedure currently used to determine masses and spins of remnants! 

Could this alternate viewpoint be viable at a fundamental level? Could it be that one just has to accept that there is a small but \emph{inherent} ambiguity in the notion of the surface of a generic dynamical BH? After all, there does not seem to be a compelling reason to assert that the boundary must be sharp. This would be a radical departure from the commonly held view that the BH boundary is sharp. While pondering over these alternatives, it is interesting to return to S. Chandrasekhar's thoughts. In spite of the quote on Kerr solutions with which we began, he did not seem to be convinced by the commonly held view { that a BH has a sharp boundary}. In discussions with mathematical relativists at Chicago in the 1970s, he asked a number of times: \emph{Is the surface of a BH paper thin?} To mathematical relativists participating in those discussions the answer seemed clear and in the affirmative, since the surface was taken to be the EH at the time, which is indeed ``paper thin". But perhaps Chandrasekhar was thinking in more physical terms. After all, surface of a star is not sharply defined; it is not ``paper thin". There is some ambiguity, and yet it does not prevent us from making and verifying predictions about physical properties of stars and their dynamics. Could the situation be similar for BHs?

\section*{Acknowledgments}

We are grateful to Lars Andersson, Beatrice Bonga, Ivan Booth, Sukanta Bose, Miguel Campiglia, Sergio Dain, Greg Galloway, Anshu Gupta, Sean Hayward, Scott Hughes, Jose Luis Jaramillo, Neev Khera, Jerzy Lewandowski, Stefano Liberati, Ariadna Ribes Metidieri, Tomasz Pawlowski, Eric Poisson, Daniel Pook-Kolb, Vaishak Prasad, B. Sathyaprakash, Jose Senovilla, Simone Speziale, and Huan Yang for valuable discussions and correspondence. {Comments made by the referees and the feed-back we received from Daniel Paraizo, Mauricio Gamonal, Samarth Khandelwal and Jonathan Shu (who took a reading course based on this review) has improved the final presentation.} This work was supported in part by the Atherton and Eberly funds of Penn State and the Distinguished Visiting Research Chair Program of the Perimeter Institute.

\bigskip\bigskip


\bibliography{Livrev}{}
\bibliographystyle{jhep}

\end{document}